\documentclass[12pt]{article}

\usepackage{hyperref}
\usepackage{amsmath,amsfonts,amssymb}
\hypersetup{dvips,dvipdfm,colorlinks=true,urlcolor=magenta,filecolor=magenta,linktoc=page,citecolor=red,linkcolor=blue,bookmarks=true}
\usepackage{graphicx,epstopdf}
\usepackage{multirow}
\usepackage{xcolor}
\begin{document}
		\title{The classical and quantum implications of the Raychaudhuri Equation in $f(T)$-gravity}
	\author{Madhukrishna Chakraborty \footnote{chakmadhu1997@gmail.com}~~and~~
	Subenoy Chakraborty\footnote{schakraborty.math@gmail.com}
	\\Department of Mathematics, Jadavpur University, Kolkata - 700032, India}
\maketitle
	\begin{abstract}
		The present work deals with the classical and quantum aspects of the Raychaudhuri equation in the framework of $f(T)$-gravity theory. In the background of homogeneous and isotropic Friedmann–Lemaître–Robertson-Walker space-time, the Raychaudhuri equation has been formulated and used to examine the focusing theorem and convergence condition for different choices of $f(T)$. Finally in quantum cosmology, the wave function of the universe has been shown to be the energy eigen function of the time-independent Schrödinger equation of a particle. Also probability measure on the minisuperspace has been examined at zero volume for singularity analysis in the quantum regime. Lastly, the Bohmian trajectory for the present quantum system has been explicitly determined for some particular choices.
	\end{abstract}
	
\small	 Keywords :  Raychaudhuri Equation ; $f(T)$ gravity ; Quantization ; Bohmian trajectories.
\section{Introduction}
Recent cosmological observations namely Cosmic microwave background anisotropies (CMBR)\cite{WMAP:2003elm}, Type Ia Supernova Hubble diagram (SNeIa)\cite{SupernovaSearchTeam:1998fmf}-\cite{SupernovaCosmologyProject:1998vns}, Large Scale Structure (LSS) \cite{SDSS:2003eyi} and Baryon Acoustic Oscillation (BAO) \cite{SDSS:2005xqv} are in favor of accelerated expansion of the universe which is quite challenging to be explained by the standard model of particle physics and general relativity. There are possibly two ways to overcome this difficulty---(i) introduction of exotic matter (known as Dark Energy (DE)) \cite{Amendola:2015ksp} in the framework of Einstein gravity. (ii) Einstein gravity is replaced by some modified gravity theory \cite{Capozziello:2019cav}, \cite{Nojiri:2006ri}. DE \cite{Amendola:2015ksp} is an unknown matter component which has a large negative pressure component. The simplest and observationally favored model of dark energy is the cosmological constant $\Lambda$ (or the vacuum energy) with a constant equation of state parameter $\omega=-1$. Although this scenario agrees with the current astronomical observations, it fails to make the small observational value of the dark energy density consistent with the estimates from quantum field theories: The Cosmological Constant Problem \cite{Nobbenhuis:2006yf}. Currently, the $\Lambda$CDM model has been shown to have an age problem \cite{Yang:2009ae}. Therefore, it is a natural search for alternative possibilities to explain the mystery of dark energy. A quite appealing approach is the modification of General Relativity (GR) giving rise to a plethora of well-known extended gravity theory in literature \cite{Capozziello:2019cav}, \cite{Nojiri:2006ri}. A common generalization of Einstein's gravity is done by replacing the Ricci scalar, `$R$' in the Einstein-Hilbert action by a function of the Ricci scalar $f(R)$ resulting in $f(R)$ modified gravity theory \cite{Guo:2013swa}, \cite{Aditya:2018cmn}, \cite{Cognola:2007zu}, \cite{Elizalde:2010ts}, \cite{Nojiri:2010wj}, \cite{Sotiriou:2008rp}. However, studies of $f(R)$ theories are obstructed by the complication in the fourth order field equations in the metric framework \cite{Sotiriou:2006mu}. Although  Palatini formalism \cite{Gogoi:2021mhi}, \cite{Sotiriou:2005cd} for $f(R)$ theory leads to second order field equations, yet there is difficulty to get both exact and numerical solutions which can be compared with observations in many cases. Recently, another gravity theory known as teleparallel gravity \cite{Cai:2015emx} has attracted the interest of the literature. Einstein proposed such a gravity model with an aim to unify gravity and electromagnetism over Weitzenböck non-Riemannian manifold. Hence the Levi-civita connection is replaced by Weitzenböck connection in Riemann-cartan space-time. Although teleparallel gravity and GR differ conceptually, yet both of them have equivalent dynamics at classical level \cite{Unzicker:2005in}. Analogous to $f(R)$-gravity theory, a generalization to teleparallel gravity \cite{Cai:2015emx} has been done by replacing $T$, the torsion scalar by a  generic function $f(T)$ leading to $f(T)$-gravity theory  \cite{Cai:2015emx}, \cite{Darabi:2019qpz}, \cite{Li:2018ixg}, \cite{Golovnev:2018wbh}, \cite{Aviles:2013nga}, \cite{Bose:2020xdz} and Linder coined the name \cite{Linder:2010py}. In these theories, torsion \cite{Hayashi:1979qx} (instead of curvature) is a driving force for the late time accelerated expansion, and the field equations are of second order. \\

 On the other hand, detection of gravitational waves \cite{LIGOScientific:2017vwq} has proved the acceptance of Einstein's General Theory of Relativity as a universal theory of gravity to describe physical reality \cite{Wald:1984}. Moreover the major ingredient of modern cosmology is the theory of gravity by Einstein \cite{Wald:1984}, \cite{Weinberg:1972kfs}. Starting from big-bang singularity to the present era of evolution is nicely described by Einstein gravity. Nevertheless the `Singularity' of spacetime is the biggest limitation of Einstein gravity. The celebrated Singularity theorems by Hawking and Penrose \cite{Hawking:1973uf}, \cite{Penrose:1964wq}, \cite{Hawking:1970zqf} has proved the generic existence of singularity in Einstein Gravity. The singularity theorems \cite{Hawking:1973uf}, \cite{Penrose:1964wq}, \cite{Hawking:1970zqf} use the notion of geodesic incompleteness as a stand-in for the presence of infinite curvature. It is interesting to note that the Raychaudhuri equation (RE) \cite{Raychaudhuri:1953yv}-\cite{Dadhich:2005qr} is the main ingredient behind these singularity theorems. RE dictates the dynamical evolution of the universe, describing the time evolution of the expansion scalar. The RE is a geometric equation and has nothing to do with a gravity theory. Similar to Einstein's equation, the RE becomes a physical equation showing equivalence between geometry and matter when a gravity theory is imposed through the Ricci tensor ($R_{\mu\nu}$) term. In Einstein gravity with the assumption of Strong Energy Condition (SEC) on matter, the Raychaudhuri equation leads to Focusing Theorem which proves the inevitable existence of singularity in Einstein gravity and the corresponding condition on matter is termed as Convergence Condition (CC). The CC hints at the attractive nature of gravity and is responsible for focusing of geodesics. Therefore it is a natural instinct to look for a model (other than Einstein gravity) or rather some probable conditions under which singularity may be avoided. Since in other extended theories of gravity the field equations are different (than those of Einstein gravity), there may arise some possibilities for the avoidance of singularity. This motivates us to formulate the modified Raychaudhuri equation and the corresponding convergence condition in some $f(T)$ gravity models. Also the RE and CC have been extensively studied as a tool to avoid singularity for $f(R)$ gravity in the background of inhomogeneous Friedmann–Lemaître–Robertson-Walker (FLRW) model and some conditions were determined for the avoidance of singularity (for ref. see\cite{Chakraborty:2023ork}).  \\
 
In the present work, $f(T)$ gravity theory in the background of homogeneous and isotropic FLRW space-time has been studied. RE has been formulated in this model and the corresponding CC has been analyzed graphically for mainly two popular choices of $f(T)$ in literature \cite{Wu:2010xk}, \cite{Myrzakulov:2010vz} as a tool to avoid singularity in the respective models. Further, a quantum description of the RE in the present model has been dealt with as another probable approach of avoiding singularity where the solution of the Wheeler-Dewitt (WD) equation may be interpreted as the propagation amplitude
of the congruence of geodesics. Norm of this solution (wave function) can be interpreted
as the probability distribution of the system. If the wave packet so constructed by this
solution is peaked along the classical solution at the early era then the singularity may
be avoided so that the geodesics will never converge. Also, quantum trajectory i.e Bohmian trajectory has been explicitly determined for some particular choices and their behavior has been studied. It is generally speculated that quantum effects which become dominant in the strong gravity regime may alleviate the singularity problem at classical level. Hence in quantum description, canonical quantization and formulation of Bohmian trajectories are used to analyse the classical singularity at quantum level. Thus the paper aims to show two possible pathways (classical and quantum) of avoiding singularity in some $f(T)$ gravity models. There lies the motivation of the work. \\

The plan of the paper is as follows:
Section II gives a brief overview of the RE; Section III deals with the construction of $f(T)$ gravity in FLRW space-time and formulation of RE in it; Section IV discusses the CC in the models under consideration and some probable conditions to avoid  classical singularity; Section V presents a canonical quantization scheme in the framework of the present model as an alternative approach to avoid singularity from quantum RE point of view; Section VI deals with the formulation of Bohmian trajectory for the present quantum system; The paper ends with some concluding remarks in Section VII.
\section{A brief overview of the Raychaudhuri equation}
	Kinematic characteristics of time-like congruence {$u^{\mu}$} are related to the Ricci tensor $R_{\mu\nu}$ via the fundamental result of the Riemannian geometry:
	\begin{equation}\label{eq1}
		R_{\mu\nu}u^{\mu}u^{\nu}=-\dot{\Theta}-\dfrac{\Theta^{2}}{3}-2\sigma^{2}+2\omega^{2}+\dot{u}^{\mu}_{;{~}\mu} {~}.
	\end{equation}
This is the famous RE \cite{Raychaudhuri:1953yv}-\cite{Dadhich:2005qr}, named after Prof. Amal Kumar Raychaudhuri and is the main ingredient for the celebrated Singularity Theorems \cite{Hawking:1973uf}, \cite{Penrose:1964wq}, \cite{Hawking:1970zqf} by Hawking and Penrose. Moreover  it is a fundamental result to study exact solutions of Einstein's equations in general relativity and has much to contribute in modern cosmology. Here, we consider a family of fundamental observers moving along a time-like 4-velocity vector field, say $u_{\mu}$. The symmetric space-like (induced) metric tensor 
\begin{equation}
	\eta_{\mu\nu}=g_{\mu\nu}+u_{\mu}u_{\nu}
\end{equation}
satisfies the orthogonality condition ,${~~~~~~~~~~~~~~~~}u^{\nu}\eta_{\mu\nu}=0$ . 
\\Define,
\begin{equation}
	D_{\mu} =\eta_{\mu}^{\nu}{~}\nabla_{\nu}
\end{equation} operating in the 3-D hyper-surface. The kinematics of the fundamental observers are expressed by four irreducible variables, found post decomposition of the gradient of $u_{\mu}$, the velocity vector field as
\begin{equation}
	B_{\mu\nu}=\nabla_{\nu}u_{\mu}=\dfrac{1}{3}\Theta{~} \eta_{\mu\nu}+\sigma_{\mu\nu}+\omega_{\mu\nu}-A_{\mu}u_{\nu}
\end{equation}
where $B_{\mu\nu}$ is called the deformation tensor,
\begin{equation}\label{eq5*}
	\Theta= B^{\mu}_{\mu}=\nabla_{\mu}u^{\mu}
\end{equation}
is the volume scalar/ expansion scalar (trace part of $B_{\mu\nu}$) .
\begin{equation}\label{eq6*}
	\sigma_{\mu\nu}=\nabla_{(\nu\mu)}u-\dfrac{1}{3}\eta_{\mu\nu}\Theta
\end{equation} is the shear tensor (symmetric i.e $\sigma_{\mu\nu}=\sigma_{\nu\mu}$, traceless i.e $\sigma^{\mu}_{\mu}=0$).  Here, $\nabla_{(\nu\mu)}u=\nabla_{\nu}u_\mu$+$\nabla_{\mu}u_\nu$.
\begin{equation}
	\omega_{\mu\nu}=D_{[\mu\nu]}u
\end{equation}
is the vorticity tensor (anti-symmetric i.e $\omega_{\mu\nu}=-\omega_{\nu\mu}$).   Here, $D_{[\nu\mu]}u=\nabla_{\nu}u_\mu$-$\nabla_{\mu}u_\nu$.
\begin{equation}
	A_{\mu}=u^{\nu}\nabla_{\nu}u_{\mu}
\end{equation} is the 4-acceleration vector field. The above kinematic variables may be interpreted as follows:
\begin{itemize}
\item[(i)]	$\Theta{~}$  describes the average separation between the worldlines of the $u_{\mu}$--congruence, precisely the average expansion / contraction of the associated observers.\item[(ii)]$\sigma_{\mu\nu}$ measures the kinematic anisotropies. \item[(iii)]$\omega_{\mu\nu}$ monitors rotational behavior of the $u_{\mu}$ - vector field.\item[(iv)] 4--acceleration vector field guarantees the presence of non-gravitational forces (hence $A_{\mu}=0$ for geodesic worldlines).
\end{itemize}
Thus, the Raychaudhuri equation (\ref{eq1}) describes the proper time ($\tau$) evolution of the volume scalar ($\Theta$) and hence a purely geometric equation while the Ricci tensor,  $R_{\mu\nu}$ carries the effect of the local gravitational field. If the time-like curves are, in particular geodesics (the 4-acceleration vector field is identically zero) and the congruence is hyper-surface orthogonal (which by virtue of Frobenius theorem implies zero rotation) then (\ref{eq1}) reduces to much simpler form as,
\begin{equation}\label{eq9}
	\dfrac{d\Theta}{d\tau}=-\dfrac{\Theta^{2}}{3}-2\sigma^{2}-R_{\mu\nu}u^{\mu}u^{\nu}\\
	=-\dfrac{\Theta^{2}}{3}-\sigma_{\mu\nu}{~}\sigma^{\mu\nu}-R_{\mu\nu}u^{\mu}u^{\nu} .
\end{equation}
\textbf{Focusing Theorem and Convergence Condition in General Relativity:}\\
If the matter satisfies the SEC i.e ,
\begin{equation}
	T_{\mu\nu}u^{\mu}u^{\nu}+\dfrac{1}{2}T\geq0 ,
\end{equation}
then Einstein's equation 
\begin{equation}\label{eq11}
	R_{\mu\nu}-\dfrac{1}{2}Rg_{\mu\nu}=T_{\mu\nu}
\end{equation}
yields, \begin{equation}\label{eq12}
	R_{\mu\nu}u^{\mu}u^{\nu}\geq0.
\end{equation}
Employing the condition (\ref{eq12}) on (\ref{eq9}) we get ,
\begin{equation}
	\dfrac{d\Theta}{d\tau}+\dfrac{\Theta^{2}}{3}\leq0.
\end{equation}
Integrating the above inequality w.r.t proper time $\tau$ we get,
\begin{equation}
	\dfrac{1}{\Theta(\tau)}\geq\dfrac{1}{\Theta_{0}}+\dfrac{\tau}{3}.
\end{equation}
Thus, one can infer that any initially converging hyper-surface orthogonal congruence of time-like geodesics must continue to converge within a finite value of the proper time $\tau\leq-3\Theta_{0}^{-1}$ which leads to crossing of geodesics and formation of a congruence singularity (may or may not be a curvature singularity). This is called the FT and the condition (\ref{eq12}) is the corresponding CC. Further, it is to be noted that the SEC causes gravitation to be attractive and hence can't cause geodesic deviation, rather it increases the rate of convergence. Thus the FT inevitably proves the generic existence of singularity as a major drawback of Einstein gravity. As clear from the above discussion, FT follows as a consequence of the RE, this is the reason why RE is regarded as the fundamental equation of gravitational attraction. Since in other relativistic theories of gravity, the field equations are different, there may arise certain possibilities for the avoidance of singularity. In the following sections, we study the RE and CC in $f(T)$ modified gravity theory in the framework of homogeneous and isotropic FLRW spacetime .
	\section{$f(T)$ gravity in Friedmann–Lemaître–Robertson-Walker model :  Raychaudhuri equation}
	Let us begin with the action for $f(T)$ gravity \cite{Chen:2010va}
	\begin{equation}\label{eq1*}
		\mathcal{A}_{m}=\dfrac{1}{2}\int\mathrm{d}^{4}x {~}e\left[T+f(T)+\mathcal{L}_{m}\right]
	\end{equation}
where $T$ is the torsion scalar , $f(T)$ is an arbitrary differentiable function of the torsion $T$, $e=\sqrt{-g}= det (e^{\mathcal{A}_{m}}_{\mu})$, and $\mathcal{L}_{m}$ corresponds to the matter Lagrangian, $\kappa=8\pi G=1$.\\
The above torsion scalar $T$ is defined as 
\begin{equation}\label{eq2}
	T=S_{\sigma}^{\mu\nu}{~} T_{\mu\nu}^{\sigma}
\end{equation}
where  $T^{\sigma}_{\mu\nu}$, the torsion tensor is defined as 
\begin{equation}\label{eq3}
	T^{\sigma}_{\mu\nu} = \Gamma^{\sigma}_{\nu\mu}-\Gamma^{\sigma}_{\mu\nu}=\\e^{\sigma}_{A}{~}(\partial_{\mu}e^{A}_{\nu}-\partial_{\nu}e^{A}_{\mu})
\end{equation}
The Weitzenbock connection $\Gamma^{\sigma}_{\mu\nu}$ is defined as 
\begin{equation}
	\Gamma^{\sigma}_{\mu\nu}= e^{\sigma}_{A} {~}\partial_{\nu}e^{A}_{\mu} ,
\end{equation}
and the super-potential, $S^{\mu\nu}_{\sigma}$ is defined as 
\begin{equation}\label{eq5}
	S^{\mu\nu}_{\sigma}=\dfrac{1}{2}\left(K^{\mu\nu}_{{~~}\sigma}+\delta^{\mu}_{\sigma}{~} T^{\alpha\nu}_{{~~}\alpha} - \delta^{\nu}_{\sigma}{~}T^{\alpha\mu}_{{~~}\alpha}\right),
\end{equation}
where the contortion tensor takes the form
\begin{equation} \label{eq6}
K^{\mu\nu}_{{~~}\sigma}=-\dfrac{1}{2}\left(T^{\mu\nu}_{{~~}\sigma}-T^{\nu\mu}_{{~~}\sigma}-T^{\mu\nu}_{\sigma}\right).
\end{equation}
Geometrically, the orthogonal tetrad components $e_{\mathcal{A}_{m}}(x^{\mu})$ (considered as dynamical variables), form an orthonormal basis for the tangent space at each point $x^{\mu}$ of the manifold i.e 
\begin{equation}\label{eq7}
	e_{i}{\cdot}e_{j}=\eta_{ij} {~~}, {~~}\eta_{ij}= diag{~}(+1,-1,-1,-1)
\end{equation}
In a coordinate basis, we may write\\ \begin{equation}
e_{i}=e^{\mu}_{i}{~}\partial_{\mu}
\end{equation} where $e^{\mu}_{i}$ are the components of $e_{i}$ , with  $\mu,{~}i=0,1,2,3$ .
The metric tensor is obtained from the dual vierbein as 
\begin{equation}
	g_{\mu\nu}(x)=\eta_{ij}{~}e^{i}_{\mu}(x){~}e^{j}_{\nu}(x)
\end{equation}
The present work deals with $f(T)$ gravity in the framework of homogeneous and isotropic FLRW space-time having line element 
\begin{equation}
	ds^{2}=-dt^{2}+a^{2}(t)\left[\dfrac{dr^{2}}{1-kr^{2}}+r^{2}(d\theta^{2}+sin^{2}\theta d\phi^{2})\right]
\end{equation}
where $a(t)$ is the scale factor, $H=\dfrac{\dot{a}}{a}$ is the Hubble parameter, '.' denotes differentiation w.r.t cosmic time $t$. \\'$k$', the curvature index dictates the model of the universe. To be precise, 
$k = \begin{cases}
	-1,  \text{ open model}\\
	{~~}	0 , {~} \text{flat model}\\
	+1,  \text{ closed model}
\end{cases}$.
\\Further it has been assumed that the universe is filled with perfect fluid having barotropic equation of state 
\begin{equation}
	p=\omega\rho\label{eq11}
\end{equation}
where $\omega=\gamma-1{~~} ({~~}0\leq\gamma\leq2{~~})$ being the equation of state parameter.\\ Now using equations (\ref{eq2}), (\ref{eq3}), (\ref{eq5}), (\ref{eq6}) and (\ref{eq7})  we have , 
\begin{equation}
	T=-6H^{2}
\end{equation}
It is to be noted that during cosmic evolution  $T$ is negative. Varying the action (\ref{eq1*}) we get the modified Einstein field equations as
\begin{equation}\label{eq13}
\left[	e^{-1}\partial_{\mu}\left(e{~}e_{A}^{\rho}{~}S_{\rho}^{\mu\nu}\right)-e_{A}^{\lambda}{~}T_{\mu\lambda}^{\rho}{~}S^{\nu\mu}_{\rho}\right]\left[1+f_{T}\right]+e_{A}^{\rho}{~}S_{\rho}^{\mu\nu}{~}\partial_{\mu}(T)f_{TT}+\dfrac{1}{4}{~}e_{A}^{\nu}\left[T+f(T)\right]=4\pi G{~}e_{A}^{\rho}{~}T^{\nu}_{\rho}
\end{equation}
where $f_{T}=\dfrac{df}{dT}$ , $f_{TT}=\dfrac{d^{2}f}{dT^{2}}$ ,  and $T^{\nu}_{\rho}$ is the energy momentum tensor of the total matter -- baryonic matter and dark energy. Thus for FLRW model, the modified Friedmann equations can be written as :
\begin{equation}
	H^{2}=\dfrac{1}{2f_{T}+1}\left[\dfrac{\rho}{3}-\dfrac{f}{6}\right]\label{eq14}\end{equation}
\begin{equation}
	2\dot{H}=\dfrac{-(p+\rho)}{1+f_{T}+2Tf_{TT}}\label{eq15}
\end{equation}
where $p$ and $\rho$ are the thermodynamic pressure and density of the matter fluid having conservation equation
\begin{equation}\label{eq16}
	\dot{\rho}+3H(p+\rho)=0.
\end{equation}
Based on our assumption (\ref{eq11}) , the solution of the differential equation (\ref{eq16}) can be written in the form 
\begin{equation}\label{eq31}
	\rho=\rho_{0}{~}a^{-3\gamma}.
\end{equation}
Finally, using the expression for $H$ and equations (\ref{eq14}), (\ref{eq15}) after some algebraic manipulation one gets the Raychaudhuri equation in $f(T)$ gravity as
\begin{equation}
	\dfrac{\ddot{a}}{a}=\rho_{0}a^{-3\gamma}\left[\dfrac{1}{3(2f_{T}+1)}-\dfrac{\gamma}{2(1+f_{T}+2Tf_{TT})}\right]-\dfrac{f(T)}{6(2f_{T}+1)}.
\end{equation}
Therefore, the Raychaudhuri equation is homogeneous and depends on $\gamma$ which is related to the equation of state parameter $\omega$ by the relation $\gamma=\omega+1$ and choice of the torsion function $f(T)$. This hints that the CC essentially depends on the function $f(T)$.\\
\begin{tabular}{ |p{3cm}||p{5cm}||p{7.5cm}|}
\hline
\multicolumn{3}{|c|}{Table showing the RE for various choices of f(T)} \\
\hline
Choice of f(T)& Gravity Theory &Raychaudhuri Equation\\
\hline
0  & Einstein Gravity    &$\frac{\ddot{a}}{a}=\rho_{0}a^{-3\gamma}\left(\frac{1}{3}-\frac{\gamma}{2}\right)$ \\
$c$, a non-zero constant&   Einstein gravity with cosmological constant  & $\frac{\ddot{a}}{a}=\rho_{0}a^{-3\gamma}\left(\frac{1}{3}-\frac{\gamma}{2}\right)-\frac{c}{6}$ \\
$f(T)=f_{0} T$, ($f_{0}$, a non zero constant) &Einstein gravity with reconstruction of gravitational constant & $\frac{\ddot{a}}{a}=\rho_{0}a^{-3\gamma}\left[\dfrac{1}{3(2f_{0}+1)}-\dfrac{\gamma}{2(1+f_{0})}\right]-\dfrac{f_{0}T}{6(2f_{0}+1)}$\\
\hline
\end{tabular}
\section{Study of Convergence Condition in $f(T)$ gravity model }
The field equations for $f(T)$ gravity given by equations (\ref{eq14}) and ( \ref{eq15}) may be expressed as :
\begin{eqnarray}
	3H^{2}=(\rho+\rho^{(e)})\\
	2\dot{H}=-(p+\rho)-(p^{(e)}+\rho^{(e)})
\end{eqnarray}where
\begin{eqnarray}\label{eq35}
	\rho^{(e)}=-\left(\dfrac{f(T)}{2}+6H^{2}f_{T}\right)
	\\ \label{eq36}p^{(e)}=2\left(\dot{H}+3H^{2}\right)f_{T}+4Tf_{TT}\dot{H}+\dfrac{f(T)}{2}
\end{eqnarray} are the energy density and thermodynamic pressure of the effective fluid. The Raychaudhuri scalar $\tilde{R}$=$R_{\mu\nu}u^{\mu}u^{\nu}$ in this modified gravity turns out to be ,
\begin{equation}
	R_{\mu\nu}u^{\mu}u^{\nu}=\left(J_{\mu\nu}u^{\mu}u^{\nu}+\dfrac{1}{2}J\right)+\left(J_{\mu\nu}^{(e)}u^{\mu}u^{\nu}+\dfrac{1}{2}J^{(e)}\right),
\end{equation}
$J$ being the trace of $J_{\mu\nu}$ i.e $J=g^{\mu\nu}J_{\mu\nu}$.
Energy momentum tensor for perfect fluid having unit time-like vector $u^{\mu}$ ( so that $u^{\mu}u_{\mu}=-1$) is given by 
\begin{equation}
	J_{\mu\nu}=(p+\rho)u_{\mu}u_{\nu}+pg_{\mu\nu}
	\end{equation}
Thus, the expression for the Raychaudhuri scalar ($\tilde{R}$) is
\begin{equation}
\tilde{R}=	R_{\mu\nu}u^{\mu}u^{\nu}=\dfrac{1}{2}\left(\rho+3p\right)+\dfrac{1}{2}\left(\rho^{(e)}+3p^{(e)}\right)
	\end{equation}
Using the equations (\ref{eq14}), (\ref{eq15}), (\ref{eq31}), (\ref{eq35}) and (\ref{eq36}), the explicit expression for $\tilde{R}$ in terms of $f(T)$, $f_{T}$ and $f_{TT}$ can be written as
\begin{equation}
\tilde{R}=	R_{\mu\nu}u^{\mu}u^{\nu}=\dfrac{3\gamma\rho_{0}a^{-3\gamma}}{2(1+f_{T}+2Tf_{TT})}+\dfrac{\left(-\rho_{0}a^{-3\gamma}+\frac{f(T)}{2}\right)}{\left(1+2f_{T}\right)}.
\end{equation}
Therefore, the expression for the Raychaudhuri scalar ($\tilde{R}$) shows that the CC essentially depends on the choice of $f(T)$. Therefore
we shall  analyze the CC for the following choices of $f(T)$, assuming power-law expansion of the universe, i.e $a=t^{m}$. The argument behind the power law choice of the scale factor is that if we choose $f(T)$ as a power law form as given in model 1 or as a linear combination of $T$ and $T^{-1}$ in model 2, then solving the field equations (\ref{eq14}) and (\ref{eq15}) one may get $a(t)$ as a power law form only for the choices $n=1$ and $\dfrac{1}{2}$. Further it should be mentioned that analytic solution for $a(t)$ is possible only for the above choices of $n$. Even if there are a plethora of models in $f(T)$--gravity cosmology, the following models are taken into consideration for simplicity in mathematical calculations and they favor accelerating universe as per recent observation (see for ref. \cite{Wu:2010xk}, \cite{Myrzakulov:2010vz}).\\

\begin{figure}[h!]
	\begin{minipage}{0.3\textwidth}
		\centering\includegraphics[height=5cm,width=5cm]{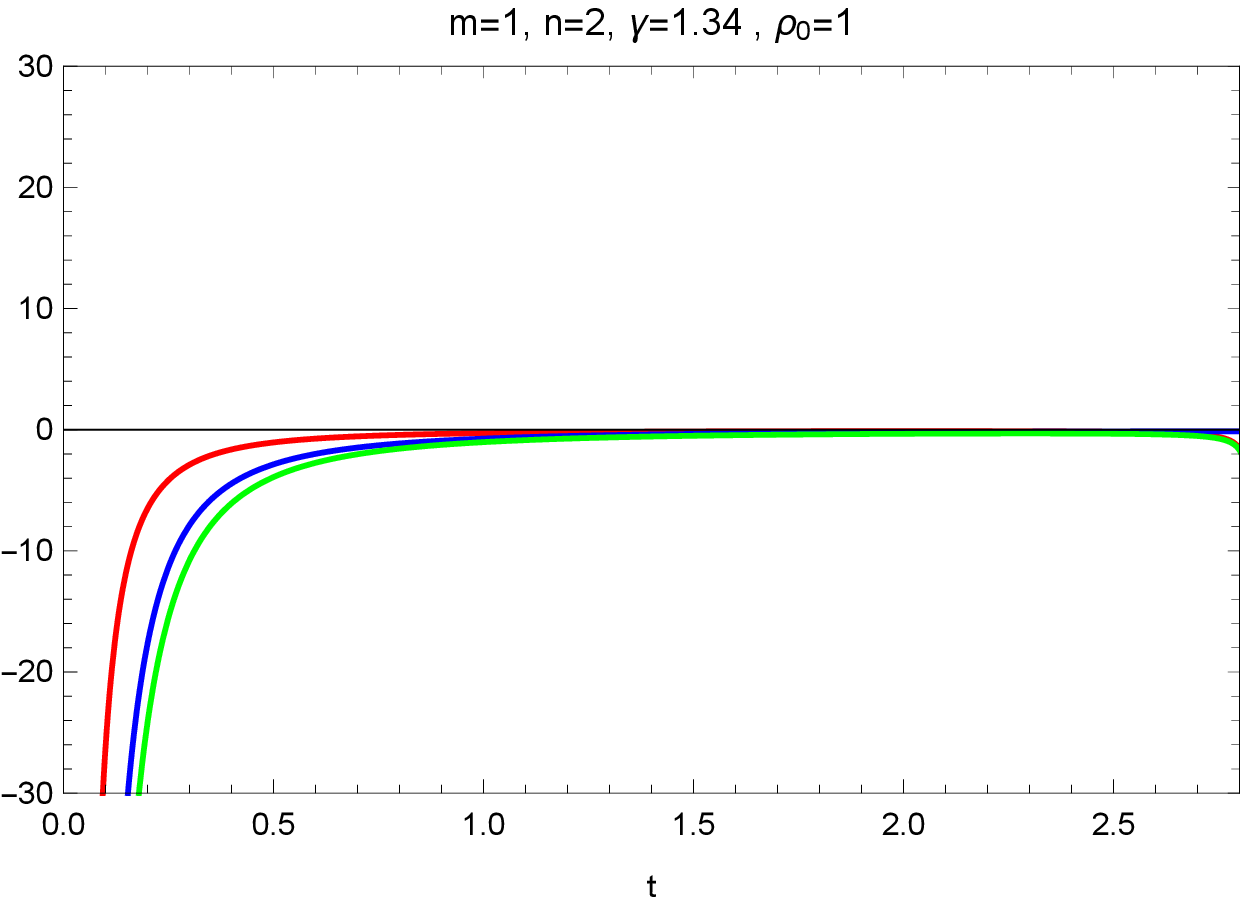}
	\end{minipage}~~~~~~~
	\begin{minipage}{0.3\textwidth}
		\centering\includegraphics[height=5cm,width=5cm]{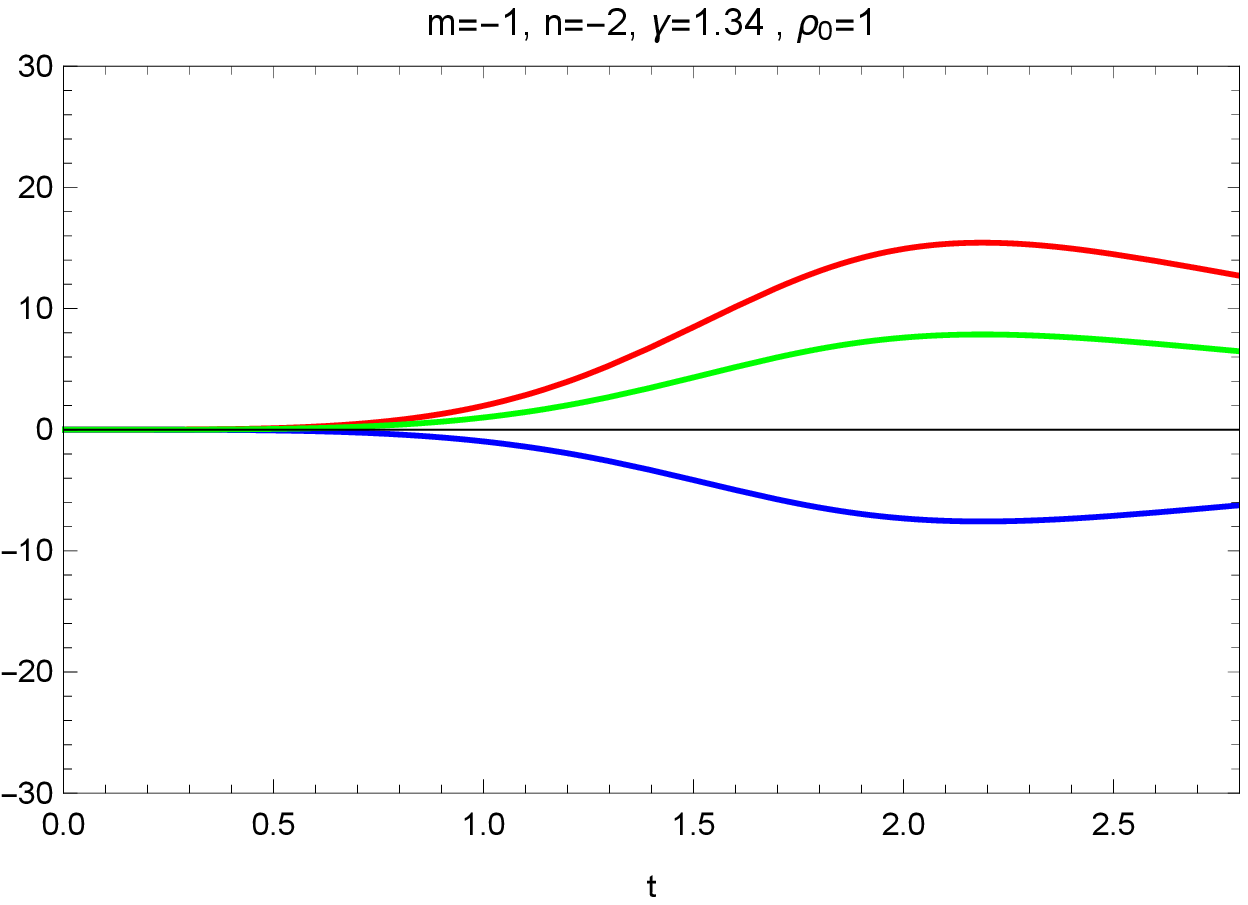}
	\end{minipage}\hfill
	\begin{minipage}{0.3\textwidth}
		\centering\includegraphics[height=5cm,width=5cm]{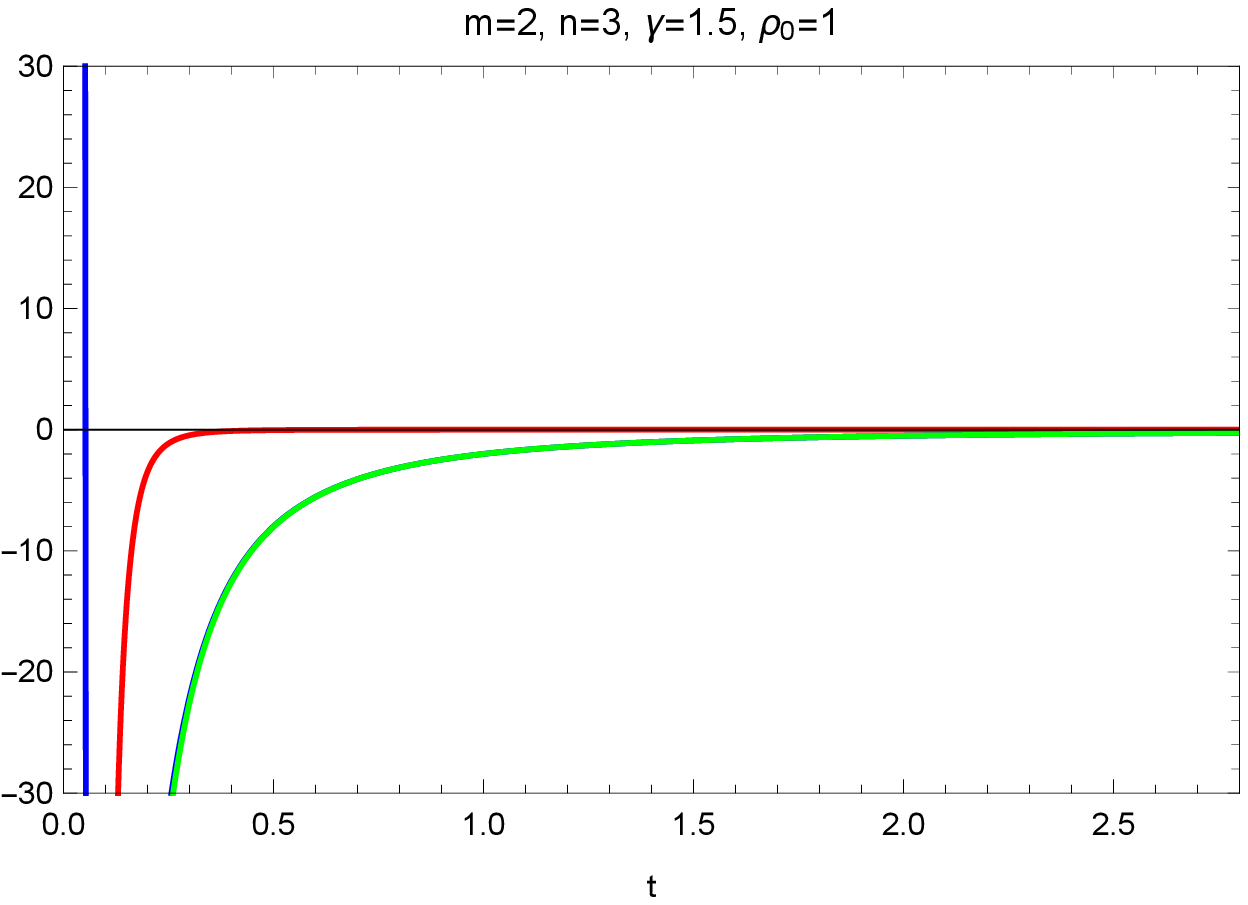}
	\end{minipage}
	\begin{minipage}{0.3\textwidth}
		\centering\includegraphics[height=5cm,width=5cm]{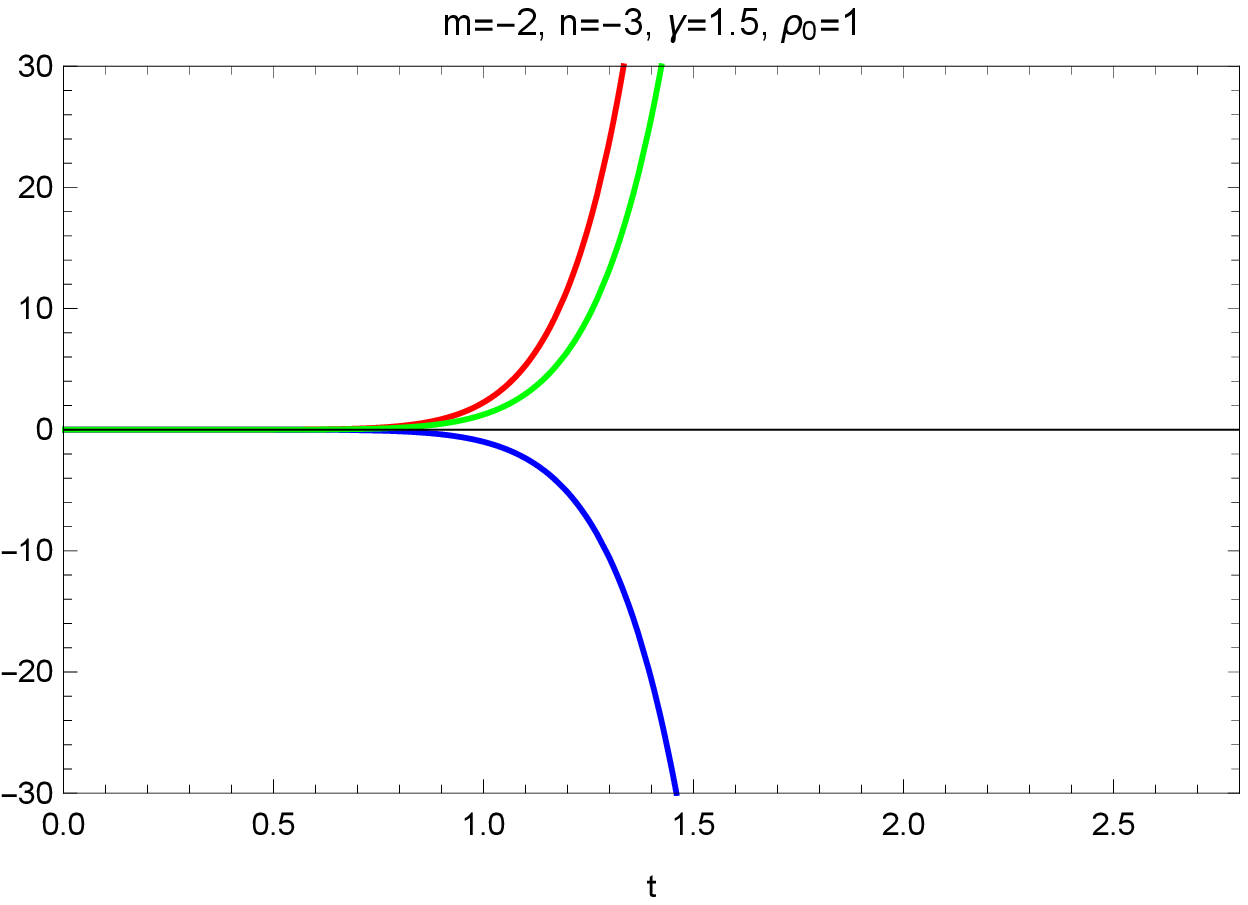}
	\end{minipage}~~~~~~~
	\begin{minipage}{0.3\textwidth}
		\centering\includegraphics[height=5cm,width=5cm]{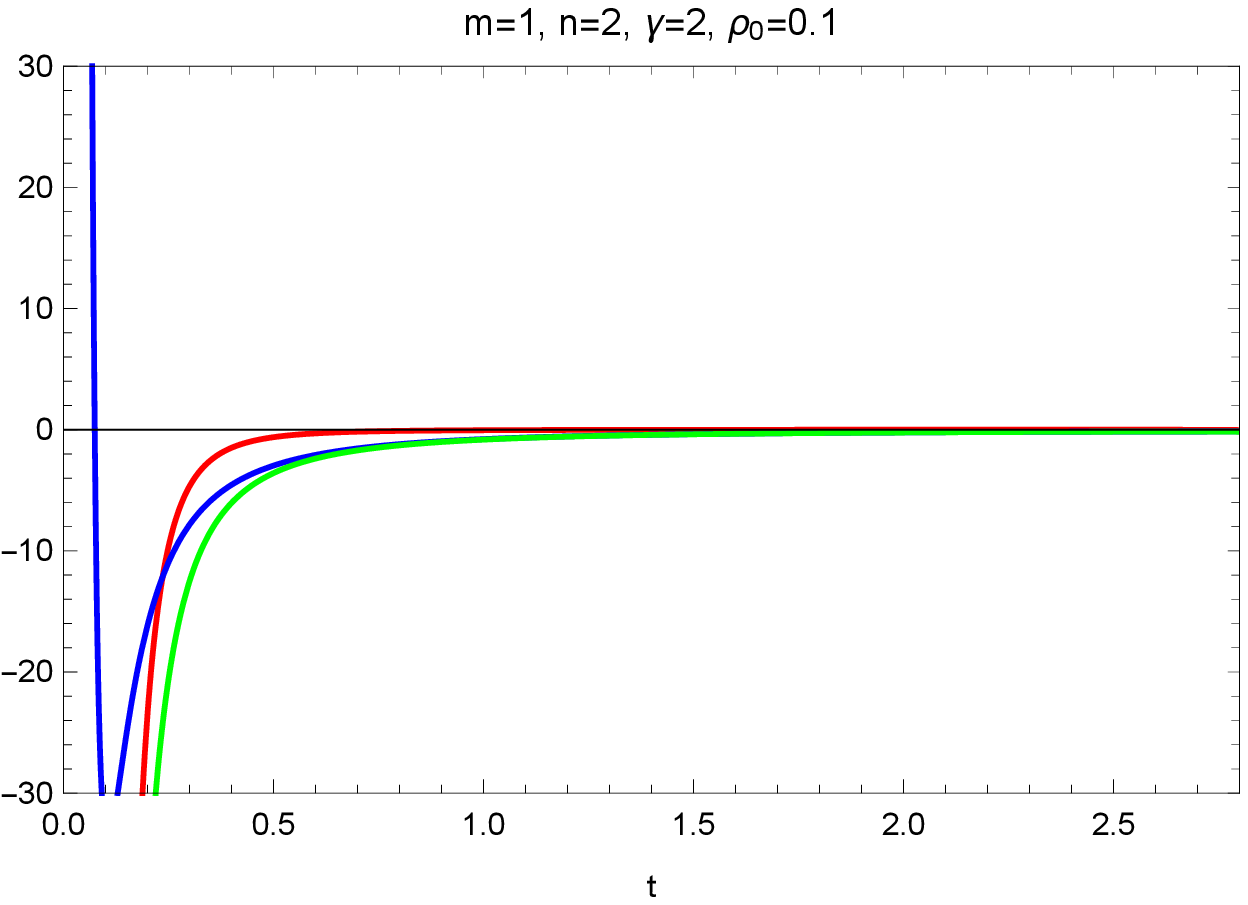}
	\end{minipage}\hfill
	\begin{minipage}{0.3\textwidth}
		\centering\includegraphics[height=5cm,width=5cm]{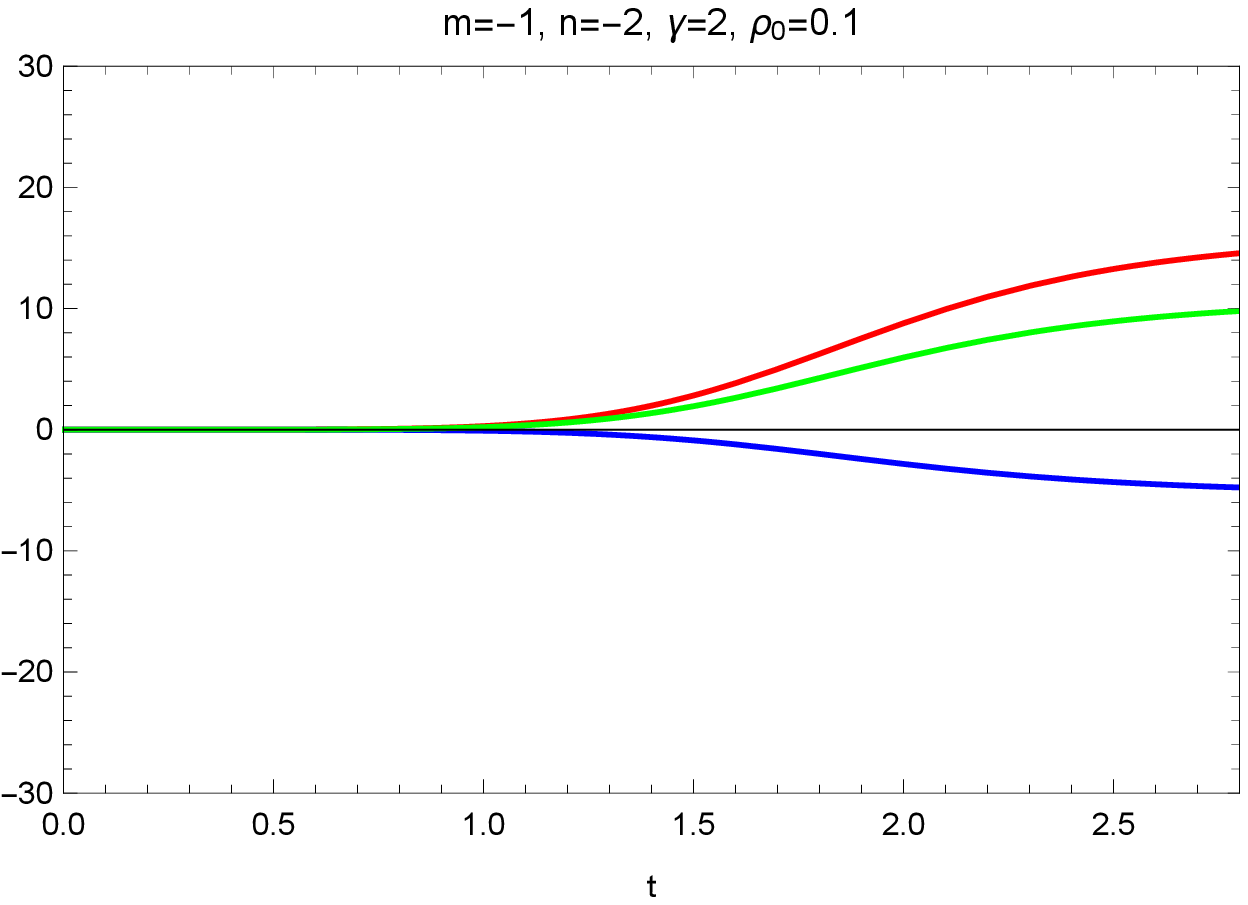}
	\end{minipage}
	\begin{minipage}{0.3\textwidth}
		\centering\includegraphics[height=5cm,width=5cm]{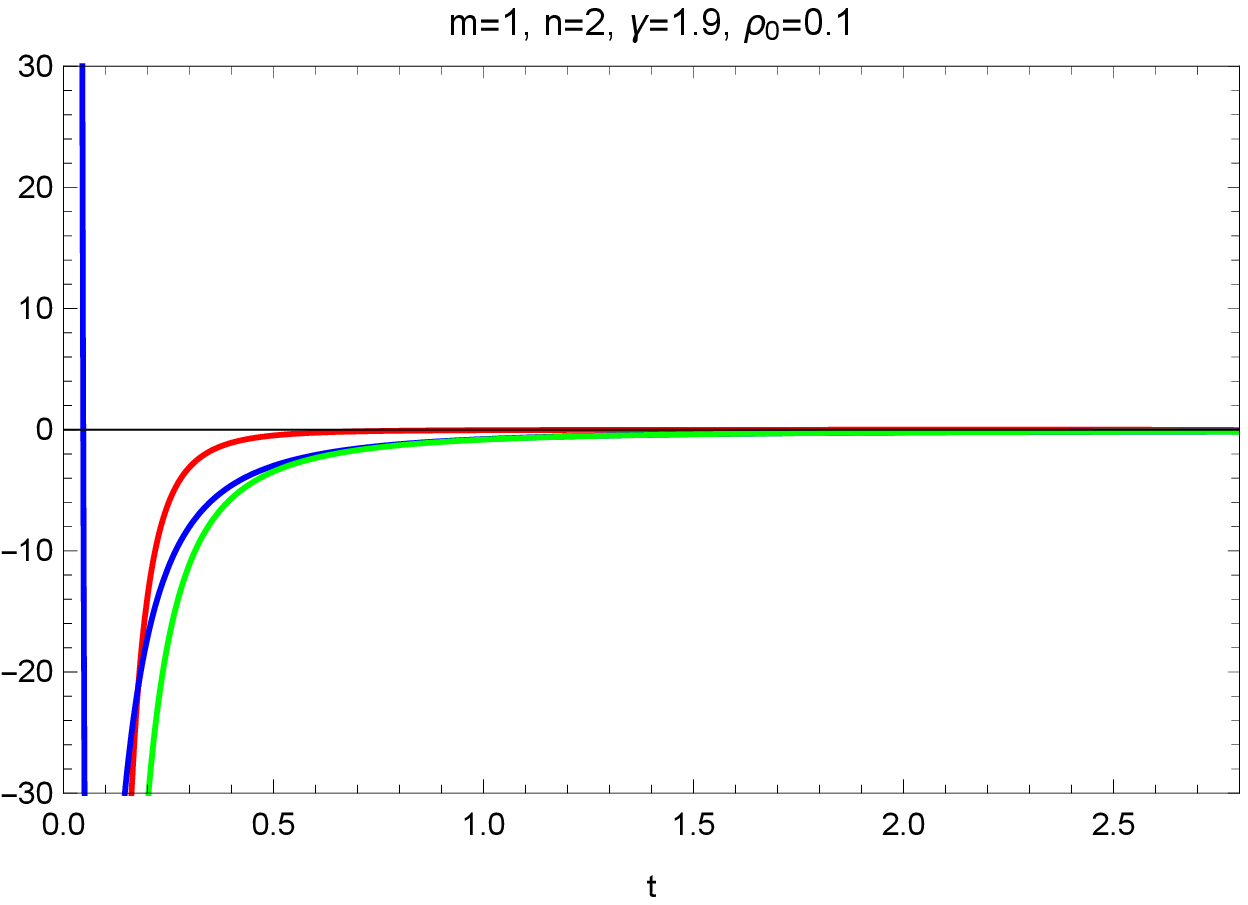}
	\end{minipage}~~~~~~~
	\begin{minipage}{0.3\textwidth}
		\centering\includegraphics[height=5cm,width=5cm]{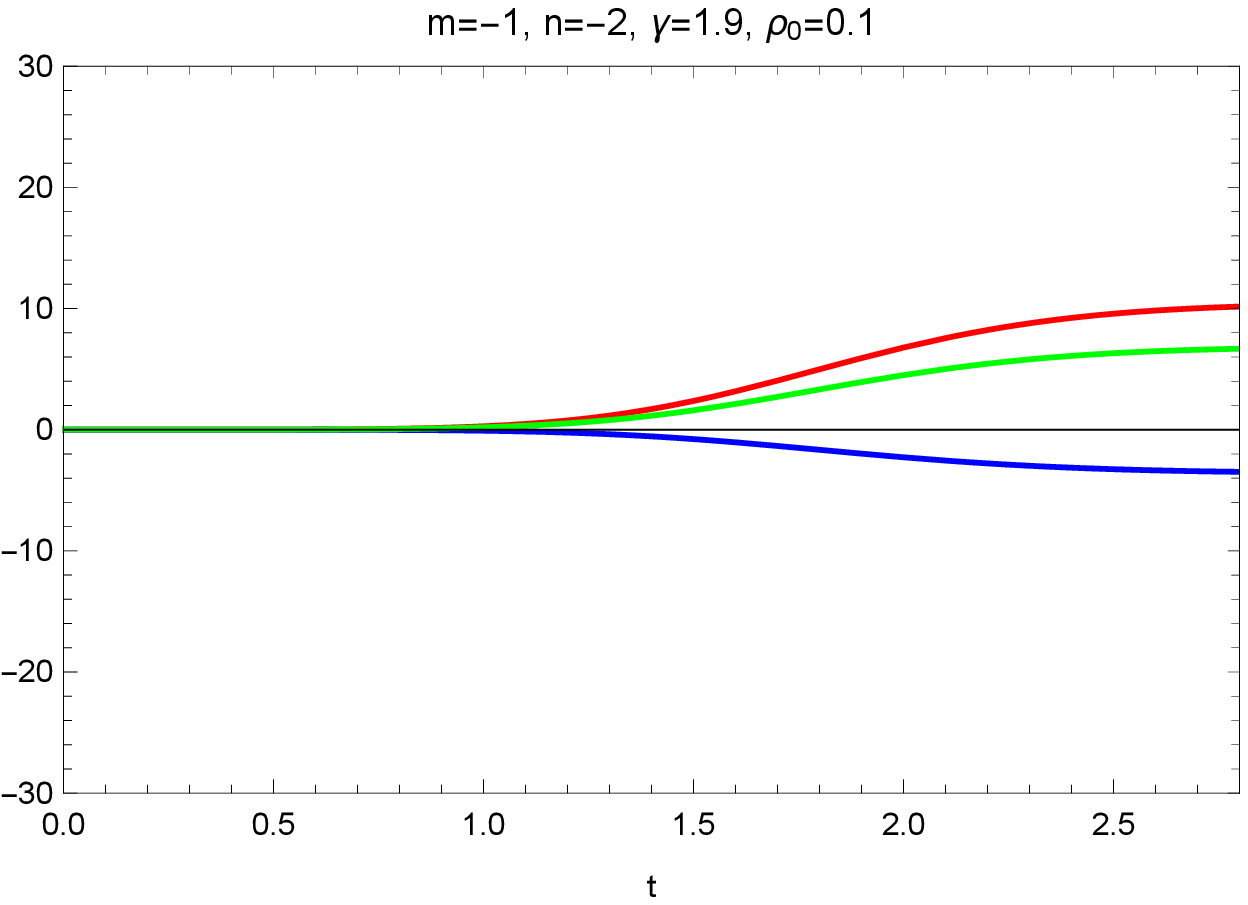}
	\end{minipage}\hfill
	\begin{minipage}{0.3\textwidth}
		\centering\includegraphics[height=5cm,width=5cm]{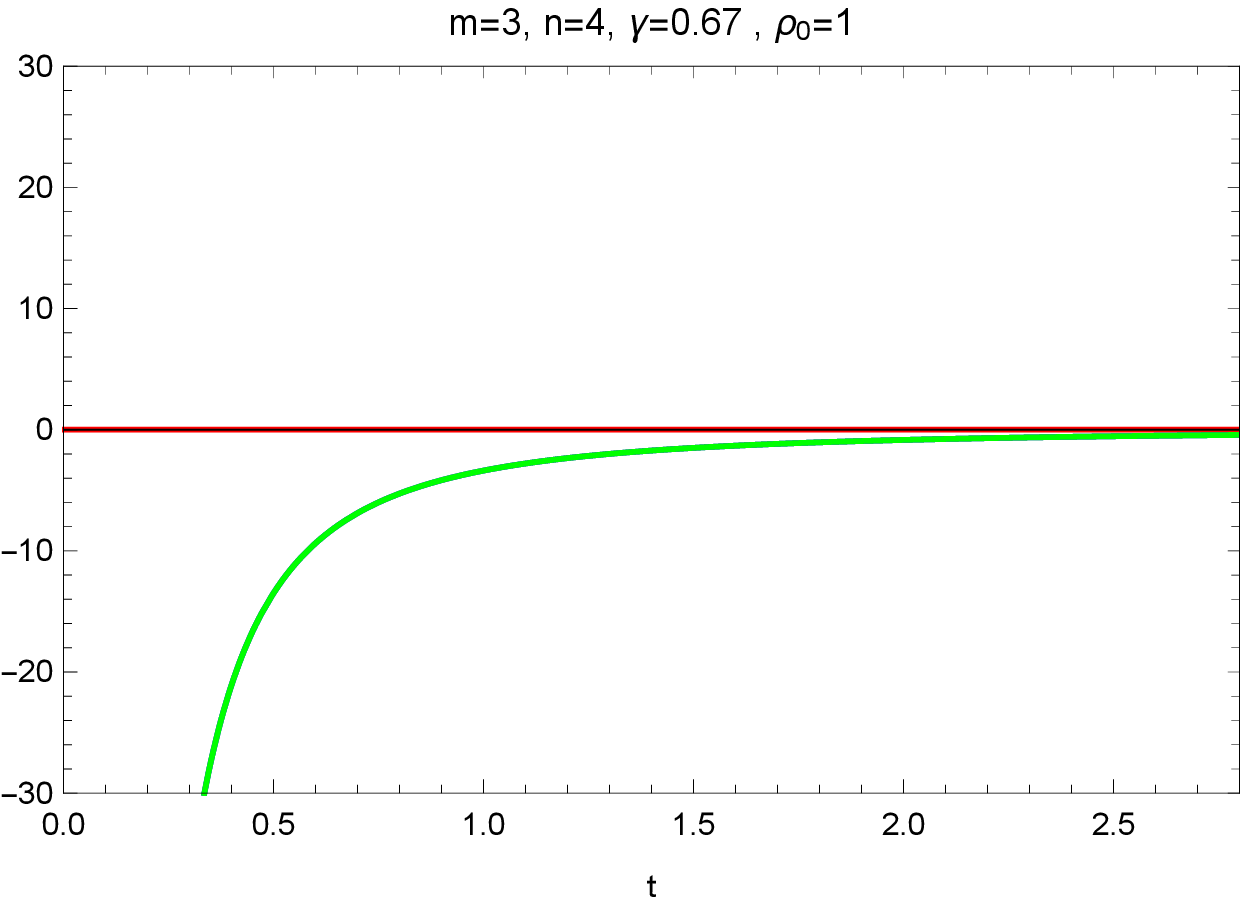}
	\end{minipage}
	\begin{minipage}{0.3\textwidth}
		\centering\includegraphics[height=5cm,width=5cm]{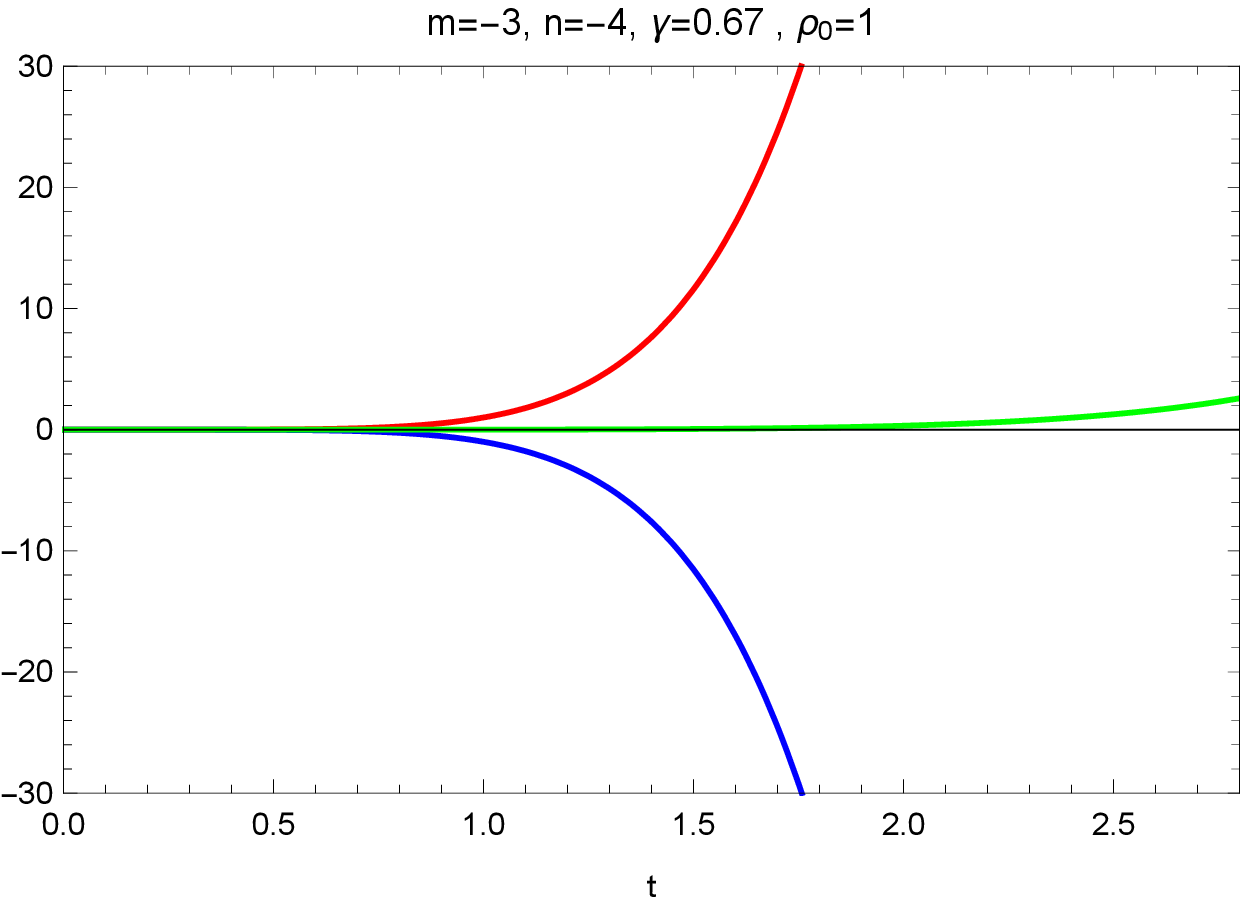}
	\end{minipage}
	\begin{minipage}{0.3\textwidth}
    	\centering\includegraphics[height=5cm,width=5cm]{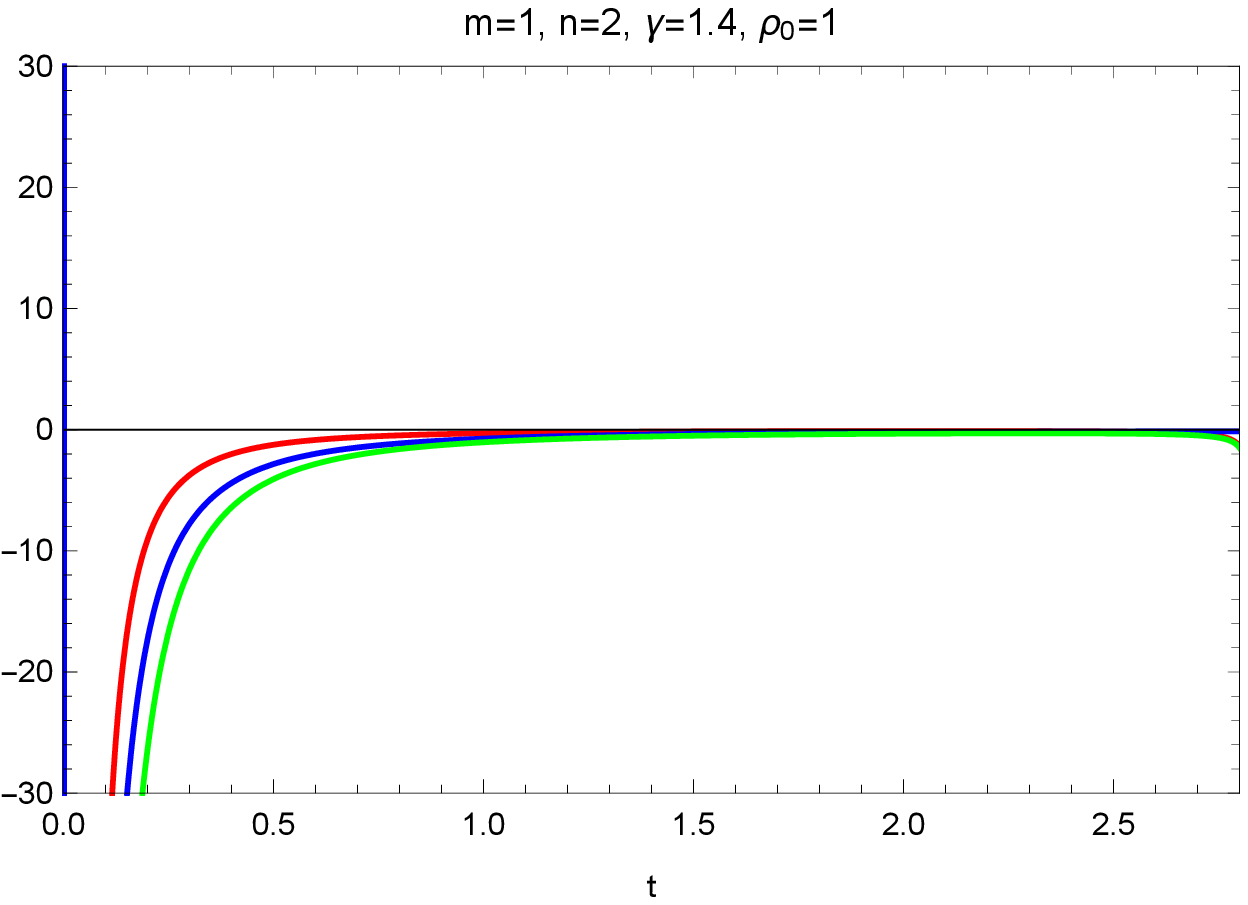}
\end{minipage}
	\begin{minipage}{0.3\textwidth}
	   \centering\includegraphics[height=5cm,width=5cm]{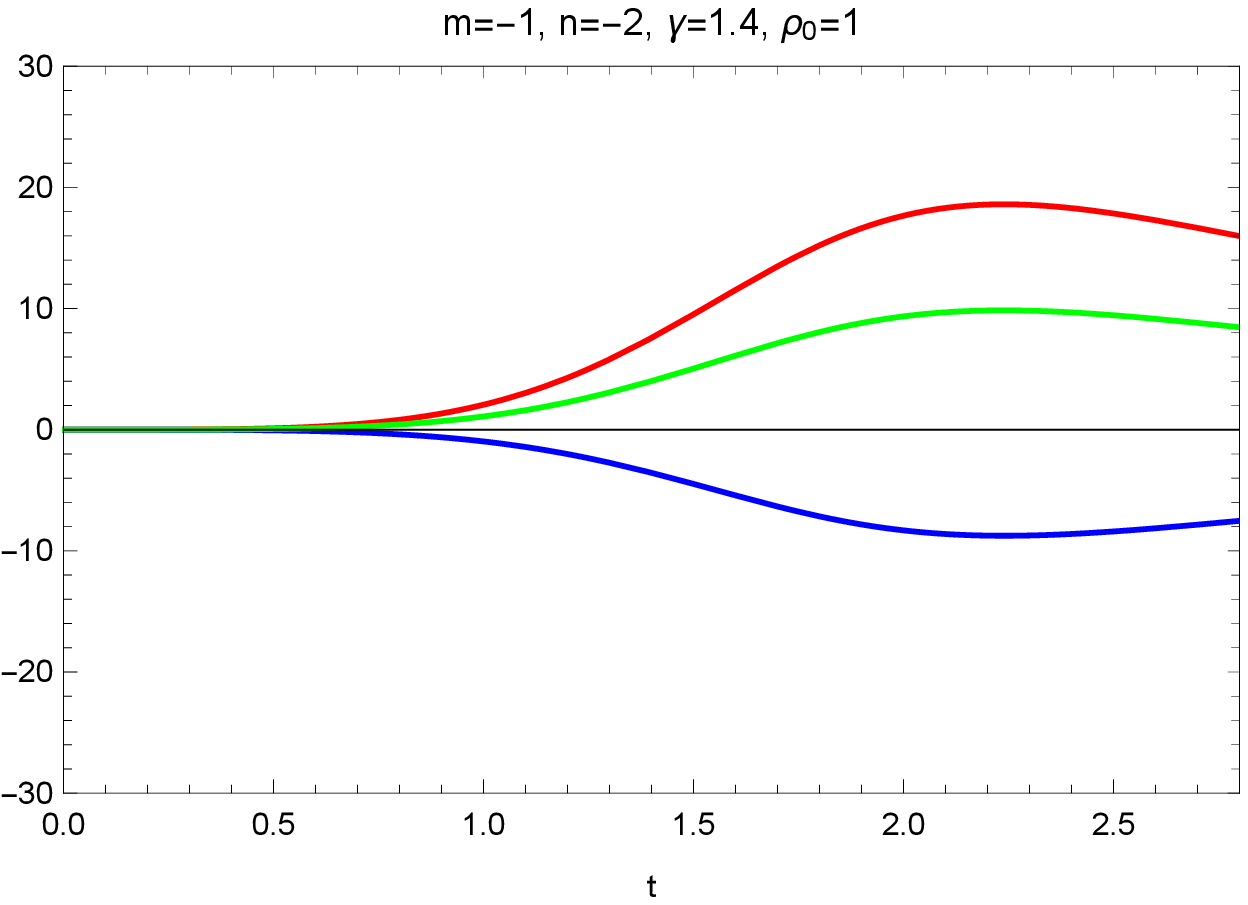}
\end{minipage}
	\begin{minipage}{0.85\textwidth}\caption{[Time variation of different terms: $R_{1}$ (red), $R_2$ (blue) and $R_{\mu\nu}u^\mu u^\nu=\tilde{R}$ (green) choosing $0\leq\gamma\leq2{~~}$, for the model $a(t)=t^{m}$, $f(T)=\alpha(-T)^{n}, \alpha=1$.]}\label{f1}
\end{minipage}
\end{figure}

\begin{figure}[h!]
	\begin{minipage}{0.3\textwidth}
		\centering\includegraphics[height=5cm,width=5cm]{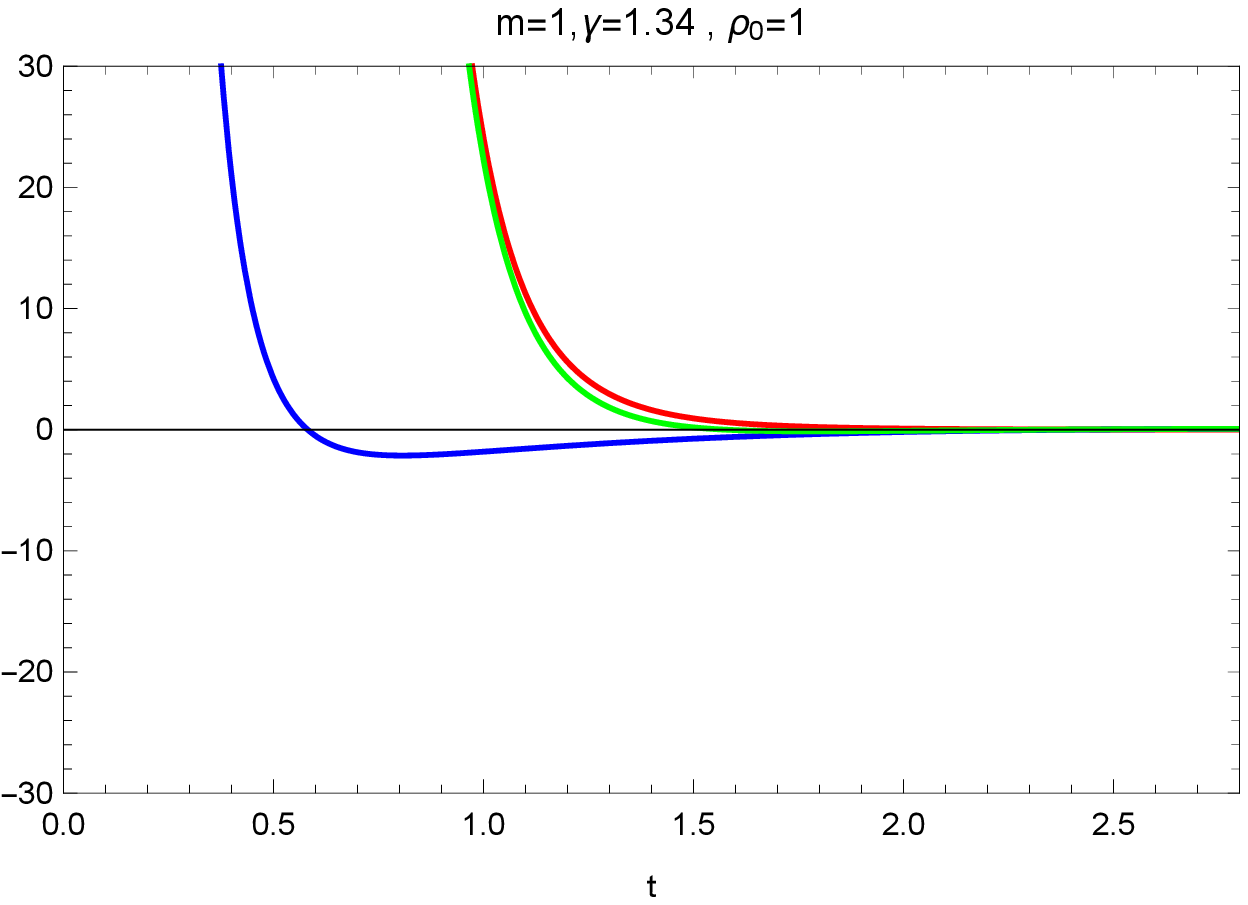}
	\end{minipage}~~~~~~~
	\begin{minipage}{0.3\textwidth}
		\centering\includegraphics[height=5cm,width=5cm]{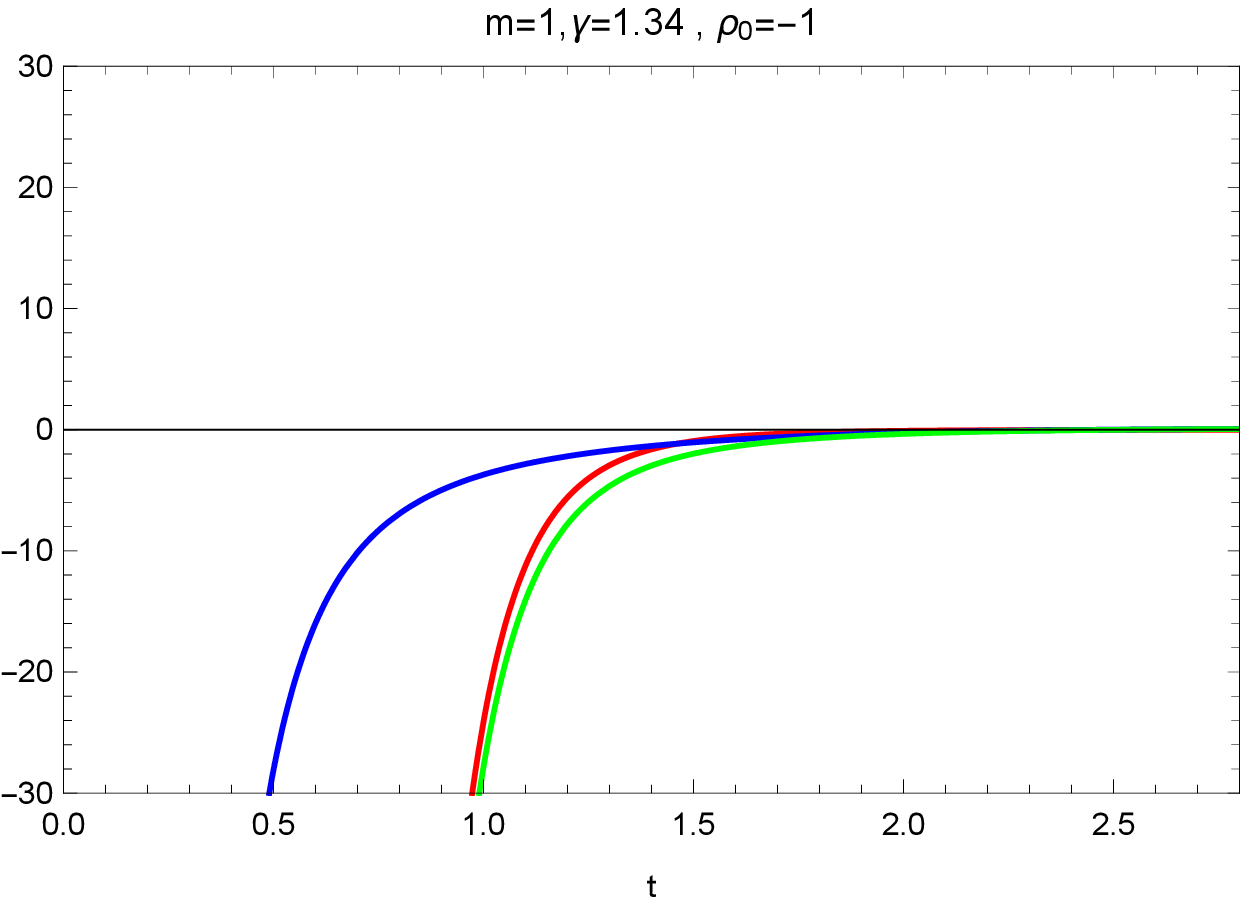}
	\end{minipage}\hfill
	\begin{minipage}{0.3\textwidth}
		\centering\includegraphics[height=5cm,width=5cm]{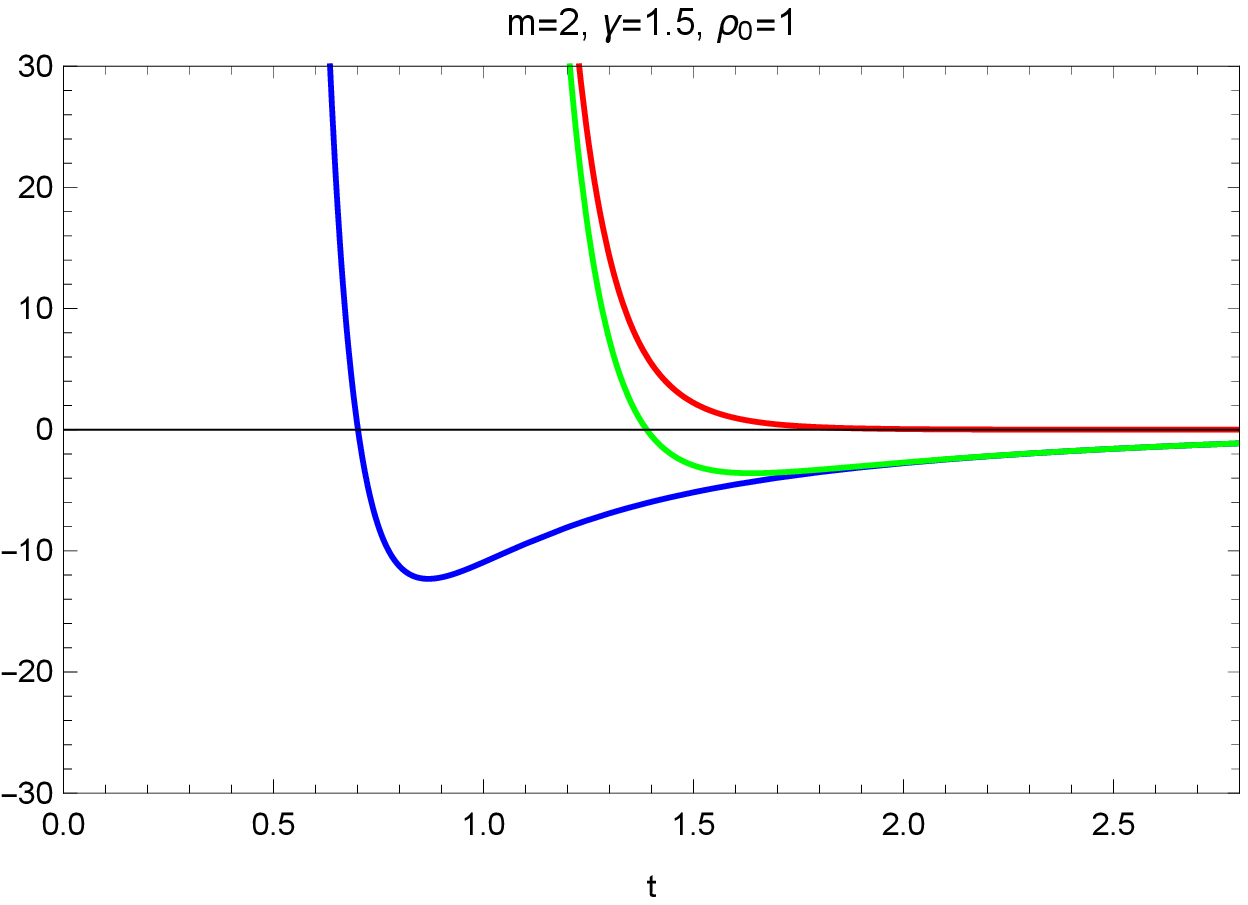}
	\end{minipage}
	\begin{minipage}{0.3\textwidth}
		\centering\includegraphics[height=5cm,width=5cm]{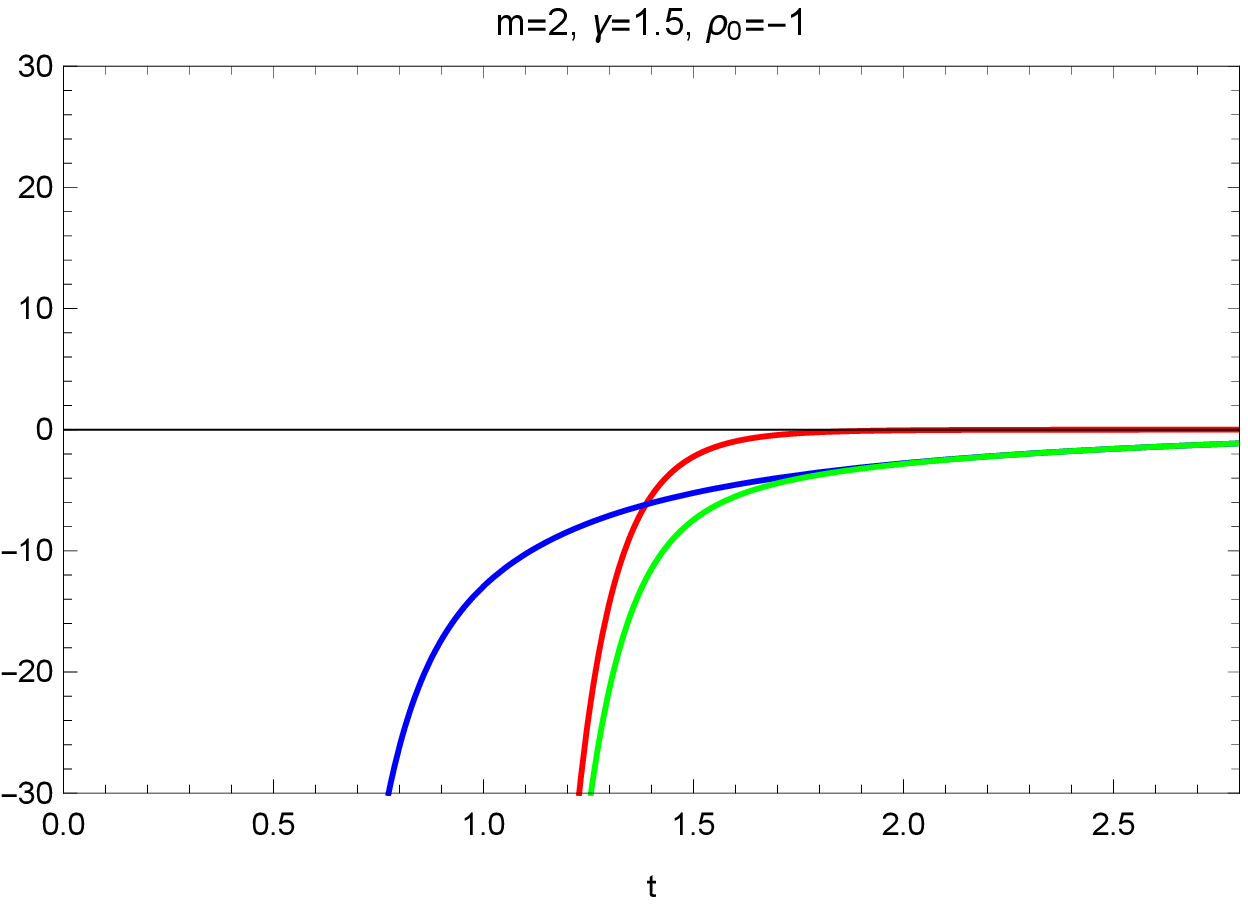}
	\end{minipage}~~~~~~~
	\begin{minipage}{0.3\textwidth}
		\centering\includegraphics[height=5cm,width=5cm]{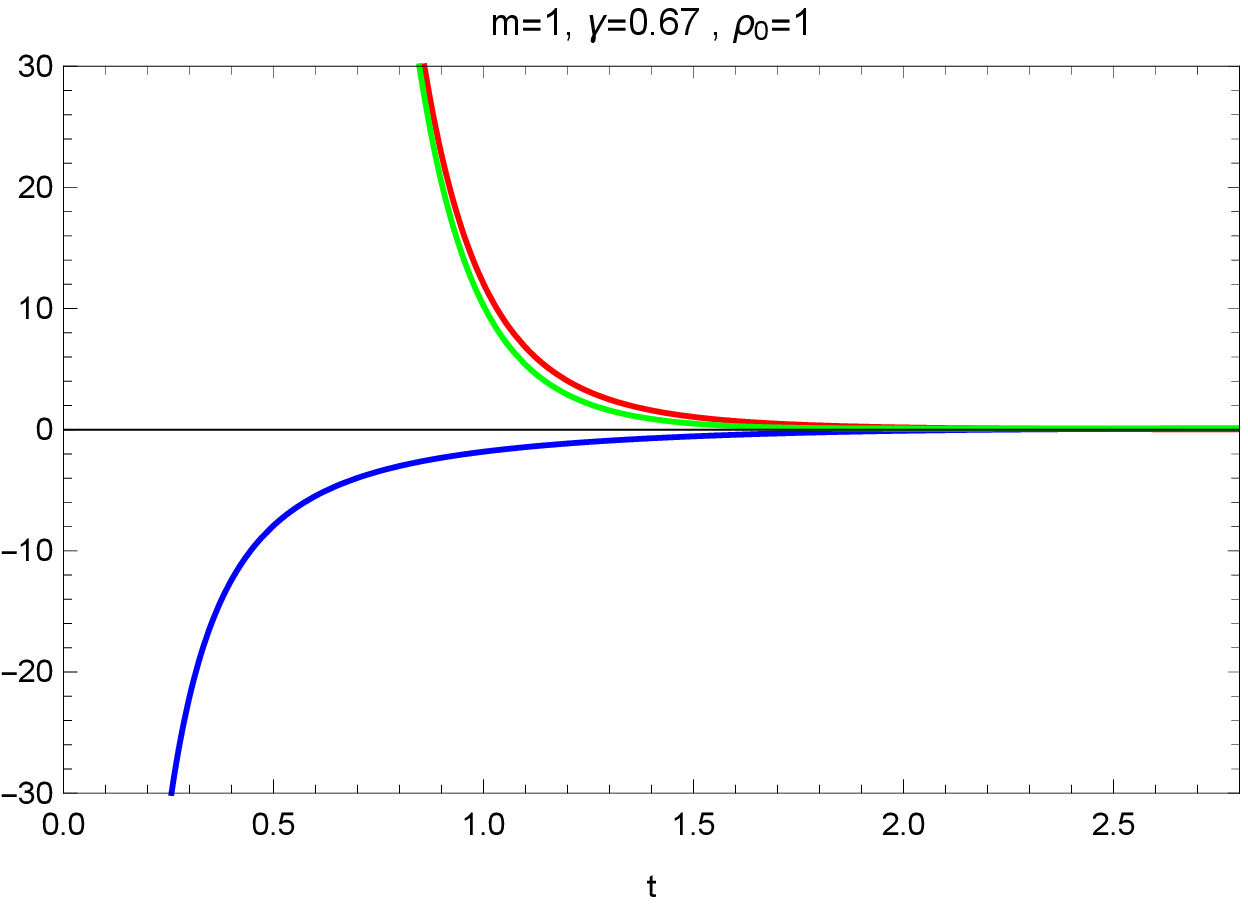}
	\end{minipage}\hfill
	\begin{minipage}{0.3\textwidth}
		\centering\includegraphics[height=5cm,width=5cm]{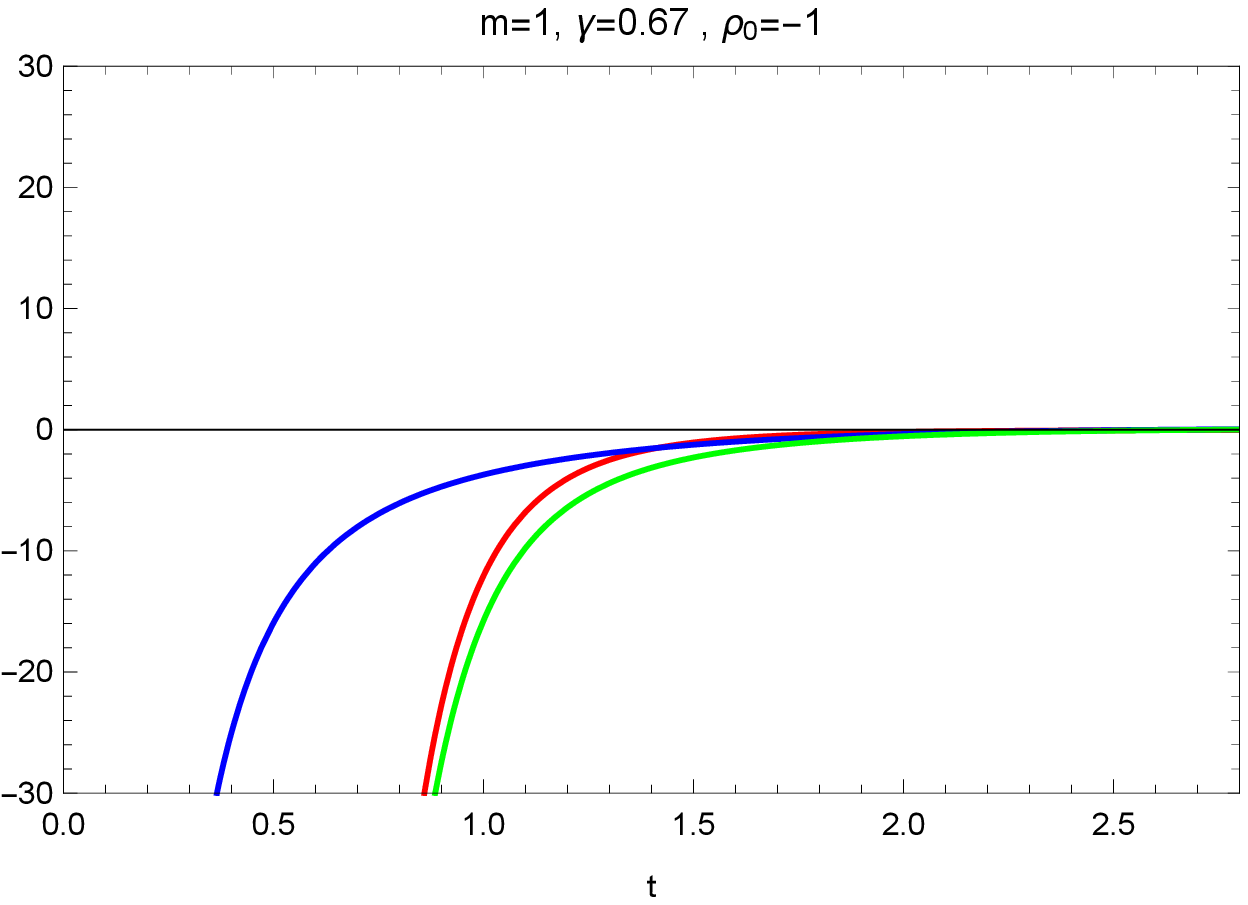}
	\end{minipage}
	\begin{minipage}{0.3\textwidth}
		\centering\includegraphics[height=5cm,width=5cm]{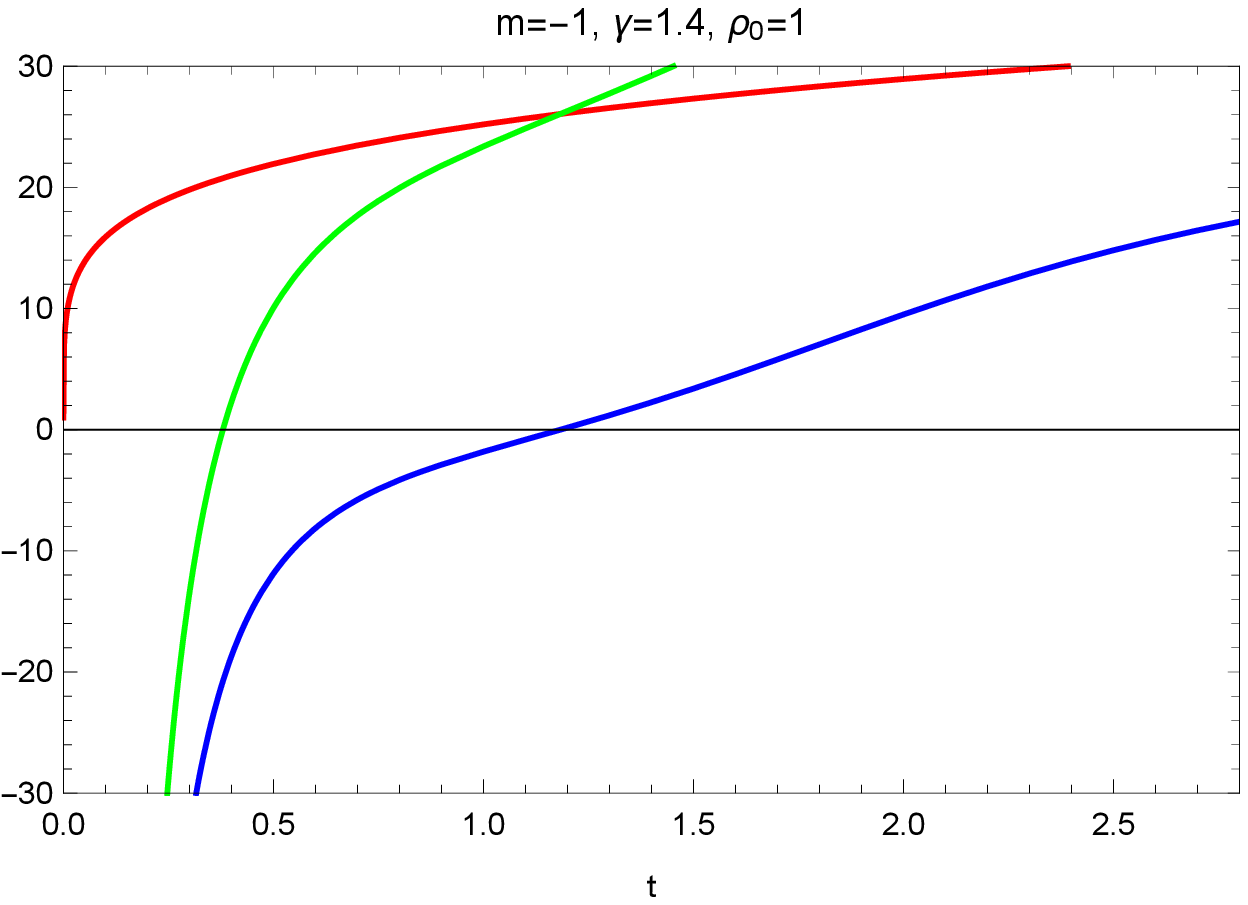}
	\end{minipage}~~~~~~~
	\begin{minipage}{0.3\textwidth}
		\centering\includegraphics[height=5cm,width=5cm]{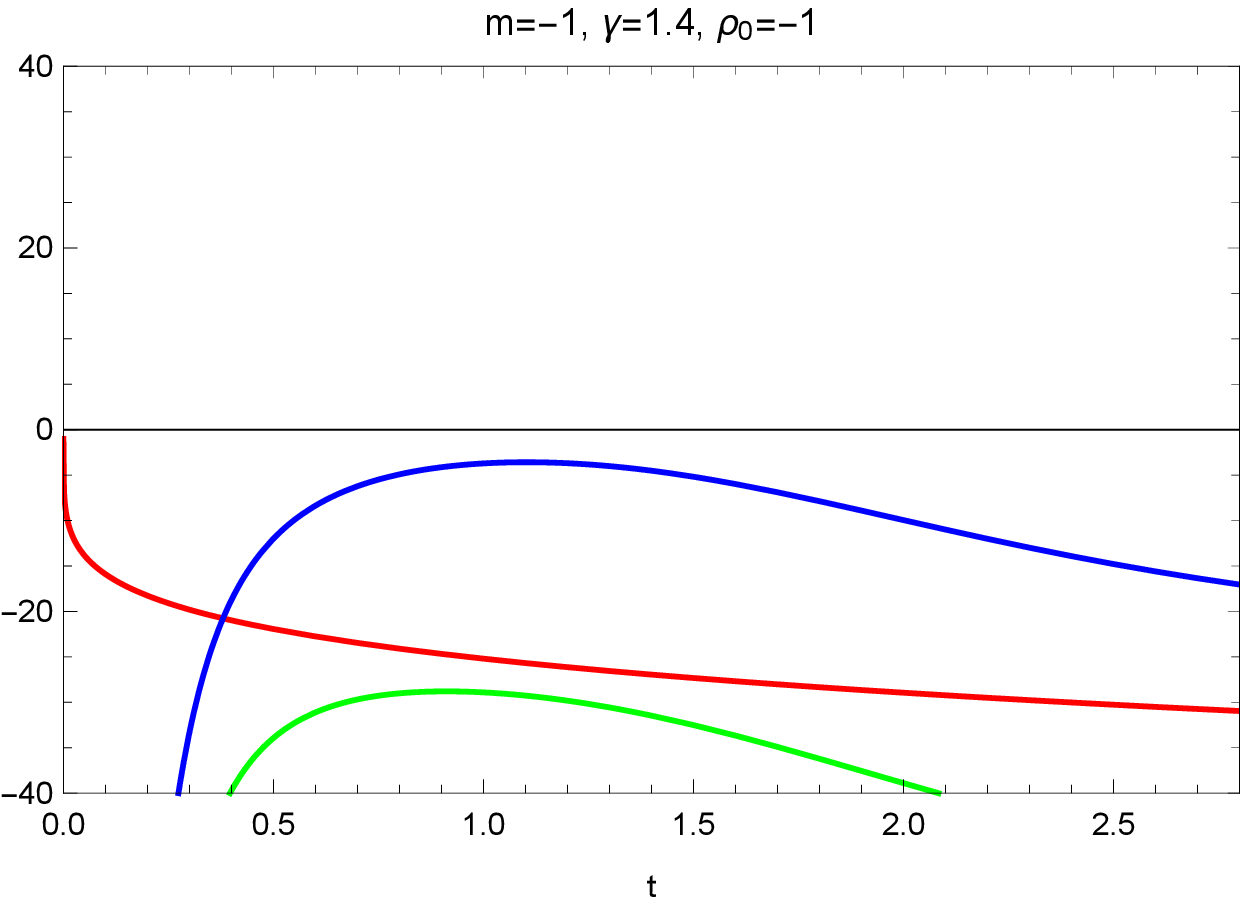}
	\end{minipage}\hfill
	\begin{minipage}{0.3\textwidth}
		\centering\includegraphics[height=5cm,width=5cm]{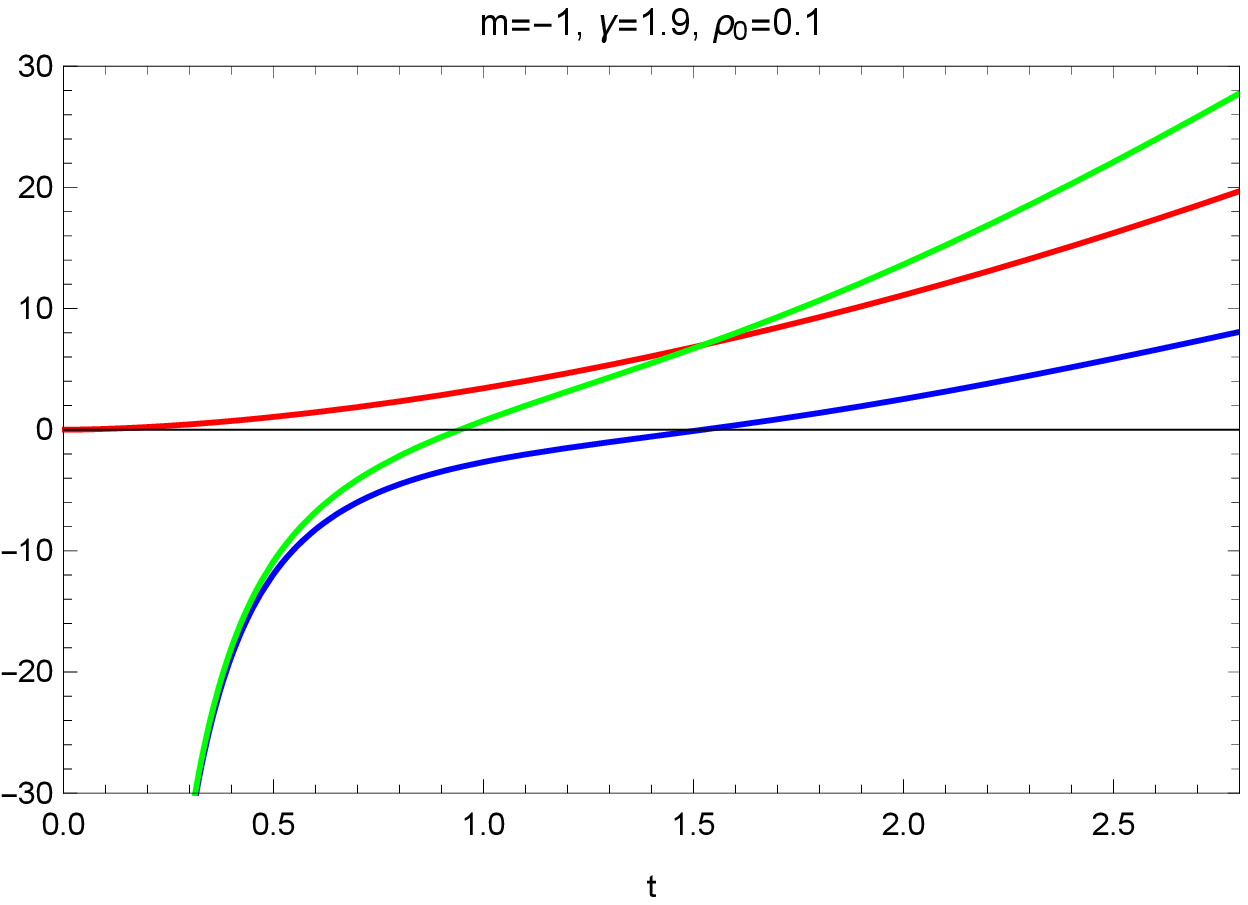}
	\end{minipage}
	\begin{minipage}{0.3\textwidth}
		\centering\includegraphics[height=5cm,width=5cm]{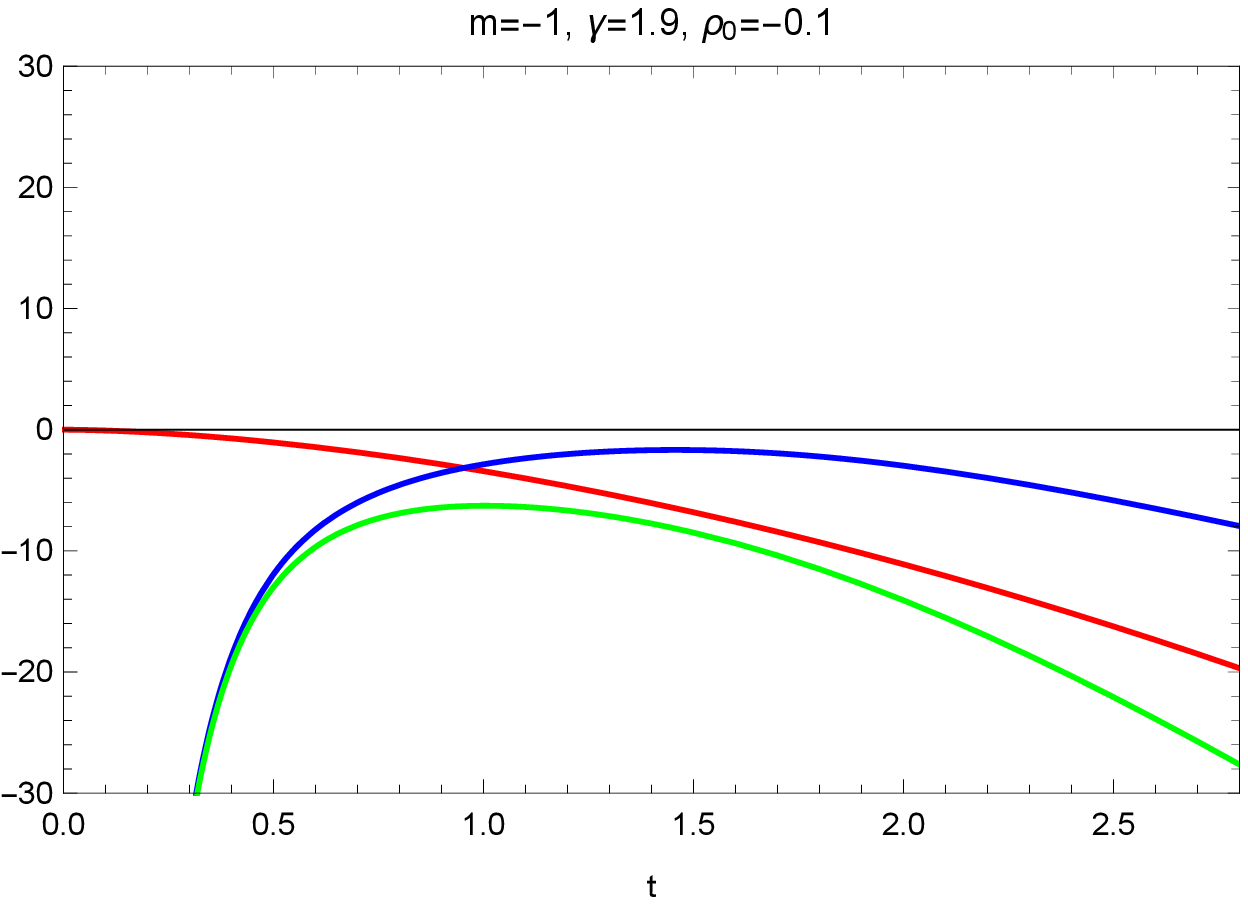}
	\end{minipage}
	\begin{minipage}{0.3\textwidth}
	\centering\includegraphics[height=5cm,width=5cm]{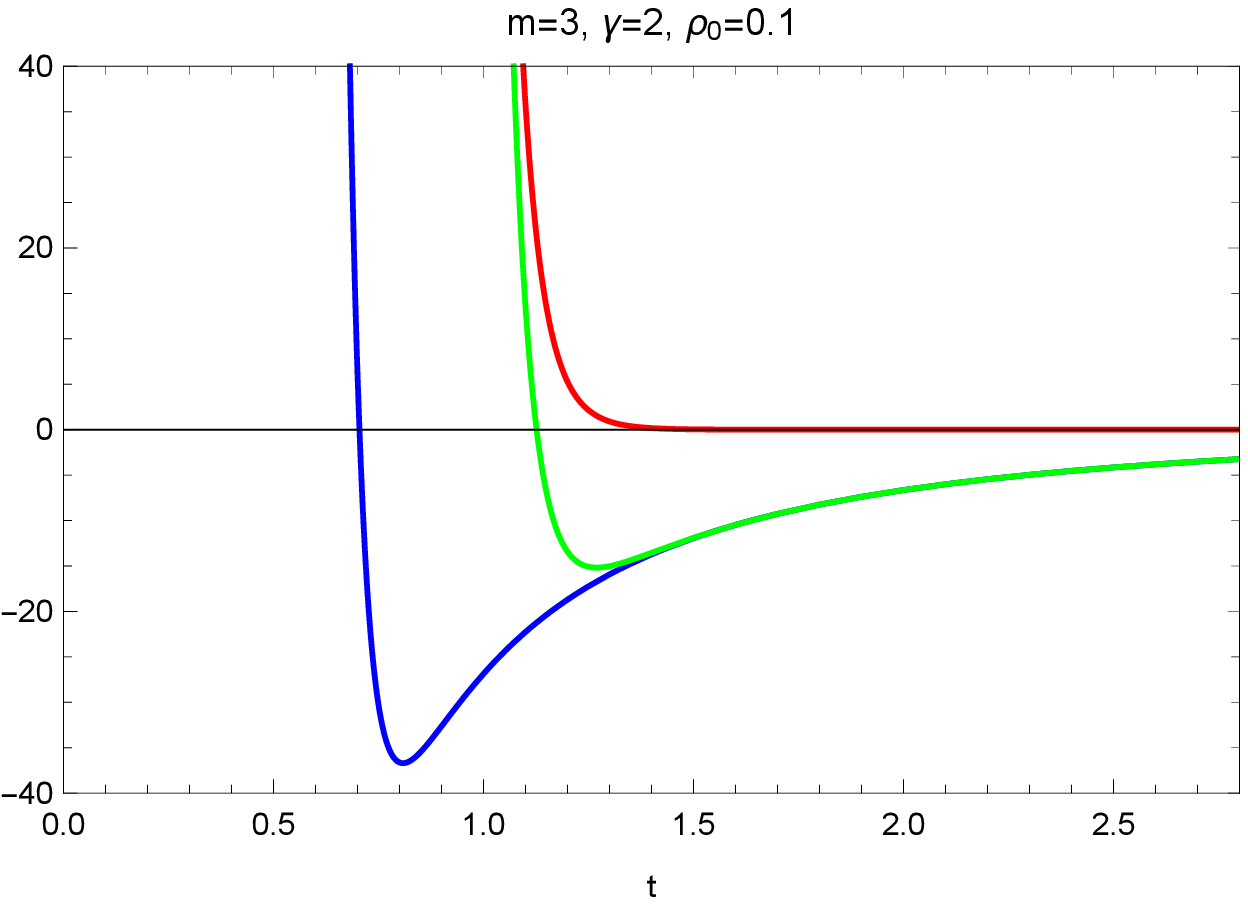}
\end{minipage}
	\begin{minipage}{0.3\textwidth}
	\centering\includegraphics[height=5cm,width=5cm]{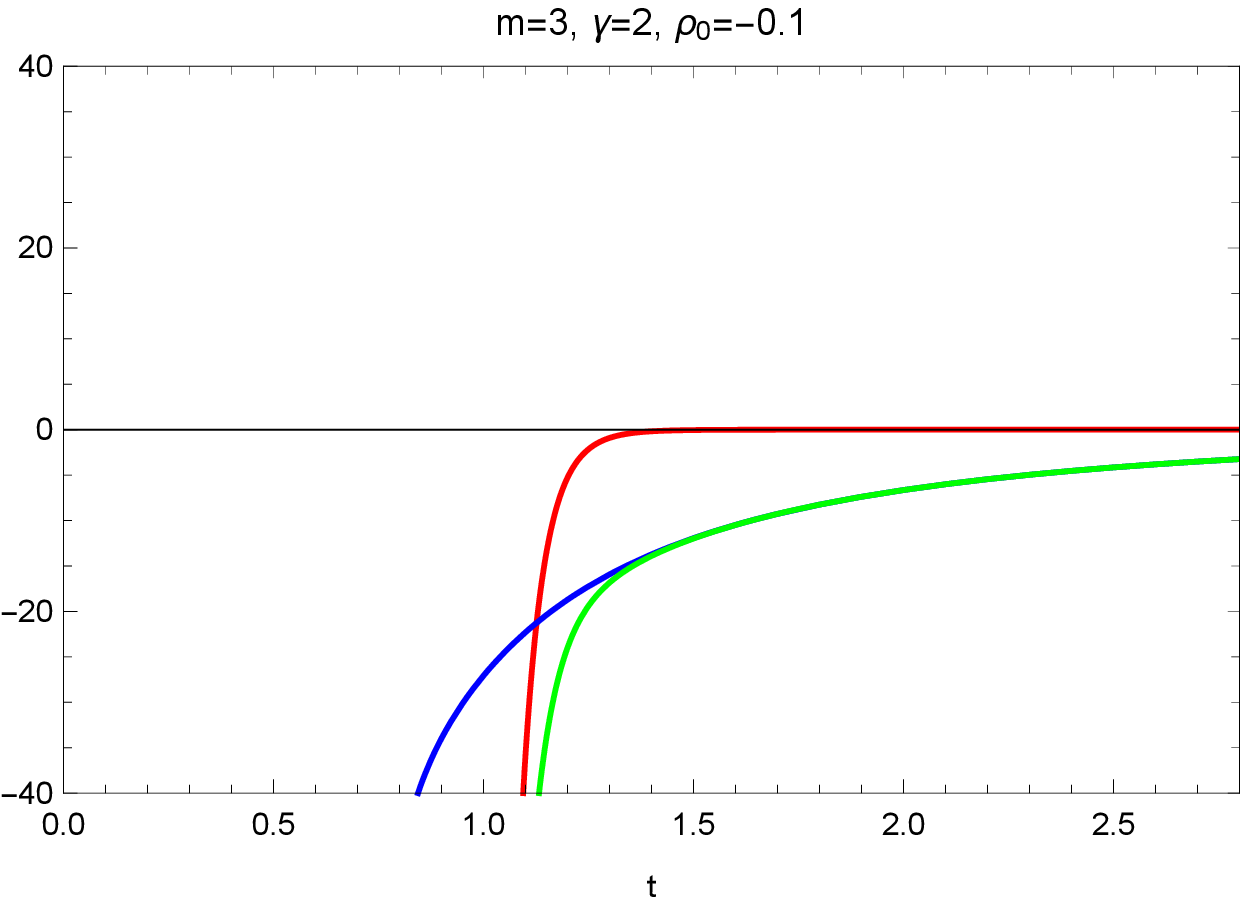}
\end{minipage}
	\begin{minipage}{0.85\textwidth}\caption{[Time variation of different terms: $R_1$ (red), $R_2$ (blue) and $R_{\mu\nu}u^\mu u^\nu=\tilde{R}$ (green) for $0\leq\gamma\leq2{~~}$. Here we choose $a(t)=t^{m}$ and $f(T)=\alpha T+\dfrac{\beta}{T}$, $\alpha=-1$, $\beta=1$.]}\label{f2}
	\end{minipage}
\end{figure}

$\underline{\boldsymbol{Model ~1 :}}$~~ $f(T)= \alpha(-T)^n$, $\alpha$ and $n$ are two model parameters \cite{Wu:2010xk}. This model has the same background evolution equation as some phenomenological
models \cite{Dvali:2003rk}, \cite{Chung:1999zs}. Further the model reduces to $\Lambda$CDM model at $n=0$ and to the DGP model \cite{Dvali:2000hr} at $n=\dfrac{1}{2}$. Thus for this model (setting $\alpha=1$ and varying $n$) one has
\begin{eqnarray}\label{eq41}	\tilde{R}={1.5\gamma\rho_{0}t^{-3m\gamma}}\left[1-n(6m^{2})^{(n-1)}t^{-2(n-1)}+2n(n-1)(-1)^{n}(6m^{2})^{(n-2)} t^{-2(n-2)}\right]^{-1}\\ \nonumber
	+\left[-\rho_{0}t^{-3m\gamma}+\frac{(6m^{2})^{n}}{2t^{2n}}\right]\left[1-\dfrac{2n(6m^{2})^{(n-1)}}{t^{2(n-1)}}\right]^{-1}
\end{eqnarray}
$\underline{\boldsymbol{Model~ 2 :}}$~~ $f(T)=\alpha T+\dfrac{\beta}{T}$,~ $\alpha$,$\beta$ are the model parameters \cite{Myrzakulov:2010vz}. The expression for Raychaudhuri scalar in this model is given by
\begin{eqnarray}\label{eq42}
\tilde{R}=1.5\gamma\rho_{0}t^{-3m\gamma}\left[1+\alpha+\dfrac{\beta t^{4}}{12m^{4}}\right]^{-1}\\ \nonumber
+\left[-\rho_{0}t^{-3m\gamma}-\dfrac{3\alpha m^{2}}{t^{2}}-\dfrac{\beta t^{2}}{12m^{2}}\right]\left[1+2\left(\alpha-\dfrac{\beta t^{4}}{36m^{4}}\right)\right]^{-1}
\end{eqnarray}
Now, $\tilde{R}$ has been split into two terms $R_{1}$ and $R_{2}$ for both the cases and time variation of $R_{1}$,$R_{2}$ and $\tilde{R}$ has been shown graphically in FIG. \ref{f1} and FIG. \ref{f2} to study the Convergence Condition (CC) in the two models.
\\
Based on the graphs one has the following findings:
\begin{itemize}
	\item For the model $f(T)=\alpha(-T)^{n}$, the CC~ i.e $\tilde{R}\geq0$ holds for  negative $m$ and $n$ while singularity may be avoided for positive exponents. It may be noted that in all cases SEC ($\rho+3p=(3\gamma-2)\rho_{0}t^{-3m\gamma}\geq0$) holds good. Therefore unlike Einstein gravity, it is possible to obviate singularity in Model 1.
\item For the model $f(T)=\alpha T+ \dfrac{\beta}{T}$, $\tilde{R}$ is either positive or indefinite in sign whenever SEC holds good, while negative $\rho_{0}$ or violation of SEC ($(3\gamma-2)\rho_{0}t^{-3m\gamma}<0$) yields $\tilde{R}<0$. Hence singularity may be avoided for any negative choice of $\rho_{0}$ irrespective of the exponent $m$.\\
\textbf{Remark:}\\
 From the above two choices of $f(T)$ we see that avoidance of singularity is very much related to the choice of $f(T)$ as well as the nature of the physical fluid considered. From the above study we find that in the first case with the positive power law choice of $f(T)$ it is possible to have avoidance of singularity even with the normal/usual fluid as the matter content of the universe, however for the second choice of $f(T)$ (a linear combination of linear power law and its inverse) it is found that there must be some ghost field that can lead to possible avoidance of singularity. Therefore the choice of $f(T)$ has a crucial role in identifying the singularity free nature of the space-time.
\end{itemize}
\section{A Quantum Description of the Raychaudhuri Equation}
We start by considering a family of hyper-surface orthogonal congruence of time-like geodesics in a $(r+1)$ dimensional spacetime-$\mathcal{M}$. Let $\eta_{\mu\nu}$ (orthogonal to time-like unit velocity vector field $u^{\mu}$ of the congruence) be the induced metric on the $r$- dimensional hyper-surface. One can define the dynamical degree of freedom as \cite{Alsaleh:2017ozf}
\begin{equation}
\Lambda(\tau)=\sqrt{\eta}
\end{equation}	
where $\eta=$det$(\eta_{\mu\nu})$ and $\tau$ is the proper time. $\Lambda$ is essentially related to the volume of the hyper-surface and $\Lambda=0$ identifies the singularity. The dynamical evolution of $\eta$ is given by \cite{Poisson:2009pwt}
\begin{equation}
	\dfrac{1}{\sqrt{\eta}}\dfrac{d\sqrt{\eta}}{d\tau}=\Theta
\end{equation}
where $\Theta$, the volume scalar of the congruence is given by (\ref{eq5*}). Thus,
\begin{equation}\label{eq47}
	\Lambda^{'}=\Lambda\Theta
\end{equation}
where ~ $'$ ~denotes differentiation with respect to $\tau$. The Raychaudhuri equation which essentially gives the evolution of the congruence can be written as
\begin{equation}\label{eq48}
	\dfrac{d\Theta}{d\tau}+~\dfrac{\Theta^{2}}{r}+~2\sigma^{2}+~\tilde{R}~=~0
\end{equation} where the expression for $\Theta$ and $\sigma$ are given by equations (\ref{eq5*}) and (\ref{eq6*}). The Raychaudhuri scalar is 
\begin{eqnarray}
\tilde{R}=R_{\mu\nu}u^{\mu}u^{\nu}.
\end{eqnarray}	
Since hyper-surface orthogonal congruence of time-like geodesics are taken into consideration, by virtue of Frobenius Theorem $\omega_{\mu\nu}=0$.
Using (\ref{eq47}) and (\ref{eq48}) one may obtain the RE as :
\begin{equation}
	\dfrac{\Lambda^{''}}{\Lambda}+\left(\dfrac{1}{r}-1\right)\dfrac{\Lambda'^{2}}{\Lambda^{2}}+2\sigma^{2}+\tilde{R}=0
\end{equation}
Let us write
\begin{equation}\label{eq53}
	\mathcal{F}=\dfrac{\Lambda^{''}}{\Lambda}+\left(\dfrac{1}{r}-1\right)\dfrac{\Lambda'^{2}}{\Lambda^{2}}+2\sigma^{2}+\tilde{R}=0
\end{equation}
The necessary and sufficient conditions which (\ref{eq53}) must satisfy for being the Euler-Lagrange equation corresponding to a Lagrangian $\mathcal{L}$ are known as the Helmholtz conditions \cite{Davis:1928}-\cite{Nigam:2016}. Therefore, with an aim to formulate a Lagrangian corresponding to which the Euler-Lagrange equation gives back the RE,  it has been found that $\mathcal{F}$ in the above form fails to satisfy all the Helmholtz conditions \cite{Davis:1928}-\cite{Nigam:2016}. However if one multiplies $\mathcal{F}$ by $\Lambda^{(\frac{2}{r}-1)}$ then
\begin{equation}
	\mathcal{\tilde{F}}=\Lambda^{(\frac{2}{r}-1)}\left[\dfrac{\Lambda^{''}}{\Lambda}+\left(\dfrac{1}{r}-1\right)\dfrac{\Lambda'^{2}}{\Lambda^{2}}+2\sigma^{2}+\tilde{R}\right],
\end{equation} satisfies all the Helmholtz conditions, since in the present model~ $2\sigma^{2}+\tilde{R}$ ~is a function of $\Lambda$ only.
The Lagrangian is given by
\begin{equation}\label{eq51*}
	\mathcal{L}=\dfrac{1}{2}\Lambda^{(\frac{2}{r}-2)}\Lambda'^{2}-V[\Lambda].
\end{equation}
A variation of the lagrangian $\mathcal{L}$ with respect to dynamical variable $\Lambda$ gives, 
\begin{equation}
	\delta\mathcal{L}=-\Lambda^{(\frac{2}{r}-1)}\left[\dfrac{\Lambda^{''}}{\Lambda}+\left(\dfrac{1}{r}-1\right)\dfrac{\Lambda'^{2}}{\Lambda^{2}}\right]\delta\Lambda-\delta V+\dfrac{d}{d\tau}\left(\Lambda^{(\frac{2}{r}-2)}\Lambda^{'}\delta\Lambda\right).
\end{equation}
Hence to get $\mathcal{\tilde{F}}=0$ from the principle of least action one needs,
\begin{equation}\label{eq55}
	\dfrac{\delta V[\Lambda]}{\delta\Lambda}=\Lambda^{(\frac{2}{r}-1)}\left(2\sigma^{2}+\tilde{R}\right).
\end{equation}
In the present model ($f(T)$ gravity constructed in the background of homogeneous and isotropic FLRW spacetime) one has  $\Lambda=a^{3}$ , $2\sigma^{2}=0$ and for the two choices of $f(T)$, the expression for $\tilde{R}$ are given by (\ref{eq41}) and (\ref{eq42}).
Finally solving the differential equation (\ref{eq55}) for $V$ one has the potential in the form\\
\begin{equation}\label{eq56}
		V=V_{0}+\dfrac{t^{\frac{2}{r}}}{12}g(t)
\end{equation} where,\\
$g(t)=
\left(\dfrac{2c_{0}t^{-\gamma}}{2l-m\gamma}\right)-\left(\dfrac{3t^{\left(\frac{-2}{3m}-\gamma\right)}~2^{F_{1}}~\left(1,-1+3l-\frac{3m\gamma}{2};3l-\frac{3m\gamma}{2};-2t^{\left(\frac{2}{3m}\right)}\right)c_{1}}{2-6l+3m\gamma}\right)+\\ ~~~~~\dfrac{3c_{2}t^{\left(-\frac{2}{3m}-\gamma\right)}}{(2-6l+3m\gamma)}-\left(\dfrac{12t^{\left(\frac{-2}{3m}-\gamma\right)}~2^{F_{1}}~\left(1,-1+3l-\frac{3m\gamma}{2};3l-\frac{3m\gamma}{2};-t^{\left(\frac{2}{3m}\right)}\right)c_{3}}{2-6l+3m\gamma}\right)+\\~~~~~~~~~~~~~~~~~~~~~~~~~~~~~~~~~~~~~~~~~~~~~~~~~~~~~~~~~~~~\dfrac{6~c_{4}~ 2^{F_{1}}~\left(1,-1+3l;~3l;-t^{\frac{2}{3m}}\right)}{(-1+3l)~t^{\frac{2}{3m}}},\\l=\dfrac{m}{r}$ and $V_{0}$,~$c_{0}$,~$c_{1}$,~$c_{2}$,~$c_{3}$, and $c_{4}$ are arbitrary constants. $2^{F_{1}}$ is the Gauss Hyper-geometric function.~\\For general $n$ in $f(T)=\alpha (-T)^n$, it is difficult to solve for $V$. Therefore $V$ has been found for $\alpha=1$ and $n=2$ for simplicity, while $f(T)=\alpha T+\dfrac{\beta}{T}$ with $\alpha=-1, \beta=1$ yields the expression for $V$ as
\begin{equation}\label{eq57}
	V=V_{0}+\dfrac{t^{\frac{2}{r}}}{6}h(t)
	\end{equation} where,\\~~ $h(t)=
\left(\dfrac{2t^{-\gamma}}{2l-m\gamma}\right)-\left(\dfrac{6t^{\left(\frac{4}{3m}-\gamma\right)}~2^{F_{1}}~\left(1,1+\frac{3l}{2}-\frac{3m\gamma}{4};2+\frac{3l}{2}-\frac{3m\gamma}{4};-t^{\left(\frac{4}{3m}\right)}k_{0}\right)k_{0}}{4+6l-3m\gamma}\right)+\\ ~~~~~~~~~~~~~~~~~~~~~~~~~~~~~~~~~~~\dfrac{3k_{2}}{(3l-1)t^{\left(\frac{2}{3m}\right)}}+~~ \dfrac{3t^{\left(\frac{2}{3m}\right)}~2^{F_{1}}~\left(1,\frac{1+3l}{2};\frac{3(l+1)}{2};-t^{\left(\frac{4}{3m}\right)}k_{4}\right)k_{3}}{1+3l}+\\~~~~~~~~~~~2k_{1}t^{-\gamma}\left(\dfrac{l}{2l-m\gamma}-\dfrac{3t^{\left(\frac{4}{3m}\right)}~2^{F_{1}}~\left(1,1+\frac{3l}{2}-\frac{3m\gamma}{4};2+\frac{3l}{2}-\frac{3m\gamma}{4};-t^{\left(\frac{4}{3m}\right)}k_{4}\right)k_{4}}{4+6l-3m\gamma}\right), l=\dfrac{m}{r}$ and $V_{0}$, $k_{0}$, $k_{1}$, $k_{2}$, $k_{3}$ and $k_{4}$ are constants. $2^{F_{1}}$ is the Gauss Hyper-geometric function.\\
Now, we aim to find the Hamiltonian in operator version. This is because in this form, it admits a self-adjoint extension quite generally and therefore the conservation of probability is ensured. The momentum conjugate to the configuration variable $\Lambda$ is given by
\begin{equation}\label{eq56*}
	\Pi_{\Lambda}=\dfrac{\partial\mathcal{L}}{\partial\Lambda}=\Lambda^{2(\frac{1}{r}-1)}\Lambda'.
\end{equation}
Hence, the Hamiltonian is given by:
\begin{equation}\label{eq57*}
	\mathcal{H}=\dfrac{1}{2}\Lambda^{2(1-\frac{1}{r})}\Pi_{\Lambda}^{2}+V[\Lambda].
\end{equation}
It may be verified that, the Hamilton's equation of motion gives the RE and definition of momentum. For canonical quantization of the system under consideration, $\Lambda$ and $\Pi_{\Lambda}$ are considered as operators acting on the state vector $\Psi(\Lambda,\lambda)$ of the geometric flow. In $\Lambda$-representation, the operators assume the form:\begin{equation}
	\tilde{\Lambda}\longrightarrow{\Lambda}
\end{equation} and\begin{equation}
	\tilde{\Pi}_{\Lambda}\longrightarrow{-i\hbar\dfrac{\partial}{\partial{\Lambda}}}.
\end{equation} One may verify,
\begin{equation}
	[\tilde{\Lambda}  , \tilde{\Pi}_{\Lambda}]=i\hbar
\end{equation}
So the evolution of the state vector is given by
\begin{equation}\label{eq62}
	i\hbar\dfrac{\partial\Psi}{\partial\lambda}=\tilde{H}\Psi
\end{equation} with ,
\begin{equation}\label{eq62}
	\tilde{H}=-\dfrac{\hbar^{2}}{2}\Lambda^{2(1-\frac{1}{r})}\dfrac{\partial^{2}}{\partial\Lambda^{2}}+V[\Lambda]
\end{equation} being , the operator version of the Hamiltonian. Equation (\ref{eq62}) is known as the quantized RE. In the context of cosmology, there is notion of Hamiltonian constraint and operator version of it acting on the wave function of the universe ($\Psi$) i.e,\begin{equation}
\tilde{\mathcal{H}}\Psi=0,
\end{equation}known as the WD equation \cite{Wheeler:1968iap}, \cite{Jalalzadeh:2016gqs}. However, there is a problem of non-unitary evolution \cite{Pinto-Neto:2013toa}-\cite{Pal:2014dya}, which can be resolved by the proper choice of operator ordering in the first term of the Hamiltonian. One may note that the operator form (\ref{eq62}) is symmetric with norm given by
\begin{equation}
	|\Psi|^{2}=\int_{0}^{\infty}d\Lambda \Lambda^{2\left(\frac{1}{r}-1\right)}\Psi^{*}\Psi ,
\end{equation}but it fails to be self-adjoint. However, one may extend it as a self-adjoint operator with the following operator ordering \cite{Pal:2016ysz}
\begin{equation}
\tilde{H}=-\dfrac{\hbar^{2}}{2}\Lambda^{(1-\frac{1}{r})}\dfrac{\partial}{\partial\Lambda}\Lambda^{(1-\frac{1}{r})}\dfrac{\partial}{\partial\Lambda}+V[\Lambda]
\end{equation}
Further, if a change of minisuperspace variable is carried out as
\begin{equation}\label{eq68*}
	v=r\Lambda^{\frac{1}{r}}
\end{equation} then the transformed WD equation is written as:
\begin{equation}\label{eq68}
	\left[\dfrac{-\hbar^{2}}{2}\dfrac{d^{2}}{dv^{2}}+V(v)\right]\Psi(v)=0,
\end{equation} with symmetric norm as
\begin{equation}
	|\Psi|^{2}=\int_{0}^{\infty}dv\Psi^{*}\Psi.
\end{equation}	
The above WD equation (\ref{eq68}) can be interpreted as time-independent Schrödinger equation of a point particle of unit mass moving along $v$ direction in a potential field $V(v)$ and it has zero eigen value of the Hamiltonian and the wave function is the corresponding energy eigen function.\\

\textbf{Possible Solutions of the Wheeler-Dewitt equation in the present model:}\\

For the first model, the expression for the potential V is given by (\ref{eq56}). Now we opt for particular choice of the arbitrary integration constants involved in the expression of potential. This is because, it has been found post calculation that other choices lead to either unrealistic cases or complicated calculations. Therefore for the choice $c_{0}=\dfrac{2l-m\gamma}{2}, c_{1}=c_{2}=c_{3}=c_{4}=0$, one has
\begin{equation}
	V=V_{0}+v_{0}v^{\delta},
\end{equation}~~~~~~~~~~~~~~~~~~~~~~~~~~~~~~~~~~~~~~~~~~~~ where $\delta=\dfrac{\left(\dfrac{2}{3}-\gamma\right)}{3l}=\dfrac{\left(\dfrac{2}{3}-\gamma\right)}{m}$, $v_{0}=\dfrac{1}{12\times3^{\frac{\left(\frac{2}{3}-\gamma\right)}{3l}}}=\dfrac{1}{12\times3^{\frac{\left(\frac{2}{3}-\gamma\right)}{m}}}$.\\ Since in the general quantum description we have considered $(r+1)$ dimensional spacetime, so in the present model $r=3$ and $3l=m$.\\

The WD equation in this case is written as
\begin{equation}
	\dfrac{d^{2}\Psi(v)}{dv^{2}}-\dfrac{2v_{0}}{\hbar^{2}}\left(\dfrac{V_{0}}{v_{0}}+v^{\delta}\right)\Psi(v)=0
\end{equation} i.e,
\begin{equation}
	\left(-\dfrac{\hbar^{2}}{2v_{0}}\dfrac{d^{2}}{dv^{2}}+v^{\delta}\right)\Psi(v)=\left(-\dfrac{V_{0}}{v_{0}}\right)\Psi(v).
	\end{equation} This form of the WD equation can be interpreted as the energy eigen value equation of a particle of mass $v_{0}$ moving in a potential field $V(v)=v^{\delta}$ with energy eigen value $\left(-\dfrac{V_{0}}{v_{0}}\right)$ and the wave function of the universe is nothing but the corresponding energy eigen function.\\
The solution to the above equation for $\delta=-1$ is,
\begin{equation}
	\Psi(v)= K_{1}~ M_{\frac{-1}{\sqrt{2V_{0}}}\frac{v_{0}}{\hbar},\frac{1}{2}}\left(\dfrac{2}{\hbar}\sqrt{2V_{0}}v\right)+ K_{2}~ W_{\frac{-1}{\sqrt{2V_{0}}}\frac{v_{0}}{\hbar},\frac{1}{2}}\left(\dfrac{2}{\hbar}\sqrt{2V_{0}}v\right),
\end{equation}
where $K_{1}$ and $K_{2}$ are arbitrary integration constants. $M$ and $W$ are the usual Whittaker functions of 1st and 2nd kind.\\

The solution for $\delta=1$ is given by
\begin{equation}
	\Psi(v)=C_{1}~ Ai\left[\left(\dfrac{2v_{0}}{\hbar^{2}}\right)^{\frac{1}{3}}\left(\dfrac{V_{0}}{v_{0}}+v\right)\right]+ C_{2}~ Bi\left[\left(\dfrac{2v_{0}}{\hbar^{2}}\right)^{\frac{1}{3}}\left(\dfrac{V_{0}}{v_{0}}+v\right)\right],
\end{equation}
where $C_{1}$, $C_{2}$ are arbitrary integration constants. $Ai$ and $Bi$ are the Airy and Bairy functions.\\

For $\delta=-2$, the solution is given by
\begin{equation}
	\Psi(v)= B_{1}~\sqrt{v}~J_{\frac{1}{2}\sqrt{1+\frac{8v_{0}}{\hbar^{2}}}}\left(\sqrt{-2\dfrac{V_{0}}{\hbar^{2}}}v\right)+ B_{2}~\sqrt{v}~Y_{\frac{1}{2}\sqrt{1+\frac{8v_{0}}{\hbar^{2}}}}\left(\sqrt{-2\dfrac{V_{0}}{\hbar^{2}}}v\right),
\end{equation}
where $B_{1}$ and $B_{2}$ are arbitrary integration constants. $J$ and $Y$ are the Bessel functions of 1st and 2nd kind.\\

\begin{figure}[h!]
	\begin{minipage}{0.4\textwidth}
		\centering\includegraphics[height=6cm,width=6cm]{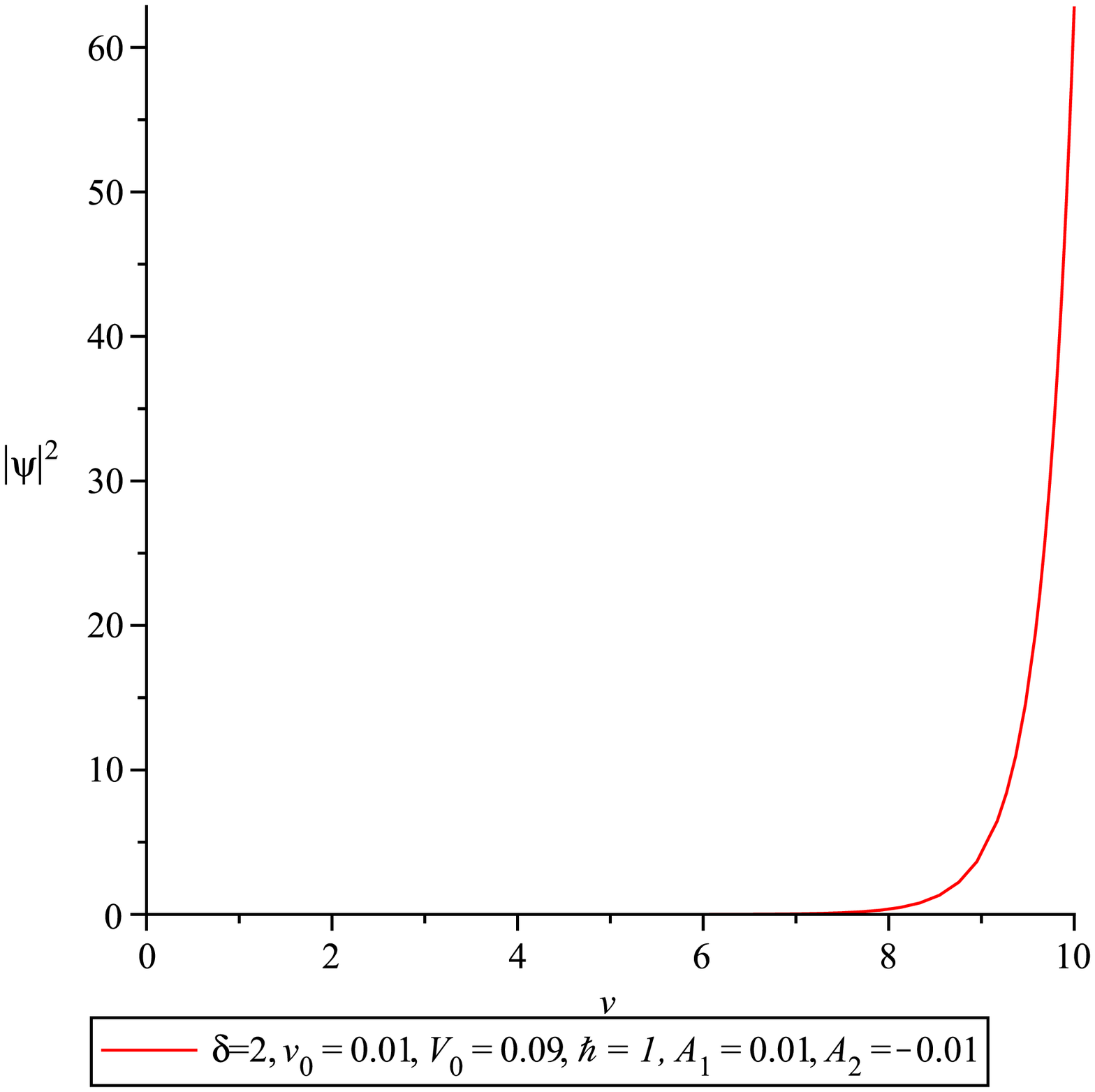}
	\end{minipage}\hfill
	\begin{minipage}{0.4\textwidth}
		\centering\includegraphics[height=6cm,width=6cm]{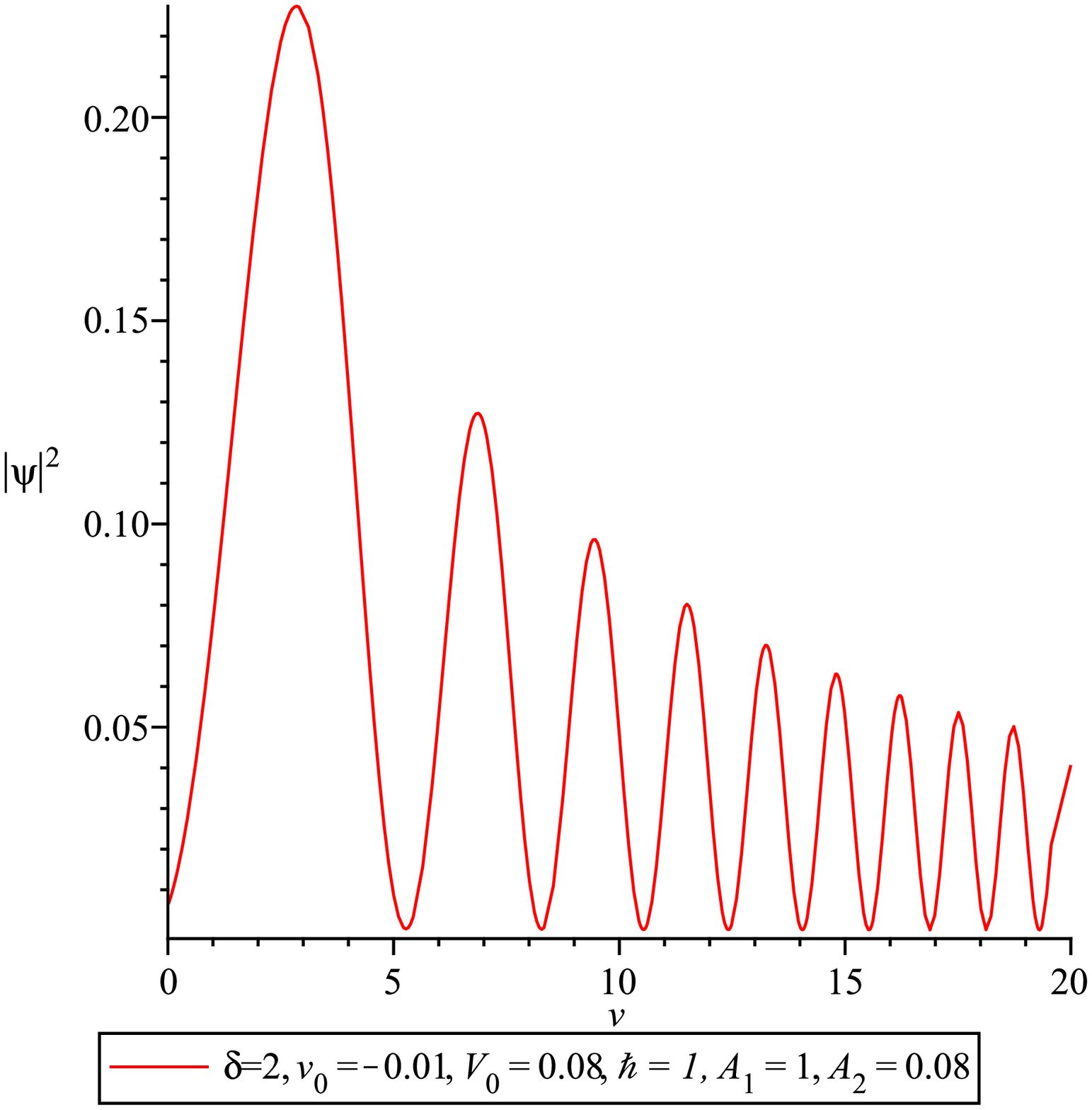}
	\end{minipage}
\caption{$|\psi|^{2}$ vs $v$ for Model 1 for various choices of parameters specified in each panel.}\label{f3}
\end{figure}

\begin{figure}[h!]
	\begin{minipage}{0.4\textwidth}
		\centering\includegraphics[height=6cm,width=6cm]{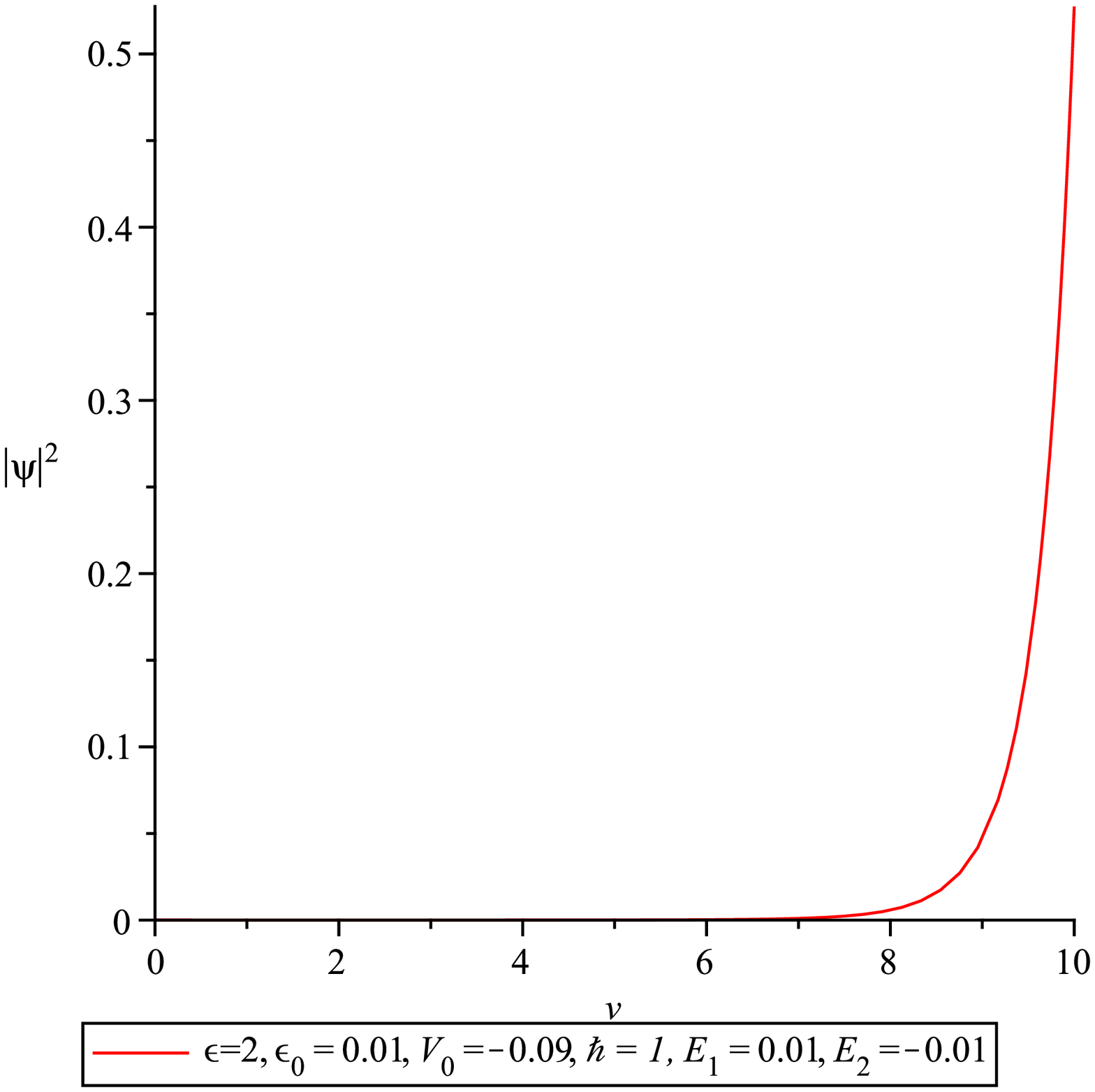}
	\end{minipage}\hfill
	\begin{minipage}{0.4\textwidth}
		\centering\includegraphics[height=6cm,width=6cm]{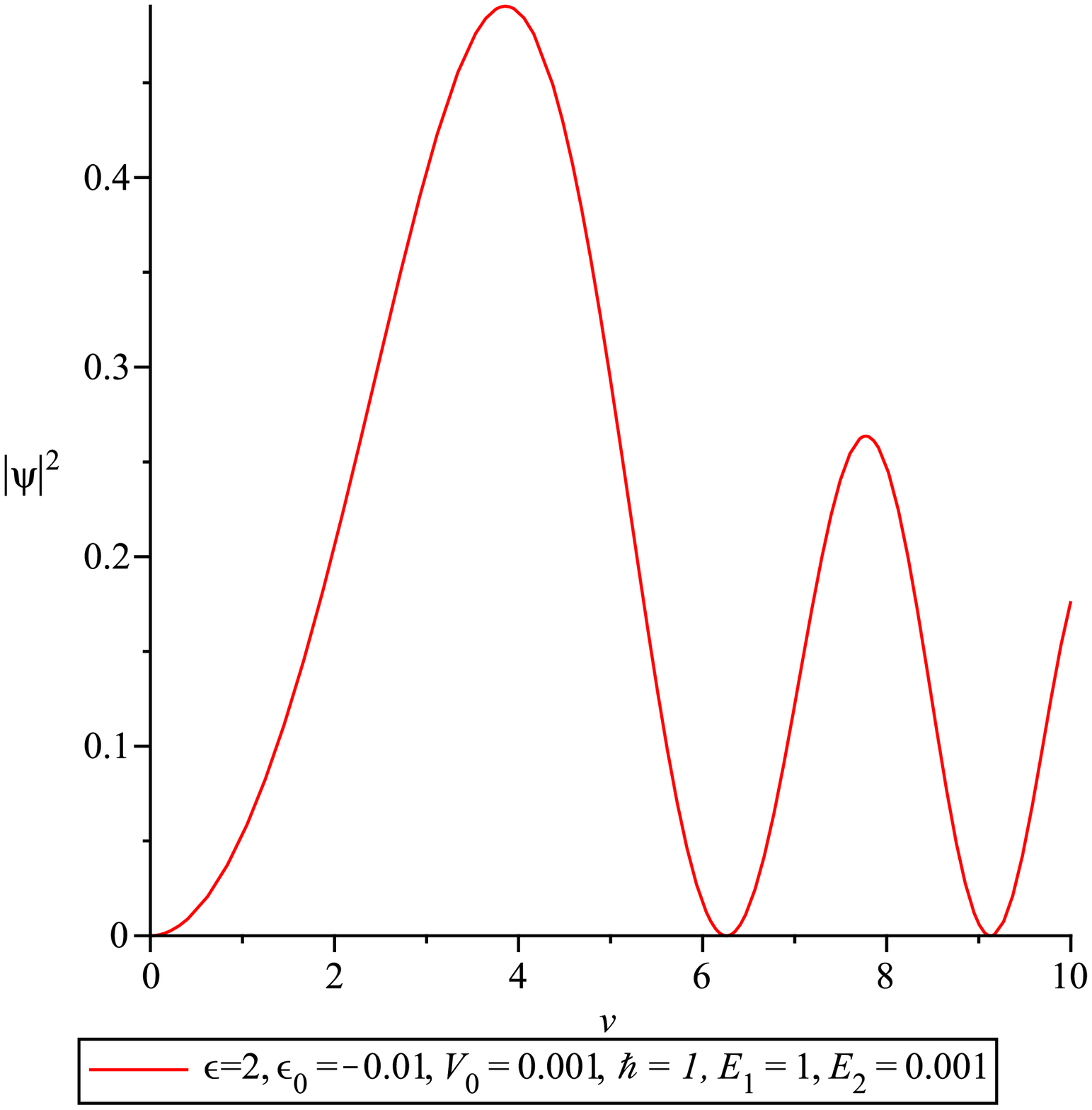}
	\end{minipage}
\caption{$|\Psi|^{2}$ vs $v$ for Model 2 for various choices of parameters specified in each panel.}\label{f4}
\end{figure}

For $\delta=2$, the solution is given by
\begin{equation}
	\Psi(v)= A_{1}\dfrac{ M_{\frac{-1}{4}\sqrt{\frac{2}{v_{0}}}\frac{V_{0}}{\hbar},~\frac{1}{4}}\left( \sqrt{2v_{0}}\dfrac{v^{2}}{\hbar}\right)}{\sqrt{v}}~+A_{2}\dfrac{ W_{\frac{-1}{4}\sqrt{\frac{2}{v_{0}}}\frac{V_{0}}{\hbar},~\frac{1}{4}}\left( \sqrt{2v_{0}}\dfrac{v^{2}}{\hbar}\right)}{\sqrt{v}},
\end{equation}
where $A_{1}$, $A_{2}$ are arbitrary integration constants. $M$ and $W$ are the usual Whittaker functions of 1st and 2nd kind.\\
For the second model, the expression for V is given by (\ref{eq57}). The choice $k_{0}=k_{1}=k_{2}=k_{3}=k_{4}=0$ yields
\begin{equation}
	V=V_{0}+~\epsilon_{0} v^{\epsilon},
\end{equation} where $\epsilon=\dfrac{\left(\dfrac{2}{3}-\gamma\right)}{m}$,~ $\epsilon_{0}=\dfrac{1}{\epsilon m^{2}\times 3^{\left(1+\frac{\epsilon}{3}\right)}}$.\\
The WD equation in this case turns out to be\\
\begin{equation}
\dfrac{d^{2}\Psi(v)}{dv^{2}}-\dfrac{2\epsilon_{0}}{\hbar^{2}}\left(\dfrac{V_{0}}{\epsilon_{0}}+ v^{\epsilon}\right)\Psi(v)=0
\end{equation} \\
or,
\begin{equation}
	\left(-\dfrac{\hbar^{2}}{2\epsilon_{0}}\dfrac{d^{2}}{dv^{2}}+v^{\epsilon}\right)\Psi(v)=\left(-\dfrac{V_{0}}{\epsilon_{0}}\right)\Psi(v).
\end{equation}\\ The above equation can be compared to energy eigen value equation of a particle having mass $\epsilon_{0}$ and moving in a potential field $V(v)=v^{\epsilon}$ with energy eigen value $\left(-\dfrac{V_{0}}{\epsilon_{0}}\right)$ and the wave function can be interpreted as the corresponding energy eigen function.\\
The solution corresponding to $\epsilon=-1$ is given by
\begin{equation}
		\Psi(v)= F_{1}~ M_{\frac{-1}{\sqrt{2V_{0}}}\frac{\epsilon_{0}}{\hbar},\frac{1}{2}}\left(\dfrac{2}{\hbar}\sqrt{2V_{0}}v\right)+ F_{2}~ W_{\frac{-1}{\sqrt{2V_{0}}}\frac{\epsilon_{0}}{\hbar},\frac{1}{2}}\left(\dfrac{2}{\hbar}\sqrt{2V_{0}}v\right),
\end{equation}
where $F_{1}$ and $F_{2}$ are arbitrary integration constants. $M$ and $W$ are the usual Whittaker functions of 1st and 2nd kind.\\
The solution for $\epsilon=1$ is given by
\begin{equation}
\Psi(v)=D_{1}~ Ai\left[\left(\dfrac{2\epsilon_{0}}{\hbar^{2}}\right)^{\frac{1}{3}}\left(\dfrac{V_{0}}{\epsilon_{0}}+v\right)\right]+ D_{2}~ Bi\left[\left(\dfrac{2\epsilon_{0}}{\hbar^{2}}\right)^{\frac{1}{3}}\left(\dfrac{V_{0}}{\epsilon_{0}}+v\right)\right],
\end{equation} where $D_{1}$ and $D_{2}$ are arbitrary integration constants. $Ai$ and $Bi$ are the Airy and Bairy functions.\\
For $\epsilon=-2$, one has the solution,
\begin{equation}
	\Psi(v)= G_{1}~\sqrt{v}~J_{\frac{1}{2}\sqrt{1+\frac{8\epsilon_{0}}{\hbar^{2}}}}\left(\sqrt{-2\dfrac{V_{0}}{\hbar^{2}}}v\right)+ G_{2}~\sqrt{v}~Y_{\frac{1}{2}\sqrt{1+\frac{8\epsilon_{0}}{\hbar^{2}}}}\left(\sqrt{-2\dfrac{V_{0}}{\hbar^{2}}}v\right),
\end{equation}
where $G_{1}$ and $G_{2}$ are arbitrary integration constants. $J$ and $Y$ are the Bessel functions of 1st and 2nd kind.\\
For $\epsilon=2$, the wave function is given by
\begin{equation}
		\Psi(v)= E_{1}\dfrac{ M_{\frac{-1}{4}\sqrt{\frac{2}{\epsilon_{0}}}\frac{V_{0}}{\hbar},~\frac{1}{4}}\left( \sqrt{2\epsilon_{0}}\dfrac{v^{2}}{\hbar}\right)}{\sqrt{v}}~+E_{2}\dfrac{W_{\frac{-1}{4}\sqrt{\frac{2}{\epsilon_{0}}}\frac{V_{0}}{\hbar},~\frac{1}{4}}\left( \sqrt{2\epsilon_{0}}\dfrac{v^{2}}{\hbar}\right) }{\sqrt{v}}
\end{equation} where $E_{1}$ and $E_{2}$ are arbitrary integration constants. $M$ and $W$ are the usual Whittaker functions of 1st and 2nd kind.\\
It follows from (\ref{eq68*}) that in the present model one has,
$v=3a(t)$~(since $\Lambda=a^{3}, r=3$). Now, to have an idea about probability measure on the minisuperspace  $|\Psi|^{2}$ has been plotted against $v$ as a tool to perform singularity analysis in the quantum regime for both the models in FIG.\ref{f3} (Model 1) and FIG.\ref{f4} (Model 2). Based on the plots one has the following findings:
\begin{itemize}
\item From FIG.\ref{f3}, it is clear that the probability of approaching singularity (i.e zero volume) is zero for the choice of the parameter $\delta=-1, -2$, while for $\delta=+1, +2$, there is a finite non-zero probability for the existence of initial big-bang singularity.
\end{itemize}
\begin{itemize}
	\item From FIG.\ref{f4}, the probability of having zero volume is zero for all the cases except for a typical choice of $\epsilon=1$. Therefore, similar to the previous model it is possible to obviate big-bang singularity by the present quantum description. 
\end{itemize}
\textbf{Remark:} In this section, canonical quantization technique has been used to study the singularity. The basic question that we have attempted to investigate is whether quantum formulation may overrule the singularity particularly the initial big-bang singularity. Essentially we have examined it by studying the probability function near the classical singularity. Similar to classical analysis we have found that probability is zero at big-bang singularity (i.e. avoidance of singularity) if the potential corresponding to the  to
	the dynamical system representing the congruence of time-like geodesics is in (i) inverse power law (linear and quadratic) form in case of model 1, and (ii) in linear, quadratic and inverse linear power law form for model 2.  Therefore we may conclude that the present quantum description may eliminate the classical singularity.
\begin{figure}[h!]
	\begin{minipage}{0.4\textwidth}
		\centering\includegraphics[height=6cm,width=6cm]{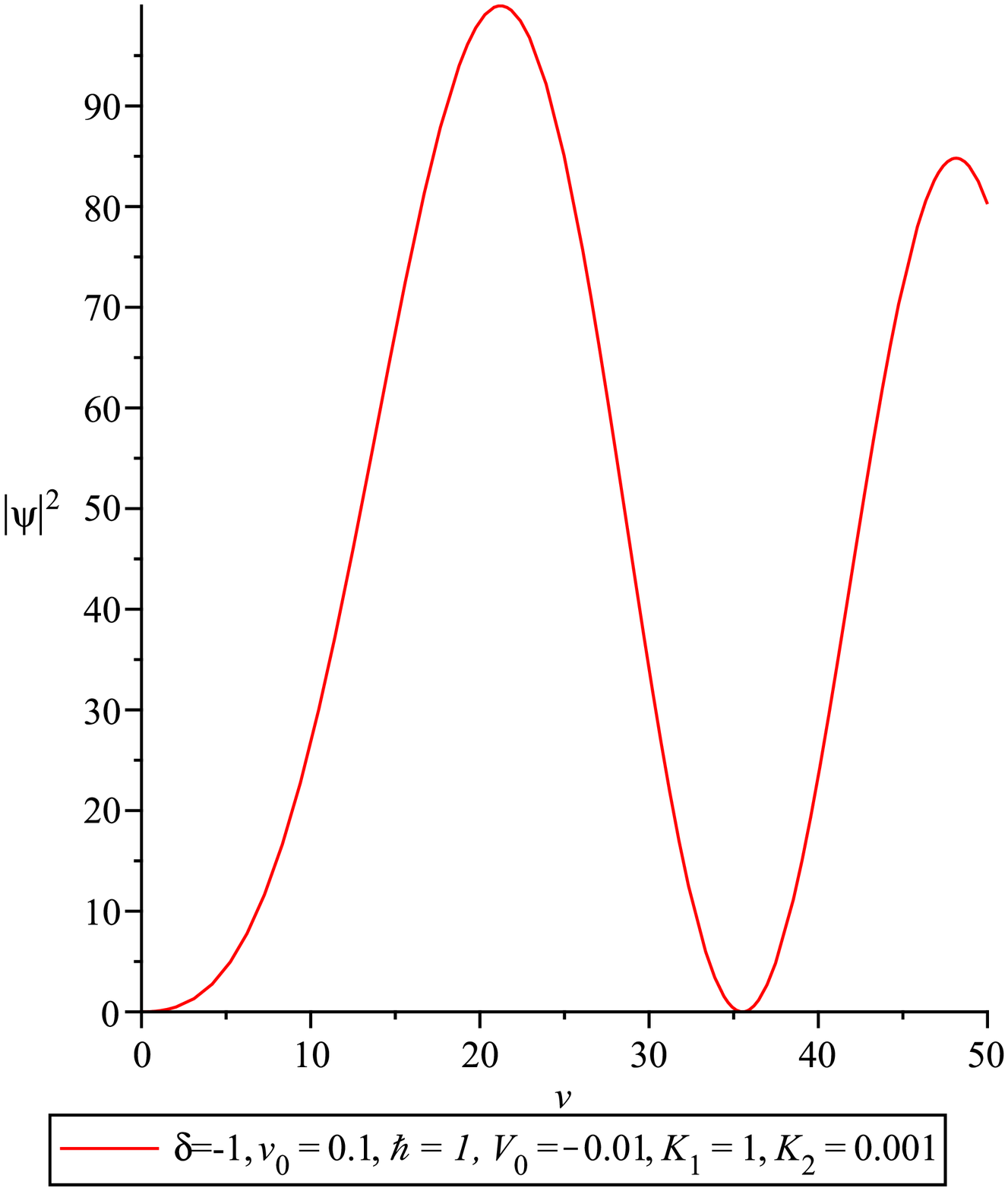}
	\end{minipage}\hfill
	\begin{minipage}{0.4\textwidth}
		\centering\includegraphics[height=6cm,width=6cm]{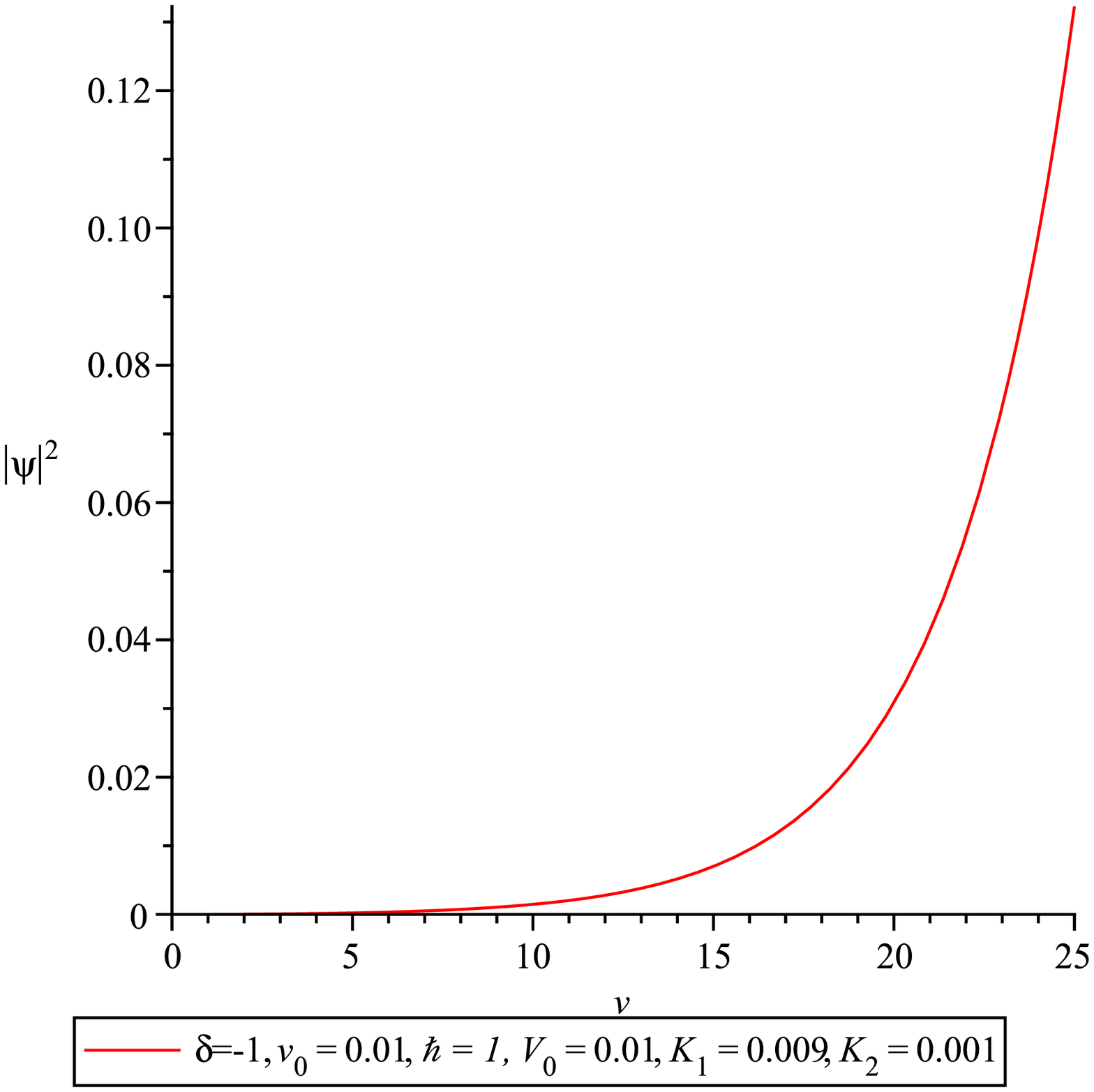}
	\end{minipage}
\end{figure}
\begin{figure}[h!]
	\begin{minipage}{0.4\textwidth}
		\centering\includegraphics[height=6cm,width=6cm]{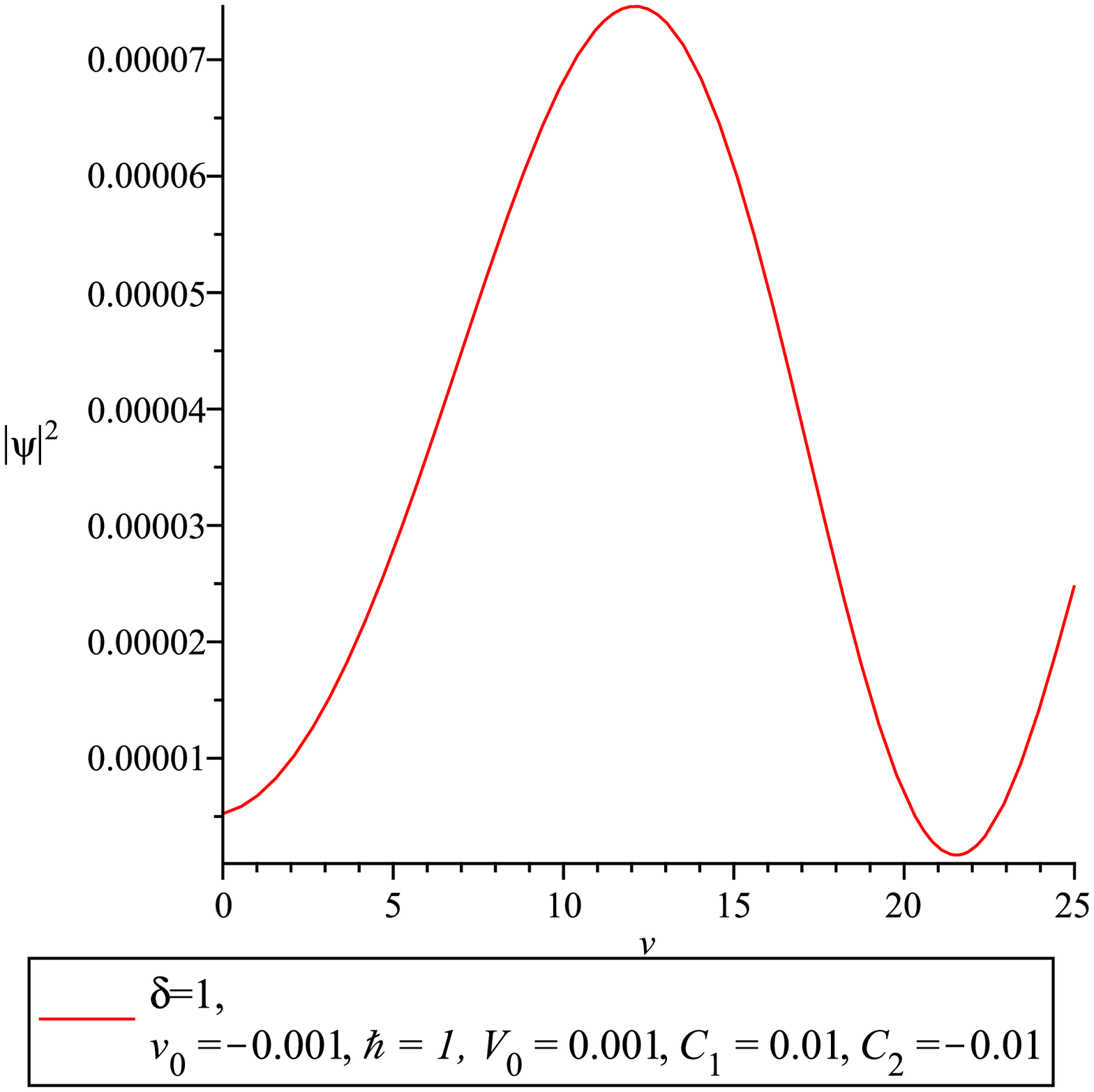}
	\end{minipage}\hfill
	\begin{minipage}{0.4\textwidth}
		\centering\includegraphics[height=6cm,width=6cm]{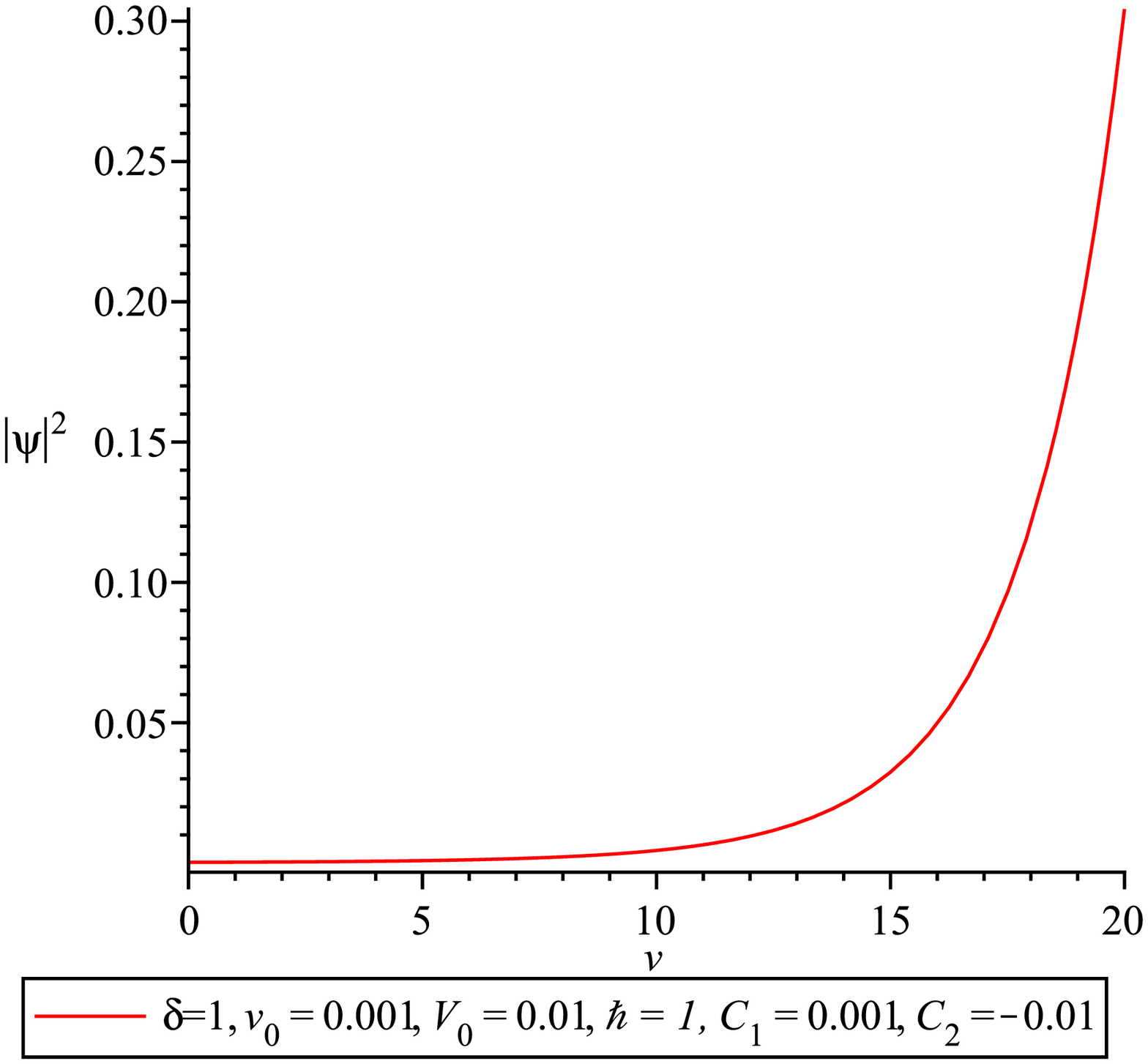}
	\end{minipage}
\end{figure}
\begin{figure}[h!]
	\begin{minipage}{0.4\textwidth}
		\centering\includegraphics[height=6cm,width=6cm]{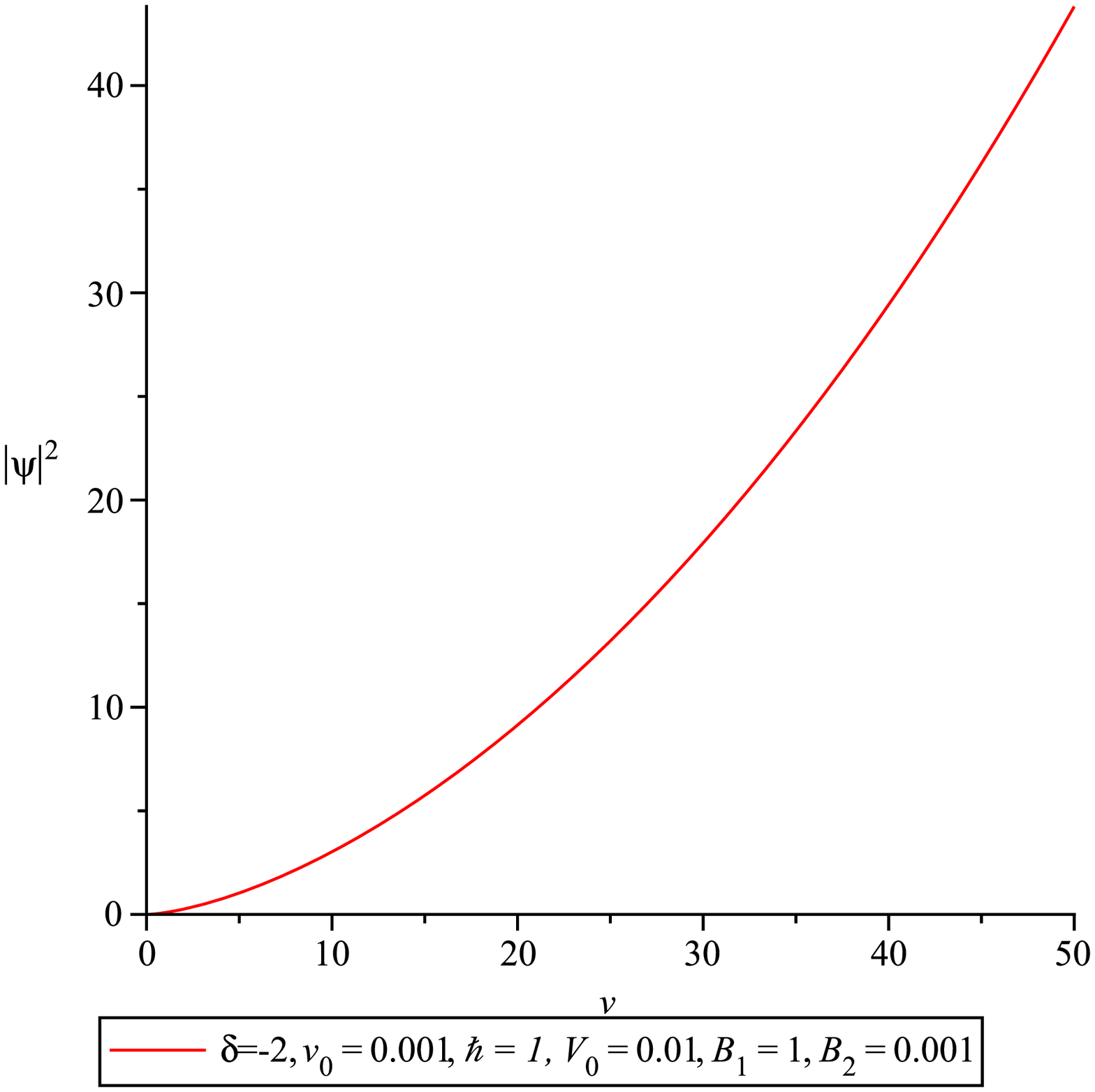}
	\end{minipage}\hfill
	\begin{minipage}{0.4\textwidth}
		\centering\includegraphics[height=6cm,width=6cm]{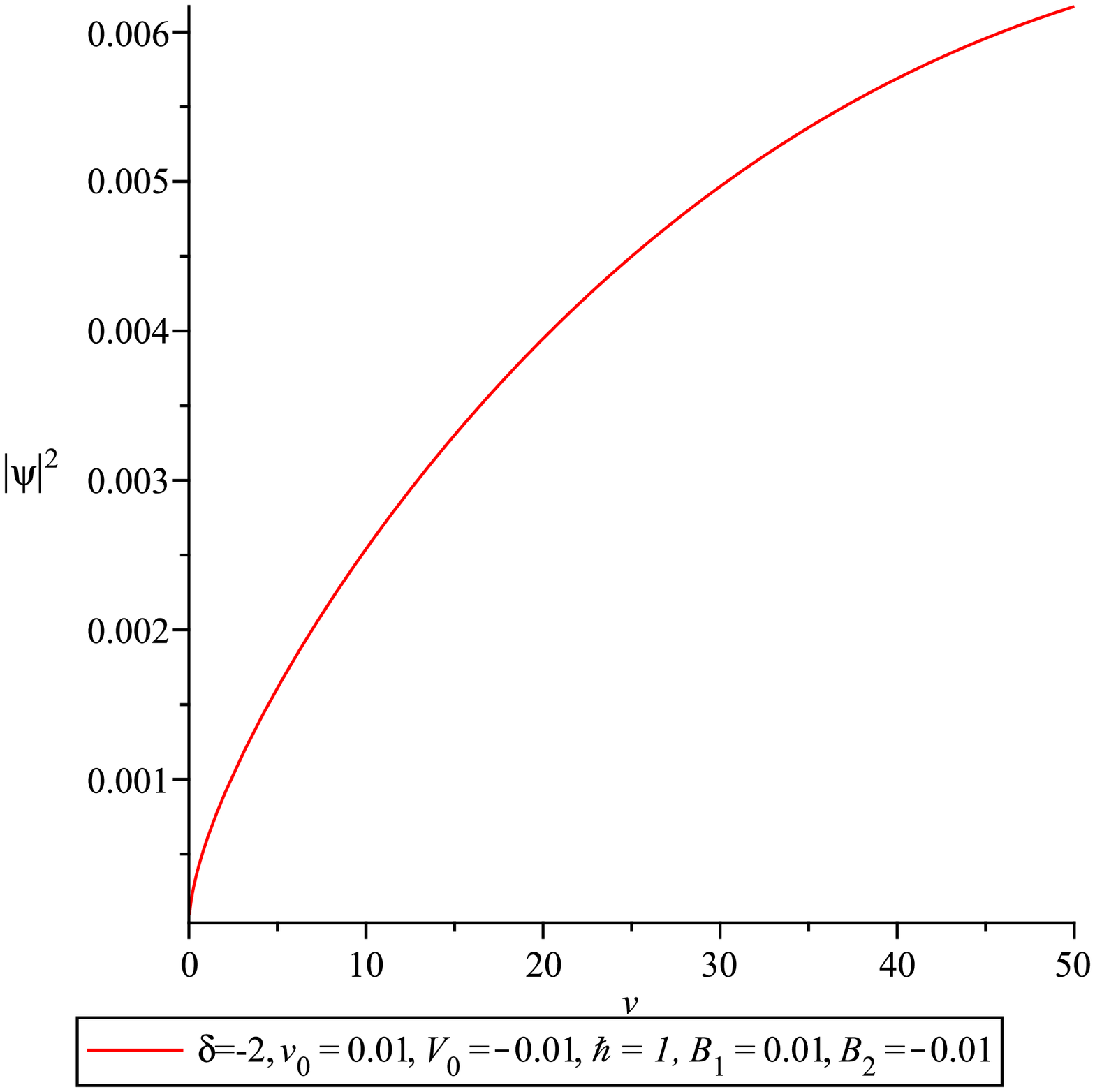}
	\end{minipage}
\end{figure}
\begin{figure}[h!]
	\begin{minipage}{0.4\textwidth}
		\centering\includegraphics[height=6cm,width=6cm]{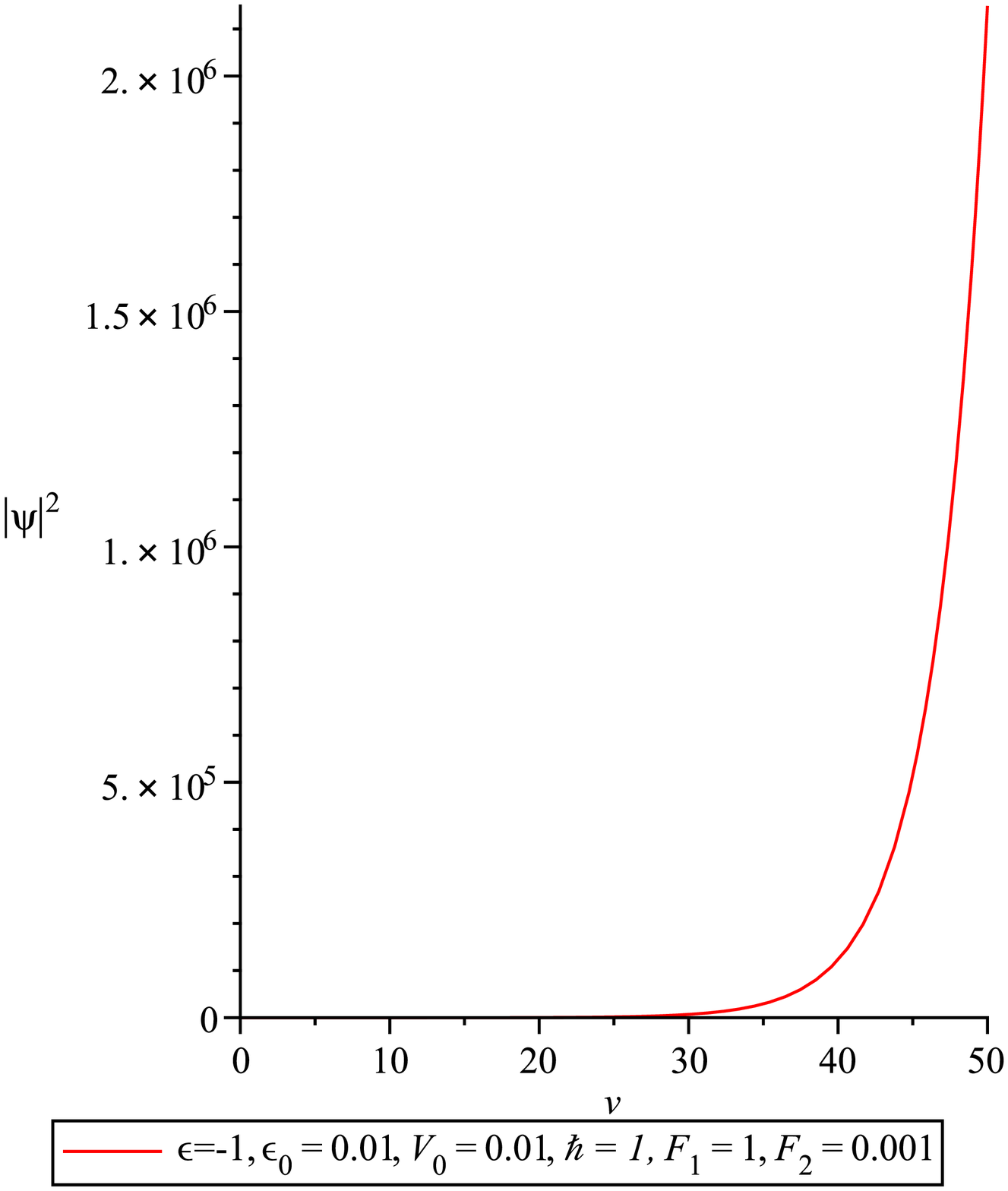}
	\end{minipage}\hfill
	\begin{minipage}{0.4\textwidth}
		\centering\includegraphics[height=6cm,width=6cm]{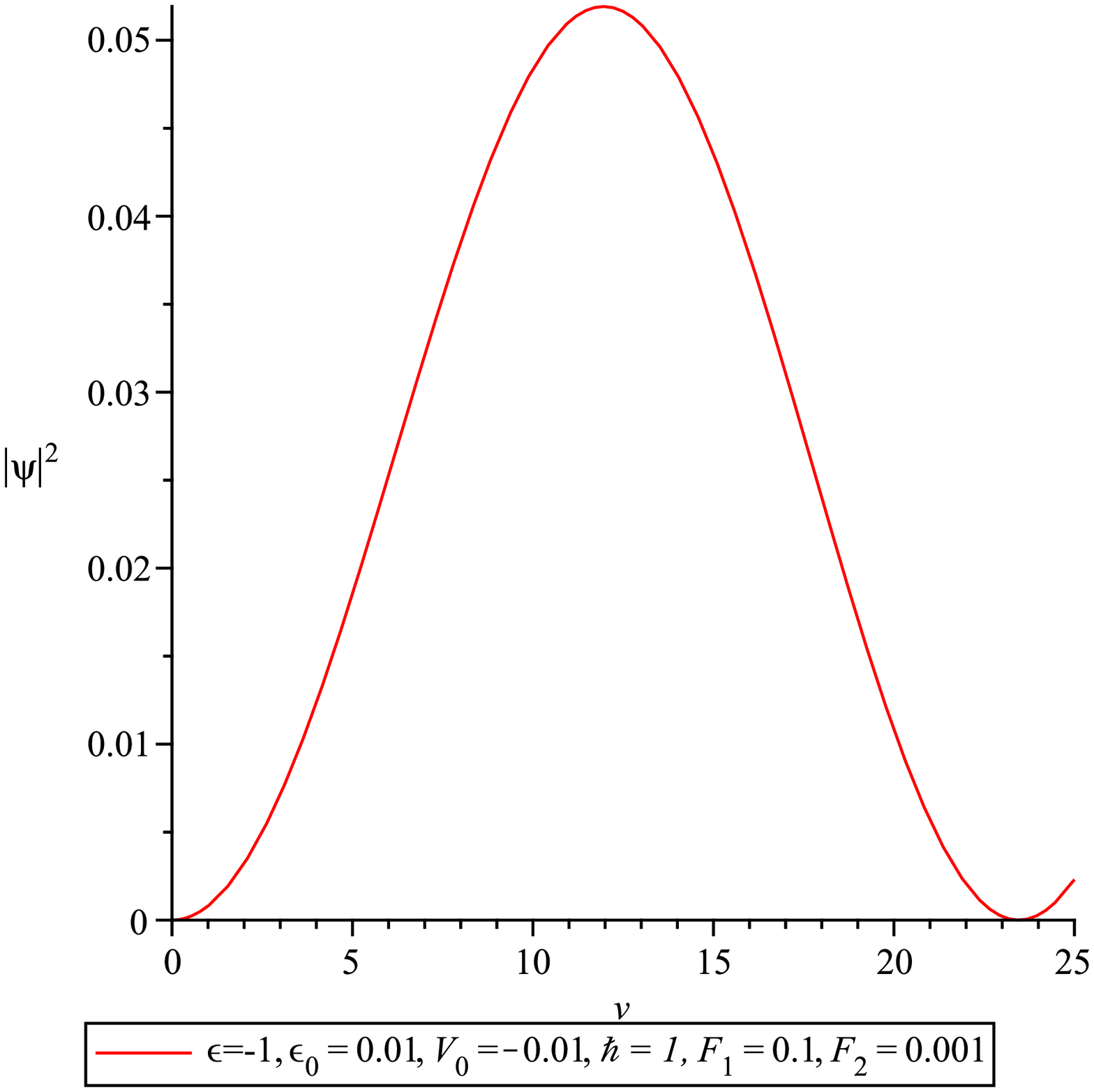}
	\end{minipage}
\end{figure}
\begin{figure}[h!]
	\begin{minipage}{0.4\textwidth}
		\centering\includegraphics[height=6cm,width=6cm]{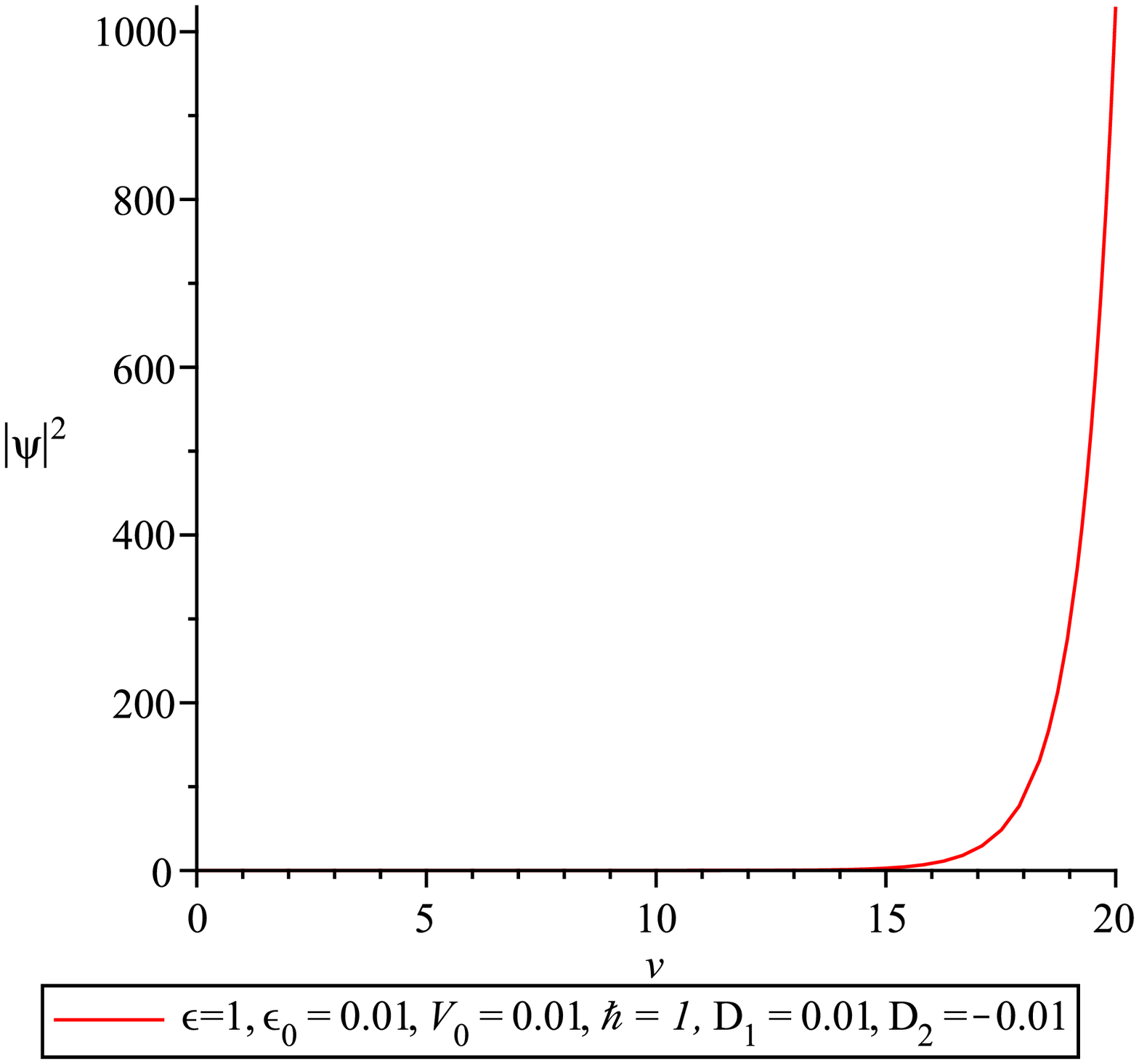}
	\end{minipage}\hfill
	\begin{minipage}{0.4\textwidth}
		\centering\includegraphics[height=6cm,width=6cm]{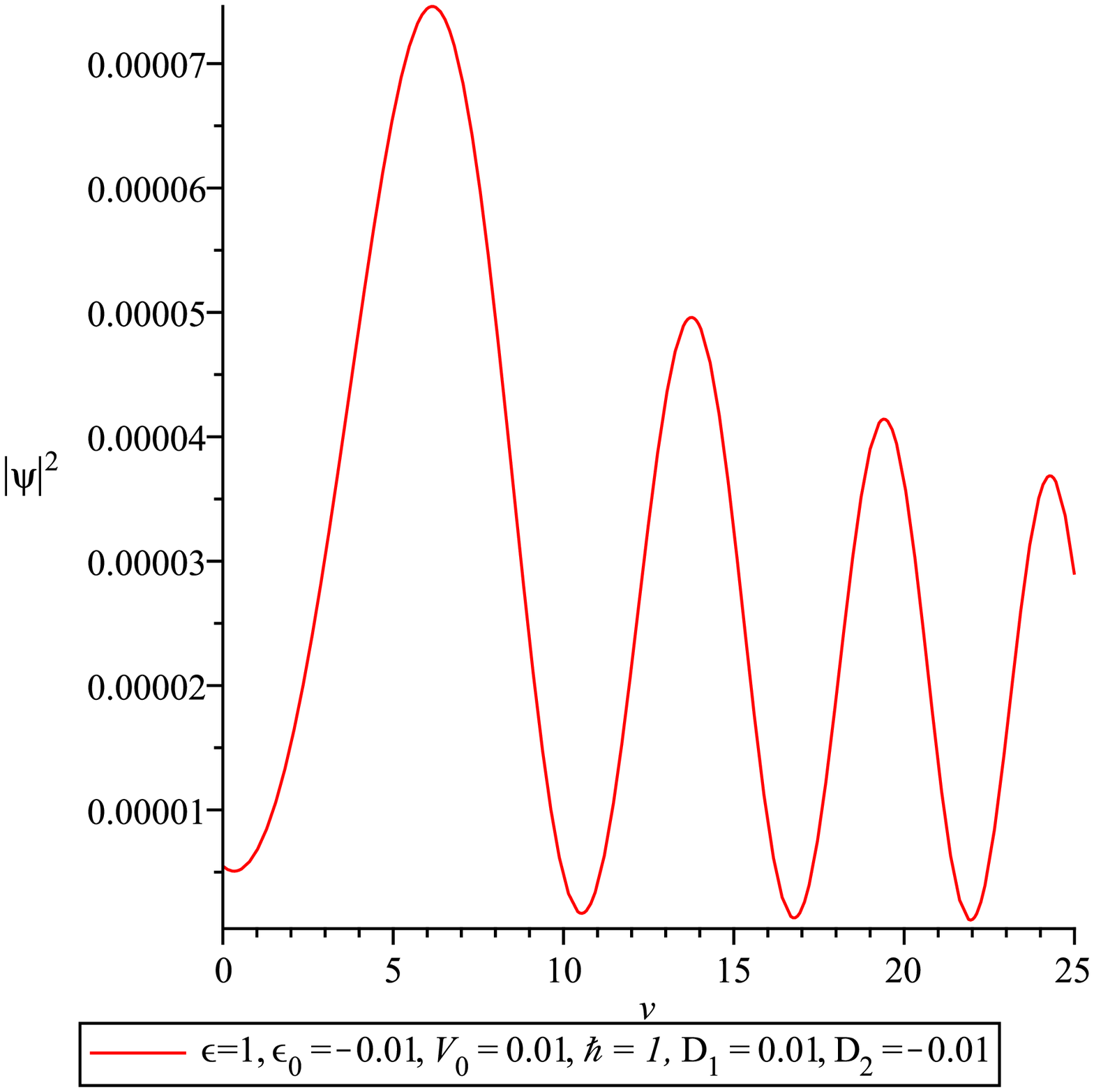}
	\end{minipage}
\end{figure}
\begin{figure}[h!]
	\begin{minipage}{0.4\textwidth}
		\centering\includegraphics[height=6cm,width=6cm]{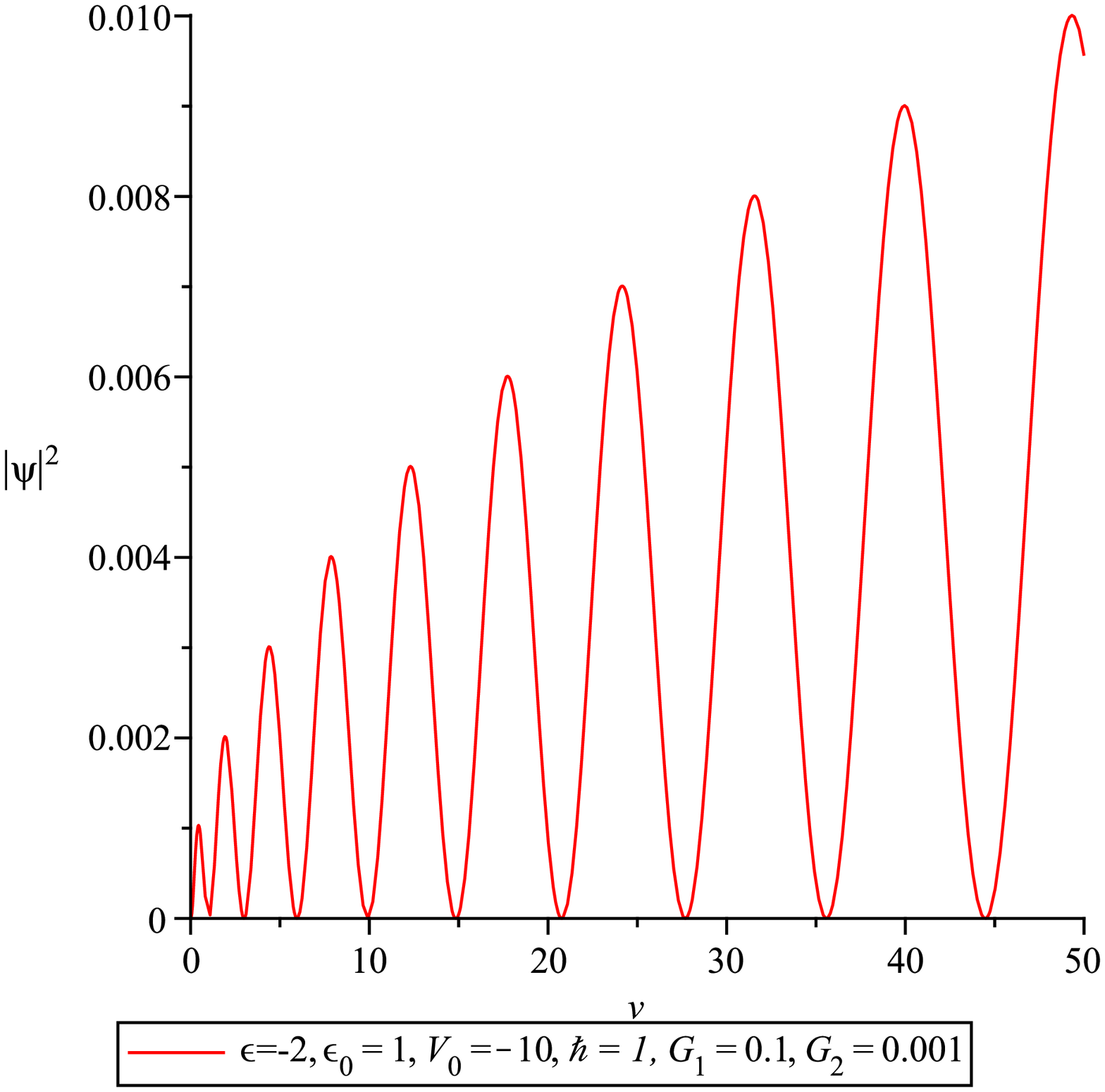}
	\end{minipage}\hfill
	\begin{minipage}{0.4\textwidth}
		\centering\includegraphics[height=6cm,width=6cm]{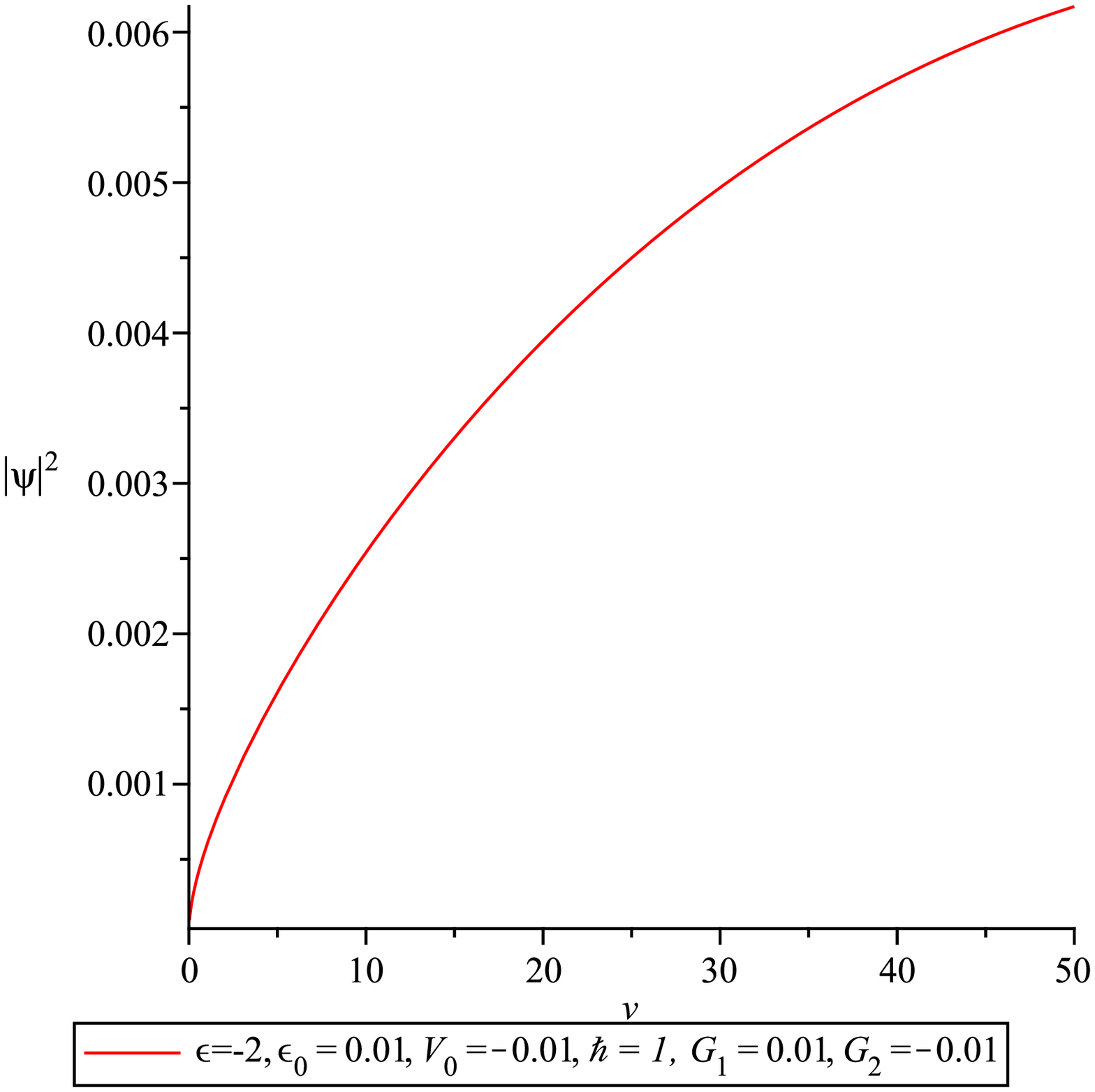}
	\end{minipage}
\end{figure}
\section{Quantum Bohmian Trajectories}
In metric formulation of Einstein gravity there are four constraints: the scalar constraint (or Hamiltonian constraint) and the vector constraint (or super momentum constraint). However, for homogeneous and isotropic minisuperspace models the vector constraint vanishes identically and the quantum version of the scalar constraint equation is termed as Wheeler Dewitt equation
\begin{equation}\label{eq83}
	\tilde{H}[\tilde{q_{\mu}(t)}, \tilde{p^{\mu}(t)}]\Psi(q_{\mu})=0.
\end{equation}
Here the arguments $q_{\mu}(t)$, $p^{\mu}(t)$ are the homogeneous degrees of freedom due to the three metric $\eta_{\mu\nu}$ and $\Pi^{\mu\nu}$, the conjugate momenta. $\Psi(q_{\mu})$ is the wave function of the universe. In the present context, the Lagrangian, the Hamiltonian (both classical and operator version) are given by (\ref{eq51*}), (\ref{eq57*}) and (\ref{eq62}) respectively. The expression for the conjugate momenta is given by (\ref{eq56*}). Therefore, equation (\ref{eq83}) leads to the WD equation as
\begin{equation}\label{eq84}
	\dfrac{d^{2}\Psi(\Lambda)}{d\Lambda^{2}}=\dfrac{2}{\hbar^{2}}~\Lambda^{2\left(\frac{1}{r}-1\right)} ~ V[\Lambda]\Psi(\Lambda).
\end{equation}
For semi-classical limit with WKB approximation the wave function takes the form
\begin{equation}\label{eq85}
	\Psi=exp\left(\frac{i}{\hbar}S\right),
\end{equation} 
where the classical H-J function $S$ has the expansion in $\hbar$ as
\begin{equation}\label{eq86}
S=S_{0}+\hbar S_{1}+\hbar^{2}S_{2}+...~~~~.
\end{equation}
As a result, the wave packet 
\begin{equation}
	\Psi=\int\mu(\alpha)~exp\left(\dfrac{S_{0}}{\hbar}\right)~d\alpha,
\end{equation}
characterizes the classical solution with $\alpha$, an arbitrary separation parameter and $\mu$ has Gaussian distribution having finite mean and standard deviation.
Now using (\ref{eq85}) and (\ref{eq86}) in the WD equation (\ref{eq84}), the zeroth order in $\hbar$ gives the differential equation for $S_{0}$ as 
\begin{equation}
	\dfrac{-1}{2~\Lambda^{2(\frac{1}{r}-1)}}~\left(\dfrac{dS_{0}}{d\Lambda}\right)^{2} + V[\Lambda]=0,
\end{equation} i.e
\begin{equation}
	S_{0}=\pm \sqrt{2} \int \Lambda^{(\frac{1}{r}-1)}~\sqrt{V[\Lambda]}~d\Lambda.
\end{equation}
Now to obtain the quantum Bohmian trajectories (for causal interpretation) one may choose the ansatz for the wave function
\begin{equation}\label{eq90*}
	\Psi(\Lambda)= A(\Lambda)~exp\left(\frac{i}{\hbar}~S(\Lambda)\right).
	\end{equation}
Using this ansatz into the WD- equation (\ref{eq84}) one gets the Hamilton-Jacobi equation as 
\begin{equation}\label{eq91}
		\dfrac{-1}{2~\Lambda^{2(\frac{1}{r}-1)}}~\left(\dfrac{dS}{d\Lambda}\right)^{2} + V_{Q} + V[\Lambda]=0,
\end{equation}
 where $V_{Q}$, the quantum potential has the expression as
 \begin{equation}
 	V_{Q}=\dfrac{1}{2 A(\Lambda) \Lambda^{2(\frac{1}{r}-1)}} \dfrac{d^{2}A(\Lambda)}{d\Lambda^{2}}.
 \end{equation}
Thus the Hamilton-Jacobi function $S$ is given by 
\begin{equation}\label{eq93}
	S=s_{0} \pm \int\left(\dfrac{1}{A(\Lambda)}\dfrac{d^{2}A(\Lambda)}{d\Lambda^{2}}+ 2\Lambda^{2(\frac{1}{r}-1)}\right)^{\frac{1}{2}}~d\Lambda.
\end{equation} $s_{0}$ is the constant of integration.\\
It may be noted that the trajectories $\Lambda(t)$ due to causal interpretation should be real, independent of any observation and are classified by the above H-J equation (\ref{eq91}). In fact, the quantum trajectories i.e the Bohmian trajectories are first order differential equations characterized by the equivalence of the usual definition of momentum with that from the Hamilton-Jacobi function $S$ as 
\begin{equation}
	\dfrac{dS(\Lambda)}{d\Lambda}=-2 \Lambda^{2(\frac{1}{r}-1)} \Lambda^{'}
\end{equation}i.e
\begin{equation}
	2\Lambda^{2(\frac{1}{r}-1)}\Lambda^{'}=\mp \left(\dfrac{1}{A(\Lambda)}\dfrac{d^{2}A(\Lambda)}{d\Lambda^{2}}+2\Lambda^{2(\frac{1}{r}-1)}\right)^{\frac{1}{2}}
\end{equation} or,
\begin{equation}\label{eq96}
2~\int \dfrac{\Lambda^{2(\frac{1}{r}-1)}~d\Lambda}{\left(\dfrac{1}{A(\Lambda)}~\dfrac{d^{2}A(\Lambda)}{d\Lambda^{2}}+2\Lambda^{2(\frac{1}{r}-1)}\right)^{\frac{1}{2}}}=\mp (t-t_{0}).
\end{equation}
In particular if $A(\Lambda)=A_{0}$, a constant then from (\ref{eq93}) one has,
\begin{equation}
	S=s_{0} \pm \sqrt{2}~\int \Lambda^{(\frac{1}{r}-1)}~d\Lambda,
\end{equation} or
\begin{equation}
	S=s_{0}\pm \sqrt{2}~r~\Lambda^{\frac{1}{r}}
\end{equation} and the quantum trajectory is described as,
\begin{equation}
	\sqrt{2}~r~\Lambda^{\frac{1}{r}}=\pm (t-t_{0}).
\end{equation}
Here the quantum potential is zero and the H-J equation (\ref{eq91}) coincides with the classical one. Thus Bohmian trajectory corresponds to classical power law form of expansion and it can't avoid the initial big-bang singularity. \\
Now we study the nature of the trajectories for non-zero quantum potential. Therefore, one may choose $A(\Lambda)=\Lambda^{N}$, $N\neq0, 1$ and substituting this in equation (\ref{eq96}) one gets the quantum trajectories as,
\begin{equation}
	\Lambda^{\frac{2}{r}}=\dfrac{(t-t_{0})^{2}}{2r^{2}}-\dfrac{N(N-1)}{2}.
	\end{equation}
Hence for $N\in (0,1)$ volume is non-zero as $t\rightarrow t_{0}$. Therefore initial big-bang singularity is avoided with non-zero quantum potential and proper fractional power law choice of $A(\Lambda)$.
\section{Concluding Remarks}
An investigation of modified gravity theory namely $f(T)$ gravity has been done in the previous sections by analyzing the RE both classically and quantum mechanically. At first, a general formulation of RE has been performed for arbitrary $f(T)$. Subsequently, the existence of singularity has been examined by studying the CC in the modified gravity model with two specified choices for $f(T)$. The Raychaudhuri scalar has been plotted graphically in FIG. \ref{f1} and FIG. \ref{f2}  for various choices of the parameters involved. For Model 1, it is possible to avoid the classical singularity although SEC is satisfied while for Model 2, avoidance of singularity is possible by violation of SEC or the matter field should be ghost in nature. By choosing $\Lambda=\sqrt{\eta}$ ($\eta$ is the metric scalar of the space-like hyper-surface) it is possible to write the RE as a second order differential equation and a Lagrangian (hence a Hamiltonian) formulation can be done. Finally, quantum cosmology has been furnished and the wave function of the universe has been found to be the solution of the one dimensional time independent Schrödinger equation associating the energy eigen function to the wave function of the universe. From probabilistic description it is found that the initial big-bang singularity may be avoided for both the models with proper choice of the potential corresponding to the dynamical system representing the congruence. Further for causal interpretation quantum trajectories have been explicitly determined for the two cases--- one with vanishing quantum potential and the trajectories have been found to pass through the big-bang singularity, while the other case corresponding to non-zero quantum potential shows that big-bang singularity may be avoided with proper choice of the pre-factor $A(\Lambda)$ in the ansatz for the wave function as power law form. Therefore in the present work corresponding to the RE in $f(T)$ gravity, both classical and quantum description has been furnished and the initial big-bang singularity has been examined both at the classical as well as at the quantum level. Classically the avoidance of singularity is more realistic in model 1 even with normal fluid as the matter content of the universe but this is not possible for model 2. For quantum description the avoidance of the big-bang singularity is obtained either by probabilistic description or by examining Bohmian trajectories.
\section*{Acknowledgment}
The authors thank the anonymous referees whose comments and suggestions improved the quality of
	the paper. M.C thanks University Grant Commission (UGC) for providing the Junior Research Fellowship (ID:211610035684/JOINT CSIR-UGC NET JUNE-2021).
	

\begin{thebibliography}{100}
		\bibitem{WMAP:2003elm}
		D.~N.~Spergel \textit{et al.} [WMAP],
		Astrophys. J. Suppl. \textbf{148}, 175-194 (2003)
		\bibitem{SupernovaSearchTeam:1998fmf}
		A.~G.~Riess \textit{et al.} [Supernova Search Team],
		Astron. J. \textbf{116}, 1009-1038 (1998)
		\bibitem{SupernovaCosmologyProject:1998vns}
		S.~Perlmutter \textit{et al.} [Supernova Cosmology Project],
		Astrophys. J. \textbf{517}, 565-586 (1999)
		\bibitem{SDSS:2003eyi}
		M.~Tegmark \textit{et al.} [SDSS],
		Phys. Rev. D \textbf{69}, 103501 (2004)
		\bibitem{SDSS:2005xqv}
		D.~J.~Eisenstein \textit{et al.} [SDSS],
		Astrophys. J. \textbf{633}, 560-574 (2005)
		\bibitem{Amendola:2015ksp}
		L.~Amendola and S.~Tsujikawa,
		Cambridge University Press, 2015
		\bibitem{Capozziello:2019cav}
		S.~Capozziello, R.~D'Agostino and O.~Luongo,
		Int. J. Mod. Phys. D \textbf{28}, no.10, 1930016 (2019)
		\bibitem{Nojiri:2006ri}
		S.~Nojiri and S.~D.~Odintsov,
		eConf \textbf{C0602061}, 06 (2006)
		\bibitem{Nobbenhuis:2006yf}
		S.~Nobbenhuis,
		[arXiv:gr-qc/0609011 [gr-qc]].
		\bibitem{Yang:2009ae}
		R.~J.~Yang and S.~N.~Zhang,
		Mon. Not. Roy. Astron. Soc. \textbf{407}, 1835-1841 (2010)
		\bibitem{Guo:2013swa}
		J.~Q.~Guo and A.~V.~Frolov,
		Phys. Rev. D \textbf{88}, no.12, 124036 (2013)
		\bibitem{Aditya:2018cmn}
		Y.~Aditya and D.~R.~K.~Reddy,
		Int. J. Geom. Meth. Mod. Phys. \textbf{15}, no.09, 1850156 (2018)
		\bibitem{Cognola:2007zu}
		G.~Cognola, E.~Elizalde, S.~Nojiri, S.~D.~Odintsov, L.~Sebastiani and S.~Zerbini,
		Phys. Rev. D \textbf{77}, 046009 (2008)
		\bibitem{Elizalde:2010ts}
		E.~Elizalde, S.~Nojiri, S.~D.~Odintsov, L.~Sebastiani and S.~Zerbini,
		Phys. Rev. D \textbf{83}, 086006 (2011)
		\bibitem{Nojiri:2010wj}
		S.~Nojiri and S.~D.~Odintsov,
		Phys. Rept. \textbf{505}, 59-144 (2011)
		\bibitem{Sotiriou:2008rp}
		T.~P.~Sotiriou and V.~Faraoni,
		Rev. Mod. Phys. \textbf{82}, 451-497 (2010)
		\bibitem{Sotiriou:2006mu}
		T.~P.~Sotiriou and S.~Liberati,
		J. Phys. Conf. Ser. \textbf{68}, 012022 (2007)
		\bibitem{Gogoi:2021mhi}
		D.~J.~Gogoi and U.~D.~Goswami,
		Int. J. Mod. Phys. D \textbf{31}, no.06, 2250048 (2022)
		\bibitem{Sotiriou:2005cd}
		T.~P.~Sotiriou,
		Class. Quant. Grav. \textbf{23}, 1253-1267 (2006)
		\bibitem{Cai:2015emx}
		Y.~F.~Cai, S.~Capozziello, M.~De Laurentis and E.~N.~Saridakis,
		Rept. Prog. Phys. \textbf{79}, no.10, 106901 (2016)
		\bibitem{Unzicker:2005in}
		A.~Unzicker and T.~Case,
		[arXiv:physics/0503046 [physics]].
		\bibitem{Darabi:2019qpz}
		F.~Darabi and K.~Atazadeh,
		Phys. Rev. D \textbf{100}, no.2, 023546 (2019)
		\bibitem{Li:2018ixg}
		C.~Li, Y.~Cai, Y.~F.~Cai and E.~N.~Saridakis,
		JCAP \textbf{10}, 001 (2018)
		\bibitem{Golovnev:2018wbh}
		A.~Golovnev and T.~Koivisto,
		JCAP \textbf{11}, 012 (2018)
		\bibitem{Aviles:2013nga}
		A.~Aviles, A.~Bravetti, S.~Capozziello and O.~Luongo,
		Phys. Rev. D \textbf{87}, no.6, 064025 (2013)
		\bibitem{Bose:2020xdz}
		A.~Bose and S.~Chakraborty,
		Mod. Phys. Lett. A \textbf{35}, no.36, 2050296 (2020)
		\bibitem{Linder:2010py}
		E.~V.~Linder,
		Phys. Rev. D \textbf{81}, 127301 (2010)
		\bibitem{Hayashi:1979qx}
		K.~Hayashi and T.~Shirafuji,
		Phys. Rev. D \textbf{19}, 3524-3553 (1979)
		\bibitem{LIGOScientific:2017vwq}
		B.~P.~Abbott \textit{et al.} [LIGO Scientific and Virgo],
		Phys. Rev. Lett. \textbf{119}, no.16, 161101 (2017)
			\bibitem{Wald:1984}
		R.~M.~Wald,
		``General Relativity ,''
		Chicago University Press, 1984
			\bibitem{Weinberg:1972kfs}
		S.~Weinberg,
		John Wiley and Sons, 1972
			\bibitem{Hawking:1973uf}
		S.~W.~Hawking and G.~F.~R.~Ellis,
		Cambridge University Press, 2011,
		\bibitem{Penrose:1964wq}
		R.~Penrose,
		Phys. Rev. Lett. \textbf{14}, 57-59 (1965)
		
		\bibitem{Hawking:1970zqf}
		S.~W.~Hawking and R.~Penrose,
		Proc. Roy. Soc. Lond. A \textbf{314}, 529-548 (1970)
			\bibitem{Raychaudhuri:1953yv}
		A.~Raychaudhuri,
		Phys. Rev. \textbf{98}, 1123-1126 (1955)
		\bibitem{Burger:2018hpz}
		D.~J.~Burger, N.~Moynihan, S.~Das, S.~Shajidul Haque and B.~Underwood,
		Phys. Rev. D \textbf{98}, no.2, 024006 (2018)
		
		\bibitem{Kar:2006ms}
		S.~Kar and S.~SenGupta,
		Pramana \textbf{69}, 49 (2007)
		\bibitem{Ehlers:2006aa}
		J.~Ehlers,
		Int. J. Mod. Phys. D \textbf{15}, 1573-1580 (2006).
		\bibitem{Kar:2008zz}
		S.~Kar,
		Resonance J. Sci. Educ. \textbf{13}, 319-333 (2008).
		
		\bibitem{Horwitz:2021lyc}
		L.~P.~Horwitz, V.~S.~Namboothiri, G.~Varma K, A.~Yahalom, Y.~Strauss and J.~Levitan,
		``Raychaudhuri Equation, Geometrical Flows and Geometrical Entropy,''
		Symmetry \textbf{13}, no.6, 957 (2021)
		
		\bibitem{Dadhich:2005qr}
		N.~Dadhich,
		``Derivation of the Raychaudhuri equation,''18 Nov (2022)
	\bibitem{Chakraborty:2023ork}
	M.~Chakraborty, A.~Bose and S.~Chakraborty,
	Phys. Scripta \textbf{98}, no.2, 025007 (2023)
		\bibitem{Wu:2010xk}
		P.~Wu and H.~W.~Yu,
		Phys. Lett. B \textbf{692}, 176-179 (2010)
		\bibitem{Myrzakulov:2010vz}
		R.~Myrzakulov,
		Eur. Phys. J. C \textbf{71}, 1752 (2011)
		\bibitem{Chen:2010va}
		S.~H.~Chen, J.~B.~Dent, S.~Dutta and E.~N.~Saridakis,
		Phys. Rev. D \textbf{83}, 023508 (2011)
		\bibitem{Dvali:2003rk}
		G.~Dvali and M.~S.~Turner,
		[arXiv:astro-ph/0301510 [astro-ph]].
		\bibitem{Chung:1999zs}
		D.~J.~H.~Chung and K.~Freese,
		Phys. Rev. D \textbf{61}, 023511 (2000)
		\bibitem{Dvali:2000hr}
		G.~R.~Dvali, G.~Gabadadze and M.~Porrati,
		Phys. Lett. B \textbf{485}, 208-214 (2000)
		\bibitem{Alsaleh:2017ozf}
		S.~Alsaleh, L.~Alasfar, M.~Faizal and A.~F.~Ali,
		Int. J. Mod. Phys. A \textbf{33}, no.10, 1850052 (2018)
			\bibitem{Poisson:2009pwt}
		E.~Poisson,
		``A Relativist's Toolkit: The Mathematics of Black-Hole Mechanics,''
		Cambridge University Press, 2009
		\bibitem{Davis:1928}
		D.~R.~Davis,
		Trans. Amer. Math. Soc. \textbf{30} (1928), 710-736
		
		\bibitem{Davis:1929}
		D.~R.~Davis,
		Bull. Amer. Math. Soc. \textbf{35} (1929), 371-380
		
		\bibitem{Douglas:1941}
		J.~Douglas,
		Trans. Amer. Math. Soc. \textbf{50} (1941), 71-128
		
		\bibitem{Casetta:1941}
		L.~Casetta, C.~P.~Pesce
		Trans. Amer. Math. Soc. \textbf{50} (1941), 71-128
		
		\bibitem{Crampin:2010}
		M.~ Crampin, T.~ Mestdag and W.~ Sarlet
		Z. Angew. Math. Mech. \textbf{90} (2010), 502-508
		
		\bibitem{Nigam:2016}
		K.~ Nigam, K.~ Banerjee
		``A Brief Review of Helmholtz Conditions" 
		\bibitem{Wheeler:1968iap}
		J.~A.~Wheeler,
		Adv. Ser. Astrophys. Cosmol. \textbf{3}, 27-92 (1987)
		\bibitem{Jalalzadeh:2016gqs}
		S.~Jalalzadeh, T.~Rostami and P.~V.~Moniz,
		Int. J. Mod. Phys. D \textbf{25}, no.03, 1630009 (2016)
			\bibitem{Pinto-Neto:2013toa}
		N.~Pinto-Neto and J.~C.~Fabris,
		Class. Quant. Grav. \textbf{30}, 143001 (2013)
\bibitem{Halliwell:1989myn}
J.~J.~Halliwell,
``INTRODUCTORY LECTURES ON QUANTUM COSMOLOGY,''
\bibitem{Pal:2014xsa}
S.~Pal and N.~Banerjee,
Phys. Rev. D \textbf{91}, no.4, 044042 (2015)
\bibitem{Alvarenga:2003kx}
F.~G.~Alvarenga, A.~B.~Batista, J.~C.~Fabris and S.~V.~B.~Goncalves,
Gen. Rel. Grav. \textbf{35}, 1659-1677 (2003)
\bibitem{Pal:2014dya}
S.~Pal and N.~Banerjee,
Phys. Rev. D \textbf{90}, no.10, 104001 (2014)
\bibitem{Pal:2016ysz}
S.~Pal and N.~Banerjee,
J. Math. Phys. \textbf{57}, no.12, 122502 (2016)
		\end{thebibliography}
\end{document}